%% file: main.tex
\documentclass[a4paper,twoside]{article}
\usepackage{epsfig}
\usepackage{subcaption, lipsum}
\usepackage{calc}
\usepackage{amssymb}
\usepackage{amstext}
\usepackage{amsthm}
\usepackage{multicol}
\usepackage{pslatex}
\usepackage{apalike}
\usepackage{algorithm2e}
\usepackage[bottom]{footmisc}
\usepackage{SCITEPRESS}     % Please add other packages that you may need BEFORE the SCITEPRESS.sty package.

\usepackage{graphicx} % Required for inserting images
\usepackage{bigints}

\usepackage{nicefrac}

\usepackage{capt-of}
\usepackage{dsfont}
\usepackage{amsmath,amssymb,amsfonts, comment,tikz}
\usepackage{inputenc, a4wide}
\usepackage{multirow}
\usepackage{mathptmx}
\usepackage{epsfig} % for postscript graphics files
\usepackage{bm}
\usepackage{graphicx}
\usepackage{textcomp}
\usepackage{xcolor}
\usepackage{hyperref}
\usepackage{cleveref}

\usepackage{graphicx} % Required for inserting images
\usepackage{bigints}

\usepackage{nicefrac}

\newcommand{\Rev}[1]{{\color{black}  #1}}
\usepackage{capt-of}
\usepackage{dsfont}
\usepackage{amssymb,amsfonts, comment,tikz}
\usepackage{inputenc, a4wide}
\usepackage{multirow}
\usepackage{mathptmx}

\usepackage{geometry}
\usepackage{epsfig} % for postscript graphics files
\usepackage{bm}
\usepackage{graphicx, nicefrac}
\usepackage{textcomp}
\usepackage{xcolor}
\usepackage{hyperref}
\usepackage{cleveref}

\usepackage{enumerate}
\usepackage{ifthen}
\newcommand{\Canticipate}[1]{}
\newcommand{\ignore}[1]{}
\newcommand{\ENMremoved}[1]{}
\newcommand{\Details}[1]{}
\newcommand{\extra}[1]{}
\newcommand{\jlt}[1]{}%Imp material, must be included, just to get estimate of paper
\newcommand{\TR}[2]{#2}

\renewcommand{\nu}{w}

\newcommand{\y}{{\bf y}}

\usepackage{graphicx,nicefrac} % Required for inserting images

\usepackage{array,booktabs}

\newtheorem{lemma}{Lemma}
\newtheorem{theorem}{Theorem}
\newtheorem{cor}{Corollary}
\newtheorem{definition}{Definition}
\newcommand{\eop}{{\hfill~$\Box$}}
\newcommand{\tus}[1]{{\color{red}{#1}}}

\renewcommand{\a}{{\bf a}}
\newcommand{\C}{{\mathbb C}}

  \renewcommand{\H}{\mathbb{H}} 
  
\newcommand{\U}{{\mathcal U}}
\newcommand{\N}{{\mathcal N}}

\renewcommand{\S}{{\mathcal S}}

\newcommand{\A}{{\mathbb A}}
\renewcommand{\P}{{\mathbb P}}
\newcommand{\M}{{\mathbb M}}
\newcommand{\G}{{\mathbb G}}
\newcommand{\V}{{\mathbb V}}
\renewcommand{\S}{{\mathbb S}}

\newcommand{\Aset}{{\cal A}}

\usepackage{pslatex}
\usepackage{apalike}
\usepackage{algorithm2e}
\usepackage[bottom]{footmisc}
\usepackage{SCITEPRESS} 

\newcommand{\Mo}{{{\mathbb M}_1}}
\newcommand{\Mt}{{{\mathbb M}_2}}
\newcommand{\Mi}{{{\mathbb M}_i}}

\newcommand{\dbar}{\bar{d}}

\newcommand{\indc}[1]{\mathds{1}_{\left\{ #1 \right \}} }

\newcommand{\BR}{{\mathbb B}}

\newcommand{\zero}{0_n}
\renewcommand{\F}{{\cal F}}

\newcommand{\sG}{{\mbox{\fontsize{5.2}{5.2}\selectfont{$\G$}}}}
\newcommand{\sC}{{\mbox{\fontsize{5.2}{5.2}\selectfont{$\C$}}}}
\newcommand{\sM}{{\mbox{\fontsize{4.7}{5}\selectfont{$\M$}}}}
\newcommand{\sMi}{{\mbox{\fontsize{4.7}{5}\selectfont{$\Mi$}}}}
\newcommand{\sMminusi}{{\mbox{\fontsize{4.7}{5}\selectfont{${\mathbb M}_{-i}$}}}}
\newcommand{\sS}{{\mbox{\fontsize{5.2}{5.2}\selectfont{$\S$}}}}

\newcommand{\sMo}{{\mbox{\fontsize{5.2}{5.2}\selectfont{$\Mo$}}}}
\newcommand{\sMt}{{\mbox{\fontsize{5.2}{5.2}\selectfont{$\Mt$}}}}

\newcommand{\sV}{{\mbox{\fontsize{5.2}{5.2}\selectfont{$\V$}}}}
\newcommand{\sH}{{\mbox{\fontsize{5.2}{5.2}\selectfont{$\H$}}}}
\newcommand{\sA}{{\mbox{\fontsize{5.2}{5.2}\selectfont{$\A$}}}}

\renewcommand{\C}{{\mathbb C}}
\newcommand{\x}{{\bf x}}
\newcommand{\bst}{{I}}

\newcommand{\GC}{{\P_{\sG}}}
\newcommand{\CHC}{{\P_{\sH}}}
\newcommand{\VC}{{\P_{\sV}}}
\newcommand{\ALC}{{\P_{\sA}}}

 \newcommand{\pCHC}{\CHC}%{{\P_{\hspace{-0.45mm}{\mbox{\fontsize{3.5}{3.5}\selectfont{$\H$}}}}}}

 \newcommand{\pGC}{\GC}
 \newcommand{\pVC}{\VC}%{{\P_{\hspace{-0.45mm}{\mbox{\fontsize{3.5}{3.5}\selectfont{$\V$}}}}}}

\newcommand{\pALC}{\ALC} %{{\P_{\hspace{-0.45mm}{\mbox{\fontsize{3.5}{3.5}\selectfont{$\A$}}}}}}
\newcommand{\RB}[1]{}

\begin{document}
\title{Partition-form Cooperative Games  
in Two-Echelon Supply Chains}

\author{
\authorname{Gurkirat Wadhwa,  Tushar Shankar  Walunj and Veeraruna Kavitha}
\affiliation{IEOR,  Indian Institute of Technology,  Bombay, India}
\email{\{gurkirat, tusharwalunj, vkavitha\}@iitb.ac.in}
}


\abstract{Competition and cooperation are inherent features of any multi-echelon supply chain. The  interactions among the agents across the same echelon and that across various echelons influence the percolation of market demand across echelons. The agents may want to collaborate with others in pursuit of attracting higher demand and thereby improving their own revenue. We consider one supplier (at a higher echelon) and two manufacturers (at a lower echelon and facing the customers) and study the collaborations that are `stable'; the main differentiator from the existing studies in supply chain literature is the consideration of the following crucial aspect -- the revenue of any collaborative unit also depends upon the way the opponents collaborate. Such competitive scenarios can be modeled using what is known as partition form games.\\
Our study reveals that the grand coalition is not stable when the product is essential and the customers buy it from any of the manufacturers without a preference. The supplier prefers to collaborate with only one manufacturer, the one stronger in terms of market power; further, such collaboration is stable only when the stronger manufacturer is significantly stronger. Interestingly, no stable collaborative arrangements exist when the two manufacturers are nearly equal in market power.} 
  \maketitle \normalsize
  
\setcounter{footnote}{0} \vfill

\input{Introduction}

%\onecolumn

\ignore{ 

\section*{Goals with two manufacturers} 
\begin{itemize}
    \item  Theoretical results near equal $\alpha$'s,  with $\gamma$ close to one and $\varepsilon$ either near 0 or 1   ---    
    Conjecture ---  Either GC or VC or ALC are stable at these asymptotic 

    \item Numerical representation of other results 
\end{itemize}

\section*{Notations} 
Agents are represented by $M_i$ 
 $S$ like that, while coalitions (basically sets) are represented by mathbb letters, e.g., 
\begin{eqnarray*}
 \Mi &=& \{ M_i\} \mbox{ (manufactuerer $M_i$ operating alone ) }, \M = \{M_1, \cdots,  M_n\} \mbox{ (set of all manufacturers) }, \\
 \S &=& \{S\} \mbox{ set of the supplier operating alone },  \\
 \C &=& \{S, M_{i_1}, \cdots, M_{i_k}\} \mbox{ (supplier and a subset of manufacturers operating together )} \\
 \G &=& \M \cup \S = \{S, M_1, \cdots, M_n\} \mbox{ ( grand coalition )}
\end{eqnarray*}
} % end ignore
\ignore{ %end ignore at 434
\section{Problem description and Preliminaries} \label{sec:Prob_des}

%In this section we describe our model for the two echelon supply chain consisting of suppliers and manufacturers and state some preliminary results.

We examine the interplay between cooperation and competition within a two-echelon supply chain consisting of many manufacturers at the lower echelon and one supplier at the upper one. The manufacturers directly supply the final product to the customers and obtain the required raw material from the supplier. The customers purchase the product from one (or none) of the manufactures based on the prices quoted, reputation of the agents involved, importance of the product etc. 
Similarly the manufacturers select to procure the raw material from the supplier at the quoted price and operate or can select not to operate.

Any manufacturer  can interact with the supplier for procurement of  the   raw material or can chose to collaborate completely with  the supplier. In the later case, all the collaborating agents  share amicably the total profit earned by all the members of the coalition, while, in the former case a per-unit price is agreed upon   for the chosen service.  The manufacturers can also collaborate  among themselves, or a subset of manufacturers and the supplier can collaborate.   

Thus any agent has a choice to collaborate with some or none of the other agents, or select an agent for some particular service   (supply/procurement of raw material), or quote an appropriate price or finally choose not to operate. 
As a resultant of all the choices made by various agents of the system, any given agent derives some utility/revenue. 
The agents are selfish and aim  to maximize their individual rewards which drives  their choices including the collaboration attempts.

The aim of this paper is to study the `stable operating configurations' (sets of agents operating together and the individual revenues derived by them) that emerge at some appropriate equilibrium and the prices and other selection choices made by the agents at that equilibrium. Towards this goal, we begin with detailed description of the problem.

\vspace{-2mm}
\subsection{Partitions and coalitions}
\vspace{-2mm}
All the agents or a  subset of them can operate together by forming coalitions. One may have more than one coalition in the system. 
Basically  the agents within a coalition   operate together and face the competition from other coalitions/agents. Any partition,  say~$\P = \{\C_1, \cdots, \C_k \}$, represents the operating arrangement of agents into distinct coalitions and satisfies the following:
$$\cup_m \C_m = \{M_1, \cdots, M_n, S\} \mbox{ and } \C_m \cap \C_l = \emptyset \mbox{  when } m \ne l.$$ 
The aim of this paper is to study the interactions between such coalitions and the emergence of stable   partition(s) (if there exists one); prior to that one needs to understand   the criterion that can be used declare  a partition stable (this will be considered in section \ref{sec_coalition_formation_game}). For now, we discuss few important and interesting partitions/coalitions.  When two or more   manufacturers operate together, we have a coalition with horizontal cooperation at lower echelon, e.g., $\C = \{M_{i_1}, \cdots, M_{i_k}\}$;  when the supplier and a manufacturer operate together we have  a coalition with vertical cooperation~$\C = \{M_i, S\}$;  when a manufacturer operates alone we have a coalition with single player~$\Mi = \{M_i\}$; when all the agents operate together we have grand coalition (GC) represented by $\G = \{S, M_1, \cdots, M_n\}$; on the other hand when all agents operate alone we have a partition with all alone coalitions (ALC),  $\ALC = \{\{S\}, \{M_1\}, \cdots, \{M_n\} \}$ and so on. 
At this point we would like to mention that when agents of the same nature operate together (e.g., two or more manufacturers), the coalition chooses the best candidate for any given aspect (e.g,, influence, reputation, capacity etc.). 
\vspace{-2mm}
\subsection{Market Segmentation}
\label{sec_market_seg}
\vspace{-2mm}
In an SC, the manufacturers satisfy the demands of (interested) customers  and   rely on the supplier for raw materials or {\color{red}intermediate products}. The customers choose one among the  manufacturers (or none) depending  upon the prices quoted by them, their reputation factors, loyalty etc. The segmentation of the demand is further influenced by the essentialness of the product, which we represent using a parameter~$\gamma$ and a cross-linking factor $\varepsilon$ that is linked to essentialness as well as the affinity to switch their loyalties. 
Thus when the manufacturers are not operating together, the   market   (of the customers) is segmented between them  depending  upon  the selling prices~$(p_1,p_2)$,    and essentialness factors~$(\gamma, \varepsilon)$ as below%
(see~\cite{Sustainability_paper} for similar models and~$x^+ = \max\{0, x\}$):
\begin{eqnarray}
\label{Eqn_Demand_Function}
 D_{\sMi}  =    \bigg ( {\dbar}_{\sMi}-\tilde\alpha_\sMi(1-\gamma)p_{\sM_i}+\sum_{j \ne i} \varepsilon_{\sMi,\sM_j} \tilde\alpha_{\sM_j} (1-\gamma)p_{\sM_j} \bigg )^+  \mbox{ for each } i,  \mbox{ with } \alpha_{\sM_j} :=  \tilde\alpha_{\sM_j} (1-\gamma) 
\end{eqnarray}
where~$\dbar_{\sMi}$ is the maximum possible market demand that can be attracted by~$M_i$,  the fraction $\tilde\alpha_\sMi(1-\gamma)p_{\sMi}$ is lost by manufacture $M_i$ because of its price $p_\sMi$  with  (essentialness-dependent) sensitivity parameter $\alpha_{\sM_j} := \tilde\alpha_\sMi(1-\gamma)$ and  the fraction $\varepsilon_{\sMi,\sM_j} (1-\gamma){\tilde \alpha}_{\sM_j}p_{\sM_j}$ is  the customer base of the opponent $M_{j}$ that is attracted by $M_i$ (depends upon the cross linking factor $\varepsilon_{\sMi,\sM_j}$) --- recall the last  fraction is resultant of two aspects,   repulsion by the price of opponent $p_{j}$  and  the essentialness of the product, both of which are jointly captured by $\varepsilon_{i,j}$.  

Observe here that when~$\gamma = 1$   the product is completely essential and     each customer buys the product from one of the manufacturers, in particular $d_\sMi$ is the demand attracted by $M_i$ -- it is  in this regard that one can interpret $\dbar_\sMi$ as the dedicated customer base  of $M_i$.  Further  the sum of demands attracted by all the manufacturers equals the total market size when the product is essential.

{\bf One example:} Before we proceed, we present one example scenario that can be modeled by \eqref{Eqn_Demand_Function}. 
Consider a scenario with two manufacturers  where the manufacturers have dedicated customer bases, say~$d_1$ and~$d_2$ respectively, and with a common pool~$2d_c$. Thus the total market size available to either of the manufacturers equals,~$d_\sM = d_1+d_2+2d_c$.  The customers among dedicated fraction~$d_i$    buy products only from~$M_i$, and that too only if they are comfortable with price~$p_i$ quoted by~$M_i$ -- as in \cite{} one can model this fraction by~$d_i - \alpha_{i} p_i$; recall here that  $\alpha_i = (1-\gamma) {\tilde \alpha}_i$  (as explained previously) to model essentialness via factor $\gamma$. The customers among common pool~$d_c$ can buy either from~$M_1$ or~$M_2$ based on prices, essentialness factor $\gamma$ and the cross-linking factor $\varepsilon$:  as in \cite{} one can model the proportion for manufacturer $M_i$ using ~$d_{c_i} =  ( d_c - \alpha_{i} p_i +  \varepsilon  \alpha_{{-i}} p_{-i})$. 
 Then demand attracted by manufacturer 1 equals (\cite{Sustainability_paper}):
$$
\dbar (p_i, p_{-i}) = 
 d_i - \alpha_{i} p_i +  (d_c - \alpha_{i} p_i +  \varepsilon  \alpha_{-i} p_{-i}) = \dbar_{\sMi} - 2 \alpha_{i}  p_i  + \epsilon \alpha_{-i} p_{-i}, \mbox{ with }  d_\sMi = d_i + d_c.
$$which is of the form as in  \eqref{Eqn_Demand_Function}.
In this case,   the total Market size that can be attracted by joint coalition~$\M$ equals~$\dbar_\sM = d_1+d_2 + 2d_c$.  
Basically the common pool definitely buys a product either when the cross-linking is good ($\varepsilon \approx 1$) or when the essentialness is high (when $\gamma \approx 1$ in fact here $d_{c_i} = d_c$ and the common pool is equally divided). In this example,  in some sense the cross-linking factor is also linked to essentialness: when product is  essential only for the common pool (can be now be represented by $\varepsilon \approx 1$), then the customers from this pool of size $2d_c$, buy either from $M_1$ or $M_2$.

 {\bf Demand attracted by a coalition:}
 When  two or more manufacturers operate together in coalition $\C$ (e.g.,  $\C = \M$ or  $\C = \{S, M_1, M_2\}~$), they quote a common price~$p_\sC$ by projecting themselves to the customers as a single unit mainly belonging to the manufacturer with best reputation (indexed by $\bst(\C) := \arg \max_{M_i \in \C} \alpha_{\sM_i}$), they can attract the customers of each of their members, etc. Further,  the market demand attracted by the coalition also depends upon other operating coalitions exactly as above. To be more precise if $\P = \{\C_1, \C_2, \cdots, \C_k\}$ is the operating partition then the market attracted by a coalition $\C_i$ is given by:  

 \vspace{-4mm}
 {\small\begin{eqnarray}
\label{Eqn_Demand_Function_coalition}
 D_{\sC_i} \hspace{-3mm}  & =&   \hspace{-4mm}  \left ( {\dbar}_{\sC_i}-\tilde\alpha_{\sC_i}(1-\gamma)p_{\sC_i}+\sum_{j \ne i} \varepsilon_{\sC_i,\sC_j}  \tilde\alpha_{\sC_j} (1-\gamma)p_{\sC_j} \right )^+ \hspace{-2mm},     \nonumber     \\   {\dbar}_{\sC_i} \hspace{-3mm} &=& \hspace{-3mm}  \sum_{k \in \sC_i} \dbar_{\sM_k }   \ 
 {\tilde \alpha}_{\sC_i}  \  :=  \  {\tilde \alpha}_{\sM_{\bst(\sC_i)}}, \mbox{ for any coaltion } \C_i,  \nonumber \\   \bst(\C) \hspace{-3mm}  &:=& \hspace{-3mm} \arg   \min_{M_k \in \C} \tilde\alpha_{\sM_k} ,  \mbox{ and }  \varepsilon_{\sC_i, \sC_j} := \varepsilon_{\bst(\sC_i), \bst(\sC_j) }. 
\end{eqnarray}}

 {\color{red}
 ---  thus the combined market size   attracted by~$\M$  is given by (in similar lines as in \eqref{Eqn_Demand_Function}):
 \begin{eqnarray}
 \label{Eqn_Both_demand}
D_{\sM} =  {\dbar}_{\sM}-(1-\varepsilon) \alpha_{f_\sM}p , \mbox{ with }  \alpha_{f_\sM} = g(\alpha_\sM, \gamma_{f_\sM} ) ,  \alpha_\sM := \min \{ \alpha_\sMo, \alpha_\sMt\}, 
 \end{eqnarray}
 where~$\dbar_\sM$  is the maximum possible combined market size that can be captured.  We assume,
 $$\max\{\dbar_\sMo, \dbar_\sMt \} \le \dbar_\sM \le \dbar_\sMo + \dbar_\sMt,$$ i.e., together they can  attract better market size than anyone of them, but not more than the sum of the two market sizes. 
 }

\vspace{-2mm}
\subsection{Operating choices of a given Coalition } 
\label{sec_operating_choices}
\vspace{-2mm}
The manufacturer choice $\bst(\C)$  defined in equation \eqref{Eqn_Demand_Function_coalition}  
is the one with best reputation among the manufacturer-members of the coalition $\C$. This choice is utilized to project the coalition/group to the public so as to attract maximum possible demand.  On the other hand, for manufacturing purposes they use the methods of the manufacturer with least manufacturing cost, that of  $\arg \min_{M_i \in \C} C_{\sMi}$, where $C_\sMi$ is the per-unit manufacturing cost of manufacturer $M_i$.

 {\color{red}

We now discuss the sequence of actions of agents.

The supplier quotes  its per unit wholesale price $q$ (for raw materials) and the manufacturers select to operate at a certain price $p$ (for the product) or not-operate depending upon  the anticipated or expected profit, given  the  supplier-quoted-price. 
To model this scenario, we use a Stackelberg game where suppliers act as leaders, quoting raw material prices, and manufacturers are followers  by selecting 
one among the suppliers.
After procuring raw materials, the manufacturers produce the product and quote a selling price to customers. Customers, in turn, select manufacturers based on selling prices and the reputations of manufacturers and their selected suppliers. 

Our research aims to explore the potential for horizontal (in same echelon) and vertical cooperation (across the echelons) in an SC, using cooperative game theoretic tools such as coalition formation games. We investigate the possibilities for different coalitions among the manufacturers and suppliers within an SC.

We now define the ingredients of the game precisely. }

\vspace{-2mm}
\subsection{Costs, actions and the utilities}\label{sec_costs}
\vspace{-2mm}
%We now describe the actions of various agents when they operate alone. The combined actions are described later while discussing various partitions/operating configurations. 

To begin with,
there is a fixed cost of operation for any agent, supplier or manufacturer or for any coalition --  the agent/coalition would incur negative revenue when sufficient profits are not generated and hence can choose  not to operate. Let~$n_o$ represent the choice of not operating. The utility of any agent/coalition is obviously zero, when its choice is $n_o$.

The supplier   can decide not to operate  or can quote a price~$q \in [0, q_h]$ (for some~$q_h < \infty$). %;  the former action, represented by~$n_s$,  is chosen when  it is  difficult to generate profit.
Thus the action~$a_\sS$ of supplier  while operating alone, is in the action set,~$a_\sS  \in \Aset_\sS :=\{n_o\} \cup [0, q_h]$.

%The suppliers quote a wholesale price between~$[0, q_{h}]$ (where~$q_h$ is the maximum permissible price) to the manufacturers based on market conditions; and if the supplier do not find incentive to participate, that is the fixed cost of supplying is more than the revenue derived by participation, then the suppliers choose the action~$n_s$ of not supplying. Thus the vector of actions of suppliers is~$\b=(b_1,b_2)$, where~$b_j \in \{n_s\} \bigcup \{ ( q_j )\}.$

When  manufacturer~$M_i$ decides to  operate alone, 
it   quotes a selling price~$p_{\sMi} \in [0, p_{h}]$  to the customers; here~$p_h$ is the maximum permissible price. 
Thus  the action~$a_\sMi$ of  $M_i$,~$a_\sMi \in \Aset_\sMi : = \{ n_o\} \bigcup  \  [0, p_h] $.

In the grand coalition $\C = \{S, M_1, \cdots, M_n\}$ involving vertical as well as horizontal (full) cooperation,    the coalition directly quotes a price $p_\sC$ for customers and makes a combined effort to produce the final product. If there are more manufacturers outside the coalition like in $\C = \{S, M_{i_1}, \cdots, M_{i_2}\}$,  then  $\C$ also quotes price $q$ for providing raw materials to the manufactures outside $\C$.

\vspace{-2mm}
\subsection*{All Alone (ALC) partition}
\vspace{-2mm}
We now describe the utility of various agents when all of them operate alone, i.e, when the operating configuration is represented by partition~$\P_{AA} = \left \{  \Mo, \Mt,  \cdots, \M_{n}, \S  \right \}~$ with~$\Mi := \{ M_i\}$ for any~$i$, and~$\S := \{S\}$  representing the coalitions with single agents.  The utilities in other arrangements are described later.  Let~$\a := (\a_\sM,a_\sS)$,~$\a_\sM :=  (a_\sMo\cdots a_{\sM_{n}})$ and $a_\sS $ represent the  action (vector) of all agents, all manufacturers and the supplier respectively.

We conclude this discussion with an important assumption, applicable throughout the paper. 

\begin{enumerate}[{\bf A}.1]
    \item When any agent finds $n_o$ as good  as an operating choice $\a_\sM \ne n_o$ or $\a_\sS \ne n_o$, then we assume the agent prefers operating choices.
\end{enumerate}
The the action of any agent (say of of agent $M_i$) is either  given by $a_\Mi = n_o$ or is represented by the price it quotes when it decides to operate in which case $\a_\sMi  = p_\sMi$. Similar notation is used for supplier or any coalition.

{\bf Manufacturers' Utility:} 
The   utility of manufacturer~$M_i$ is zero either when it chooses not to operate or when the supplier chooses $n_o$;
otherwise the utility is the total profit gained, the product of demand attracted~$D_\sMi(\cdot)$ given in~\eqref{Eqn_Demand_Function}  and the profit gained per unit:
\begin{eqnarray}\label{Eqn_UtilityofManufacturer}
    U_\sMi (\a   ) = \left ( D_{\sMi}( \a_\sM) (p_\sMi - C_{\sMi} - q) \indc { a_\sS \ne n_o} - O_\sMi \right ) \indc{ a_\sMi \ne n_o} ,
\end{eqnarray}
where~$C_{\sMi}$ denotes the {\color{blue}per unit} production cost incurred by~$M_i$,~$O_\sMi$ is the fixed operating/setup cost, and~$q$ denotes the wholesale price quoted by the supplier. Note that the  above  expression  is the utility of~$M_i$
when it operates alone. 
We would discuss later the utility derived in a coalition, i.e.,  
when it collaborates with other agents. 

{\bf Suppliers' Utility:}
The demand percolates from the lower echelon (i.e,manufacturers) to the higher echelon (i.e to the supplier) based on   choices made by the manufacturers and leads to the profits of the supplier. Thus  the utility of supplier~$S$  when it operates alone is given by:

\vspace{-3mm}
{\small\begin{eqnarray}\label{UtilityofSupplier}
    U_{\sS} (\a) = \left ( \left ( \sum_{i=1}^{n} D_\sMi (\a_\sM)   \indc{a_\sMi \ne n_o} \right ) (q- C_\sS ) - O_\sS  \right ) \indc{a_\sS \ne n_o },
\end{eqnarray}}%
 where~$C_{\sS}$ denotes the {\color{blue}per unit} raw material procurement cost of the supplier and~$O_{\sS}$ is the fixed operational cost of the supplier~$S$.

We now discuss the utility of various coalitions
\vspace{-2mm}
\subsection*{General Partition}
\vspace{-2mm}
This is defined as the sum of the utilities of all the agents in the coalition. As already mentioned in \eqref{Eqn_Demand_Function_coalition}, they utilize the best agent for any aspect, best reputation,  by projecting  the best manufacturer and supplier pair to the customers and best production efficiency . Thus for example the action and utility of  grand coalition $ \G = \{M_1 \cdots M_n,S\}$ are $a_\sG \in \{p \in [0, p_h]\} \cup \{n_o\} $ and 
$$
U_\sG (a_\sG) =\left ( ( \dbar_\sM  - \tilde \alpha_\sG (1-\gamma) p )
(p - C_\sG ) -O_\sG \right )   \indc{ a_\sG \ne n_o}
$$%
where combined operational cost is  
$O_\sG =  \min \{ O_\sMo \cdots O_{\sM_{n}}\} + O_\sS. 
$
and $C_\sG = \min\{C_\sMo\cdots C_{\sM_{n}}\} + C_\sS$ while the price sensitivity is $\tilde\alpha_\sG = \min\{\tilde\alpha_\sMo \cdots \tilde\alpha_{\sM_{n}}\}$.  One should also observe that all the agents share the revenue generated,  which pricesly equals $U_\sG$ given above,  and there is no price per item to be paid between any subset of agents, when the operating partition is GC. 
The definition of utilities for other partitions follow similar logic and will be discussed at the respective place.

We conclude this discussion with an important assumption, applicable throughout the paper:

\begin{enumerate}[{\bf A}.1]
    \item If any agent, either supplier or manufacturer, is indifferent between the action~$a = n_o$ or $a \ne n_o$, we assume that the agent prefers operating choices.
\end{enumerate}

We begin with a simple case, the study of a single manufacturer and supplier before we discuss and analyse the partition form game. The more generic 
supply-chain based partition form game is discussed after that in section \ref{sec_sing_Supp_two_manu}.
} % end ignore starting at 229

\vspace{-5mm}
\section{Single Manufacturer SC}\label{two_agent_supply_chain}
\vspace{-2mm}
In a two-echelon SC with one supplier and one manufacturer, the agents either operate together to form  GC partition $\{ \G\} $   or operate alone to form ALC partition  {$\{\S, \M \}$} (\textit{observe that $\M =\{M\}$, $\G = \{S,M\}$ are coalitions with one manufacturer in this section, while the same respectively represent $\M = \{M_1, M_2\}$ and $\G = \{S, M_1, M_2\}$ in the rest of the paper}). To completely understand the coalitional stability aspects of such a system, it is sufficient to analyze these two partitions.
Note that there is no competition among the agents in the same echelon in this case. In the absence of this competition, the market demand \eqref{Eqn_Demand_Function}  for the manufacturer simplifies to (as in \eqref{Eqn_Both_demand}):
\begin{eqnarray}\label{sing_agent_demand}
  D_{\sM}(a_{\sM}) = \left(\dbar_{\sM} -  \alpha_\sM p\right).  
\end{eqnarray}
Recall the demand decreases as the price increases, but the drop is reduced as the product becomes more and more essential ( when ~$\gamma \approx 1$) and $\alpha_\sM = \tilde\alpha_\sM(1-\gamma) $. 
We assume the following:
 
\begin{enumerate}
 [{\bf A}.2]
\item The total market size $\dbar_{\sM} > \alpha_\sM C_\sG   + 2 \sqrt{\alpha_{\sM}  O} $,
with $O := \max\{2 O_\sS, O_\sG, 4 O_\sM \}.$   
\end{enumerate}
Basically, the available market size has to be above a certain threshold so that the profit from the attracted demand surpasses the operating and manufacturing costs -- this ensures it is optimal for the agents to operate (see for e.g.,  Lemma~\ref{lem_basic_unconstrained_optimization} of the Appendix). If it is optimal for the agents not to operate, then there is nothing left to analyse.

\vspace{-2mm}
\subsection{ALC Partition}\label{AA_two_agent}
\vspace{-2mm}
In this partition, both agents operate independently and aim to maximize their respective utilities. We consider a Stackelberg game framework, where the supplier first quotes a price $q$ per bundle of raw material to the manufacturer. The manufacturer then quotes a price $a_\sM = p$ (per unit of product) to the customers; the customers, in turn, respond by generating a demand $D_\sM(a_\sM)$, as in~\eqref{sing_agent_demand}. Thus, the Stackelberg game between the supplier and the manufacturer is given by (with $\a = (a_\sS, a_\sM)$, see \eqref{Eqn_UtilityofManufacturer}-\eqref{UtilityofSupplier}):
\begin{eqnarray}\label{util_supp_single_player} 
U_{\sS}(\a  )\hspace{-2mm} &=&  \hspace{-2mm}\left(D_{\sM}(a_\sM)\F_\sM  (q- C_{\sS}) - O_\sS \right)\F_\sS,  
 \\
 \label{util_manu_single_player}
 U_{\sM}(\a)\hspace{-2mm} &=&  \hspace{-2mm}\left(D_{\sM}(a_\sM)(p-q-C_\sM)\F_\sS  - O_\sM\right)\F_\sM. \hspace{6mm}
\end{eqnarray}

We now derive the   Stackelberg equilibrium (SBE) of the above game.
\begin{theorem}\label{thm_all_alone_two_}
  Assume {\bf A}.1 and {\bf A}.2. There exists an SBE under which both the agents  operate,  which is given by $a^*_\sM = p^*$ and $a^*_\sS = q^*$, where

  \vspace{-3mm}
{\small \begin{equation*}
 p^{*} := \frac{3\dbar_{\sM} + \alpha_\sM(C_{\sS} + C_{\sM} )  }{4 \alpha_\sM},  \
 q^{*} := \frac{ \dbar_{\sM} + \alpha_{\sM}(C_{\sS} - C_\sM)  }{2\alpha_{\sM}}.   
\end{equation*}}
  \vspace{-3mm}
  
\noindent Further, the utilities at this SBE are given by
\begin{eqnarray}\label{eqn_util_opt_Aa}
(U_\sM^{*},U_\sS^{*})  &=& (\phi-O_\sM,2\phi - O_\sS),\\ \nonumber 
    \mbox{where } \phi &:= &\frac{\left(\dbar_{\sM} -\alpha_{\sM}(C_{\sS} + C_{\sM} ) \right)^{2}}{16\alpha_{\sM}}.    
\end{eqnarray}

 \end{theorem}  

\begin{proof}
 The  utility of manufacturer \eqref{util_manu_single_player} resembles  that in Lemma \ref{lem_basic_unconstrained_optimization} and thus for any $q$, the optimizer of the  manufacturer is operating (not equal to $n_o$) only when $\dbar_\sM \ge \alpha_\sM(C_\sM + q) + 2\sqrt{\alpha_\sM O_\sM}$ and then is given by 
 $p^{*}(q) = \nicefrac{\dbar_\sM}{2\alpha_\sM} + \nicefrac{(C_\sM + q)}{2} .$ 

If one neglects the operating conditions, then
 after   substituting  $p^{*}(q)$ in \eqref{util_supp_single_player} we have:

  \vspace{-3mm}
  {\small \begin{equation*}
      U_\sS(q) = U_\sS (q, p^*(q) ) =  \frac{\left(\dbar_\sM - \alpha_\sM(C_\sM + q))(q- C_\sS \right) - 2O_\sS }{2}
  \end{equation*}
  }which is again similar to that in Lemma \ref{lem_basic_unconstrained_optimization},  and then by the same lemma, the optimal $q$ would have been
  $q^{*} = \nicefrac{(\dbar_\sM + \alpha_\sM(C_\sS - C_\sM))}{2\alpha_\sM}$  -- this is true when  $q^*$ and $p^*:=p^*(q^*) = \nicefrac{(3\dbar_\sM + \alpha_\sM(C_\sS + C_\sM))}{4\alpha_\sM}$ both satisfy the required operating conditions (i.e., respective $\Delta$s are~$\ge 0$) -- these are satisfied, as by {\bf A.2}:

  \vspace{-3mm}
  {\small\begin{align*}
  \alpha_\sM (q^{*} + C_\sM) =  \frac{\alpha_\sM C_\sG +\dbar_\sM}{2}    &<  \dbar_\sM  -   2\sqrt{\alpha_\sM O_\sM}, \mbox{ and } \\
  (\dbar -\alpha_\sM C_\sM) -\alpha_\sM C_\sS - 2\sqrt{2 O_\sS} &> 0.  
\end{align*}}%
  By substituting $(q^*, p^*(q^*))$
  in \eqref{util_supp_single_player} and \eqref{util_manu_single_player}, we derive the optimal utilities.
\end{proof}

\vspace{-2mm}
\subsection{GC Partition}\label{gc_two_agent_supply_chain}
\vspace{-2mm}
Both the agents operate together, and the system directly faces the customers and quotes a common price~$p$. The per-unit cost $C_\sG = C_\sS + C_\sM$ of the system includes the procurement cost (of the raw materials) and the production cost (see Section~\ref{sec_cost_util}). Furthermore, the system also has a fixed operating cost $O_{\sG} = O_\sS + O_\sM$, when it operates.
Thus, the overall utility of the system is:
\begin{eqnarray}\label{GC_opt_prob_one_agent}
 U_{\sG} &=& 
 \left (  D_\sM   ( a_\sM )  \big (p- C_{\sG} \big ) -O_{\sG} \right ) \F_\sG  . 
\end{eqnarray}
%where $D_\sM(a_\sM)$ is the total demand attracted by the manufacturer, as in~\eqref{sing_agent_demand}. The utility is zero when the system does not operate, i.e., when $a_\sG = n_o$. 
The utility of any coalition is defined as the optimal utility that it can derive. Therefore, we have the following simple optimization problem for deriving the utility of the GC:\begin{eqnarray}\label{eqn_gc_util_opt_prob}
  \sup_{\a_\sG \in \Aset_\sG}   \left ( D_\sM   ( \a_\sM )  (p- C_{\sG})-O_{\sG} \right ) \F_\sG.  
\end{eqnarray}
This problem can be solved using basic (derivative-based) methods, and the solution is as follows:
\begin{theorem}\label{gclem}
Assume {\textbf{A}.1} and {\textbf{A}.2}. There exists an optimizer at which the system operates, and the corresponding optimal price is $a^*_\sG = p^*_\sG$, where
\begin{eqnarray}
p_{\sG}^{*} = \frac{\dbar_\sM}{2\alpha_{\sM}} +\frac{C_{\sG}}{2},
\label{eqn_GC_optimal_utility} 
\end{eqnarray}
and the corresponding optimal utility is given by:
\begin{eqnarray}
 U^{*}_{\sG} = \frac{\left( \dbar_\sM-\alpha_{\sM}C_{\sG}\right)^{2}}{4\alpha_{\sM}} - O_{\sG}.   
\end{eqnarray}
%Under assumption {\textbf{A}.1} and {\textbf{A}.2}, the optimal pricing policy of GC is given by $a^{*}_\sG = p^{*}_\sG$ and the optimal utility of GC is given by $U^{*}_\sG$.
\end{theorem}
\begin{proof}
 Again the utility function of   GC~\eqref{eqn_gc_util_opt_prob}, resembles that in Lemma~\ref{lem_basic_unconstrained_optimization}. Furthermore, by assumption {\bf{A}.2}, $\Delta > 0$ for the GC, which implies that the GC will operate. Thus the proof follows by Lemma \ref{lem_basic_unconstrained_optimization}.
%Observe again that  the structure of utility of GC in \eqref{eqn_gc_util_opt_prob} is similar to the structure in lemma \ref{lem_basic_unconstrained_optimization} and also by assumption {\bf{A}.2}, that $\Delta > 0$ for GC and thus GC will operate and thus using this lemma we get the optimizer and the corresponding utility as in theorem.
\end{proof}

\noindent {\textbf{Remarks}: By Theorems \ref{thm_all_alone_two_} and \ref{gclem}  (observe here $C_\sG = C_{\sS} + C_{\sM}$ as  in  subsection \ref{subsec:utility_general_partition}),
\begin{eqnarray}
U^{*}_\sG - (U_\sM^{*} + U_\sS^{*}) =  \phi.
\end{eqnarray}
Thus, the agents derive higher utility in GC  than the combined utility that they derive when they operate alone (i.e., ALC). This means that both agents can benefit by forming a coalition, as long as they agree to share the extra profit ($U_\sG^* - (U_\sM^*+U_\sS^*)$) in a way that benefits both of them. Consider  a configuration $(GC, (x_\sM, x_\sS) )$, where $x_i$ represents the  payoff allocation of agent $i$ and which satisfies:
\begin{eqnarray}\label{eqn_div_two_agents}
 x_\sM + x_\sS = U_\sG^*, \ \  x_\sM >  U_\sM^*, \ \ \mbox{ and }  x_\sS > U_\sS^*.    
\end{eqnarray}
When the profits are shared as above, none of the agents prefer to operate alone. Now consider a configuration  $(GC, (x_\sM, x_\sS))$ that does not satisfy \eqref{eqn_div_two_agents}. When $ x_\sM + x_\sS < U_\sG^*$, the generated revenue $U_\sG^*$ is not completely shared; if share $x_\sM < U_\sM^*$, the manufacturer would prefer to operate alone, as it would then derive $U_\sM^*$; similarly if $x_\sS < U_\sS^*$, the supplier would prefer to operate alone. Thus such configurations are `opposed' and hence are not stable.
Before we study an SC with two manufacturers, let us formally discuss the notions of stability  in partition form games.

\vspace{-5mm}
\section{Partition Form Games}\label{sec_coalition_formation_game}
\vspace{-2mm}
A partition form game is described using the tuple $\langle N,  (\nu_\sC^\P) \rangle$ where $N$ is the set of players and $\nu_\sC^\P$ is the worth of the coalition $\C$ under partition $\P$ and is defined only when $\C \in \P$. As mentioned in the introduction, here the worth $\nu_\sC^\P$ also depends upon the partition $\P$ -- basically $\nu_\sC^\P$ need not equal $\nu_\sC^{\P'}$ for two different partitions $\P \ne \P'$ both of which contain $\C$.
 %such that $\C \in \P$ and this worth depends upon the partition which coalition $\C$ is the part of.  

Given a partition $\P$ and the worths $\{\nu_\sC^\P\}$ of each coalition in $\P$, the next question is about a `pay-off' vector which defines the allocation to each agent in~$N$. A pay-off vector $\x = (x_1, \cdots, x_n)$ is defined to be consistent with respect to partition $\P$ if (see \cite{aumann1974cooperative,singhal2021coalition,shikshathesis}):
\begin{eqnarray}\label{eqn_consistency}
  \sum_{i \in \C_j} x_i = \nu_{\sC_j}^\P  \mbox{ for all }  \C_j \in  \P.    
\end{eqnarray}
The pair $(\P, \x)$ is defined to be a configuration if the latter is consistent with the former.

The quest now is to study a `solution' of the partition form game. 
The  `solution' in this context describes the configurations that are stable; in other words, it identifies the partitions and their companion consistent payoff vectors that can emerge or operate stably without being `opposed'.

To study the stability aspects, one first needs to understand if a certain coalition which is not a part of the partition can `block’ (or oppose) the given configuration -- such a blocking is possible if the coalition has an `anticipation' of the value it can achieve (irrespective of all scenarios that can result after coalition blocks) and if the anticipated value is bigger than what the members of the coalition are deriving in the current configuration. 
Basically, if there exists at least one division of this anticipated value among the members of the blocking coalition that renders all the members to achieve more than that in the given payoff vector,  then the coalition has a tendency to oppose the current configuration. The above concepts are made precise in the following definitions (\cite{singhal2021coalition,shikshathesis}):

\begin{definition}
[Blocking of a configuration by a coalition]
A configuration, the tuple of  partition and the consistent payoff vector, $(\P, \x)$,   is blocked by a coalition~$\C \notin \P$, under the pessimal anticipation rule, if the coalition derives better  than that in the original configuration irrespective of the arrangement of opponent players, i.e., if the pessimal anticipated utility 
\begin{eqnarray}\label{defn_blocking}
\nu_\sC^{pa} := \min_{\P': c \in \P'}     \nu_\sC^{\P'} > \sum_{i \in \C} x_i .
\end{eqnarray}
\end{definition}
\begin{definition}[Stability]\label{defn_stability} A configuration $(\P, \x)$ is said to be stable if there exists no coalition $\C \notin \P$ that  blocks it.
A partition $\P$ can be said to be stable if there exists at least one configuration $(\P, \x)$ involving $\P$ which is stable. 
\end{definition}

\RB{
\begin{definition}[Restricted-blocking Stability]\label{defn_restricted_blocking_stability} A configuration~$(\P, \x)$ is said to be restricted-blocking-stable, if there exists no coalition~$\C \notin \P$, which is a merger or a split of the coalitions in~$\P$, that  blocks it.
A partition $\P$ is said to be restricted-blocking-stable if there exists at least one configuration $(\P, \x)$ involving~$\P$ which is stable against mergers and splits. 
\end{definition}

The last concept is proposed in~\cite{singhal2021coalition} inspired by common rearrangements in practical scenarios. Authors in~\cite{singhal2021coalition}  consider queuing system with QoS based market segmentation and showed that none of the partitions are stable (as according to Definition~\ref{defn_stability}), while they found some duopolies (partitions with two members) to be stable against mergers and splits. We observe a similar phenomenon under certain conditions and discuss elaborately about the same in Section~\ref{sec:stability_results}.
}

\ignore{ 
We now provide some more definitions, where we directly discuss stability of a partition. As in \cite{} one can define a partition $\P$ to be stable, if configuration $(\P, \x)$ is stable for all consistent pay-off vectors $\x$. In similar lines we would define complete blocking by a coalition -- a coalition 
completely blocks a partition $\P$ if it can  block configuration $(\P, \x)$ for   payoff vector consistent  with $\P$. 

{\color{red}
\begin{definition}[Complete Blocking by a Merger]
 A partition~$\P=\{\C_{1},\C_{2},....\C_{k}\}$ is said to be blocked completely  by a merger $\C \notin \P $, using pessimal anticipation, if pessimal utility of the merger is strictly bigger than the sum total of the utilities of the coalitions involved in merger irrespective of their  payoff vectors in $\P$, i.e., if
\begin{eqnarray}    \label{Eqn_Complete_Blocking_Merger}
 \C &=& \cup_{k_l} \C_{k_l} \mbox{ for some indices } \{k_l\} \mbox { and if } \  \\ 
\nu_\C^{pa} &:=& \min_{\P' : \C \in \P' } \nu_{\sC}^{ \P'} >  \sum_{\C_k \subset \C} \nu_{\sC_k}^{ \P } . 
\end{eqnarray}
\end{definition}
Pessimal anticipation rule ensures that the blocking coalition on total would derive better than their sum total payoffs in $\P$, irrespective of the way the left over members $\N - \C$  arrange themselves.  

Thus when grand coalition GC blocks any partition, blocking via the pessimal rule \eqref{Eqn_Complete_Blocking_Merger} is equivalent to the following condition (as there is only one partition containing GC):
\begin{eqnarray}
    \label{Eqn_GC_block}
\nu_\sG^{pa} = \nu_\sG^\G > \sum_{\sC \in \P} \nu^{\P}_\sC.
\end{eqnarray}

\begin{definition}[Completely Blocked by split]\label{def_completely blocked_by_split}
A partition~$\P $ is said to be completely blocked by a split $C_1$ of $C \in \P$, under pessimal rule,   if
\begin{equation}
\inf_{\P': c \in \P'}  \nu_{\sC_1}^{\P'}  \geq \nu_{\sC}^{ \P} .
\end{equation}
This implies that the split $\C_1$ blocks all configurations, $(\P, \x) $,  involving partition $\P$  with non-zero payoff components for left over members $\C - \C_1$, i.e., with $\sum_{i \in \C - \C_1} x_i > 0.$
\end{definition}
}}

We now apply the above stability concepts to the single manufacturer case study of the previous section.
In this case, $N = \{S, M\}$  and the only possible partitions are $\pGC = \{N\}$ and $\ALC = \{ \{S\}, \{ M \} \} $. 

Clearly, the worth of grand coalition $\nu_\sG^\GC = U^*_\sG$ given in  \eqref{eqn_GC_optimal_utility}, and that of manufacturer and supplier, while operating alone,  are respectively given by $\nu_\sM^{\ALC}  = U_\sM^*$ and $\nu_\sS^{\ALC} = U_\sS^*$  of Theorem \ref{thm_all_alone_two_}. 
These complete the definition of the partition form game.
Also observe that any pay-off vector $\x = (x_\sM, x_\sS)$ is  consistent   with GC partition  if and only if $x_\sM + x_\sS = U_\sG^* = \nu_\sG^\pGC$. On the other hand, the only payoff vector consistent with the ALC partition is $\x = (\nu_\sM^{\ALC}, \nu_\sS^{\ALC} ) $.
The Theorems \ref{thm_all_alone_two_}-\ref{gclem} immediately imply the following stability result:

\begin{lemma}\label{lem_gc_blocks_alc}
In the  single manufacturer SC: i) ALC partition  $\ALC$ is blocked by grand coalition; and ii) The GC-core, the set of  consistent pay-off vectors that form stable configurations with $\GC$ ( see \eqref{eqn_util_opt_Aa}, is:

\vspace{-3mm}
{\small
$$ \left \{ \x:  x_\sS > 2\phi - O_\sS, \  x_\sM > \phi - O_\sM \mbox{ and } x_\sS + x_\sM = 4\phi - O_\sG \right \}  . $$}
\end{lemma}
\begin{proof}
From \eqref{defn_blocking}, the pessimal anticipated utilities with $|N|=2$,   clearly  equal $\nu_\sC^{pa} =  \nu_\sC^{\pGC}$ for any $\C$.
 Thus by Theorems 
 \ref{thm_all_alone_two_}-\ref{gclem},
 $\nu_\sG^{pa} =  \nu_\sG^{\pGC} > \nu_\sM^{\ALC} + \nu_\sS^{\ALC} $
 and hence part (i); part (ii) follows by direct verification. 
\end{proof}

\noindent {\textbf{Remarks:} From \eqref{Eqn_UtilityofManufacturer}- \eqref{UtilityofSupplier}, if the supplier and the manufacturer participate in a strategic form game (i.e., when they make choices simultaneously), the resultant  Nash Equilibrium (NE) is $(n_{o},n_{o})$ -- the best response of the manufacturer is $n_{o}$ for any $a_\sS = q > C_\sM$ or when $a_\sS = n_o$, while   that of the  supplier is $n_{o}$ when $a_\sM = n_o$ and equals infinity when  $a_\sM \ne n_o$.
Thus if both the agents compete at the same level, 
the SC would not operate and both of them derive 0 revenue.

On the other hand, when the supplier leads the market as in the SB game, by Theorem \ref{thm_all_alone_two_} the system operates resulting in positive revenues for both the agents. They derive even better utilities by operating together and hence GC is stable as shown in Lemma \ref{lem_gc_blocks_alc}. However the supplier gets a much  better share; again from Lemma \ref{lem_gc_blocks_alc}, the share of supplier $x_\sS$ is  at least $2\phi-O_\sS$ while that of  the  manufacturer  is at most $2\phi - O_\sM$.  
  These observations motivate us to analyse a more generic SC  with competition at the lower echelon. The aim in particular is to understand the stable configurations, the profit shares, etc., in the presence of lower echelon competition.}

\vspace{-5mm}
 \section{Two Manufacturer SC}\label{sec_sing_Supp_two_manu}
 \vspace{-2mm}
In this section, we explore the case of two echelon SC consisting of a supplier and two manufacturers. As explained in Section \ref{sec_market_seg} and equation \eqref{Eqn_Demand_Function}, the fraction of demand captured by each manufacturer is :
\begin{eqnarray}
    D_{\sMi}(\a_\sM) = (\dbar_\sMi -\alpha_{\sMi}p_{i} + \varepsilon\alpha_{\sMminusi}p_{-i}) \mbox{ for all } i.
\end{eqnarray}
If there is no cross-linking (i.e., if  $\varepsilon = 0$), the demand functions get decoupled, and each of them resemble to that of the single manufacturer SC (see \eqref{sing_agent_demand}).

One needs to derive the worths $\{\nu_\sC^\P\}$ for all possible coalitions $\C$ and partitions $\P$ to study the stability aspects. 
The worth  $\nu_\sC^\P$  can be defined as the `best' utility (the maximum sum utility) that  the members of $\C$ can derive, while facing the competition from agents outside the coalition arranged as in   partition $\P$.

The competition between various coalitions
is captured via a Stackelberg game (as in Subsection~\ref{AA_two_agent}), when at least one manufacturer is not collaborating with the supplier -- the partitions of this type are, ALC partition  $\ALC = \{ \S, \Mo,  \Mt \}$, HC partition $\CHC = \{ \S, \M \} $  and the VC partition $\VC_i = \{ \V_i, \M_{-i} \} $. In all these cases, the leader is the coalition $\C_L$ with the supplier. The coalitions with only manufactures form the followers - the followers respond optimally for any given action $\a_L$ (the quoted prices or $n_o$) of the leader. The solution of the followers is either an optimizer (when all manufacturers form a coalition) or an NE. Let ~$\a_\sM^* (\a_L)$ represent this solution in either case. The leader coalition is aware of this optimal choice, i.e.,~$\a_\sM^* (\a_L)$ for every~$\a_L$ is a common knowledge. Thus, the optimal choice of the leader is,
$$
\a_L^*  \in \arg \max_{\a_L} \sum_{j \in \C_L} U_j \left (\a_L, \a_\sM^* (\a_L))  \right ),
$$
and then $(\a_L^*, \a^*_\sM(\a_L^*) )$ represents the SBE. We then  define the worth of the leader coalition  by:
$$
\nu_{\C_L}^\P  =  \sum_{j \in \C_L} U_j \left (\a_L^*, \a_\sM^* (\a_L^*)  \right ). 
$$
The worth of the rest of the coalitions of $\P$ can be defined similarly using 
the SBE $(\a_L^*, \a^*_\sM(\a_L^*) )$.

We are just left with the GC partition $\P_\sG = \{ \G\}$, which can be analysed exactly as in Subsection \ref{gc_two_agent_supply_chain} and  is considered  in the immediate next -- we  once again assume `operating-conditions' assumption {\bf A}.2 (with terms like $C_\sG$ etc., accordingly changed);  \textit{without loss of generality, we  consider $\dbar_\sMo \ge \dbar_\sMt$.}

\vspace{-2mm}
\subsection{GC Partition}\label{sec_gc_two_agents}
\vspace{-2mm}
In GC partition $\GC$, the two manufacturers and the supplier operate together 
 as explained in Subsection \ref{subsec:utility_general_partition}. The  optimization problem is similar to that in \eqref{eqn_gc_util_opt_prob}, hence we have the following with proof exactly as in  that of Theorem \ref{gclem}:
\begin{cor}\label{cor_gc}
Assume ${\textbf{A.1}}$ and ${\textbf{A.2}}$.  The worth of $\GC$ defined using the optimizer is given by 
\begin{eqnarray*}
  \hspace{15mm} \nu_\sG^\GC = U^*_\sG = \frac{\left(\dbar_\sM - \alpha_\sM C_\sG \right)^2}{4\alpha_\sM}.      \hspace{17mm}  \mbox{\eop}
\end{eqnarray*}
    
\end{cor}
\vspace{-2mm}
\subsection{HC Partition}\label{sec_chc_two_agents}
\vspace{-2mm}
We now consider the  HC partition~$\CHC $, where both the manufacturers  $\M = \{M_1, M_2\}$ operate together. 
The coalition of manufacturers~$\M$ quotes a selling price~$p$ to the customers, and the leader (supplier)~$S$ quotes a price~$q$ to  $\M$. Recall any of them may decide not to operate (choose~$n_o$). 
The SBE $\a^* := (
\a_\sM^*, a_\sS^*)$ of the Stackelberg game satisfies the following as before:
\begin{align*}
\a_\sM^* &= \a_\sM^*( a_\sS^*) , \mbox{ and }  a_{\sS^*}  \in \arg \max_{a_\sS \in \Aset_\sS } U_\sS (a_\sS,  \a_\sM^*( a_\sS) ), 
\end{align*}
where 
the   utilities and the optimizers are given by:

\vspace{-3mm}
{\small
\begin{align*}
%\label{Eqn_Supplier_Utility_CompleteHc}
U_{\sS} (\a) &:=  \left( D_{\sM} (\a_\sM )(q-C_{\sS})\F_\sM-O_{\sS}\right)\F_\sS
\\
\a^*_\sM  (a_\sS) &:=  \arg \max_{a_\sM  \in \Aset_\sM } U_\sM (a_\sM, a_\sS ) \mbox{ with } \\
%\label{Eqn_Manufacturer_Utility_CompleteHc}
U_{\sM}(\a) &:=  \left( D_{\sM} (\a_\sM ) (p-C_{\sM}-q)\F_\sS -O_\sM \right)\F_\sM, 
 \end{align*}}%
with $D_{\sM} (a_\sM )$  defined in
\eqref{Eqn_Both_demand}.
This game is similar to that considered Subsection \ref{AA_two_agent}, and hence the following result using the proof of  Theorem \ref{thm_all_alone_two_}.
\begin{cor}
\label{cor_hc}
 Assume  {\textbf{A.1}} and
 {\textbf{A.2}}, the worths of the agents in  $\CHC$  (defined using operating SBE) equal:
 \begin{eqnarray*}
  \{ \nu_\sS^\pCHC,\nu_\sM^\pCHC \} = \{2\phi - O_\sS , \phi - O_\sM \} ,
 \end{eqnarray*}
 where $\phi   = \frac{\left(\dbar_\sM - \alpha_\sM C_\sG \right)^2}{16 \alpha_\sM}.$  \eop
\end{cor}
Using Corollaries \ref{cor_gc}-\ref{cor_hc},
as in Lemma \ref{lem_gc_blocks_alc}, it is easy to conclude that the HC partition is blocked by grand coalition $\G$ and hence is not stable.

\ignore{
From lemma \ref{lem_basic_unconstrained_optimization} when $a_\sS \ne n_o$ and $a_\sS = q$, we have:

\vspace{-4mm}
{\small\begin{eqnarray*}
\a_\sM^{*}(a_\sS) \hspace{-2mm} &:= & \hspace{-2mm} \left(\frac{\dbar_\sM}{2\alpha_\sM } + 
\frac{q + C_\sM}{2}\right)\indc{\Delta_\sM \ge 0} + n_{o}\indc{\Delta_\sM < 0}
\\
\Delta_\sM \hspace{-2mm} & := &  \hspace{-2mm}  \dbar_\sM \ge  \alpha_\sM (q + C_\sM) + 2\sqrt{ \alpha_\sM O_\sM}.
\end{eqnarray*}}
Now let the unconstrained optimizer $U_\sS^{*}$ be $a_\sS^{*} := q^{*}$. Thus for meaningful analysis, we take the following assumptions to ensure the operating conditions as in HC if any one coalition does not operate, the other will also not operate.
Thus we have the following assumptions
i) For ensuring the operating condition of the suppliers (i.e $ a_\sS^{*} \ne n_o$) from lemma \ref{lem_basic_unconstrained_optimization},we assume that   
\begin{eqnarray*}
  \dbar_{\sM} > \tilde\alpha_{\sM} (1-\gamma) (C_\sS + C_\sM)   + 2 \sqrt{2\tilde\alpha_{\sM} (1-\gamma) O_\sS}   
\end{eqnarray*}
It can be easily observed that analysis of this case is similar to ALC of single supplier-single manufacturer in subsection \ref{AA_two_agent} with $\tilde\alpha_\sM = \min\{\tilde\alpha_\sMo,\tilde\alpha_\sMt\}$ and $C_\sM = \min\{C_\sMo,C_\sMt\}$

   Under assumption {\bf A}.1 and {\bf A}.2, we have the following:
   
   i) If $\dbar_{\sM} \geq \tilde\alpha_{\sM} (1-\gamma)( C_\sS +  C_\sM)   + 4 \sqrt{\tilde\alpha_{\sM} (1-\gamma) O_\sM} $ the the operating stackelberg equilibrium is given by the following: all the agents are operating at the equilibrium,  and $a^*_\sM = p^*$, $a^*_\sS = q^*$ with
\begin{eqnarray*}
p^{*} &:=& \frac{3\dbar_{\sM} + \tilde\alpha_{\sM}(1-\gamma)(C_{\sS} + C_{\sM} )  }{4\tilde\alpha_{\sM}(1-\gamma)}  ,  \mbox{  }\\
q^{*} &=& \frac{ \dbar_{\sM} + \tilde\alpha_{\sM}(1-\gamma) (C_{\sS} - C_\sM)  }{2\tilde\alpha_{\sM}(1-\gamma)}.   
\end{eqnarray*}
Further the utilities of the two agents at equilibrium are given by:
$
 (U_\sM^{*},U_\sS^{*})  = (\phi-O_\sM,2\phi - O_\sS)$ where, 
 $$ \phi := \frac{\left(\dbar_{\sM} -\tilde\alpha_{\sM}(1-\gamma)(C_{\sS} + C_{\sM} ) \right)^{2}}{16(1-\gamma)\tilde\alpha_{\sM}}.
$$
\extra{
ii) In the other condition, the following are the quantities at SBE:

{\small\begin{eqnarray}
    p_{1}^* &=&  \frac{\dbar_{\sM}-\sqrt{\tilde\alpha_{\sM}(1-\gamma)O_{\sM}}}{\tilde\alpha_{\sM}(1-\gamma)},   q_{1}^* = \frac{ \dbar_\sM -\tilde\alpha_\sM (1-\gamma) C_\sM - 2 \sqrt{ \tilde\alpha_\sM (1-\gamma) O_\sM} }{\tilde\alpha_\sM (1-\gamma)} ,   \\
     U_\sM^{*} &=& 0 \mbox{ and }U_\sS^{*}  \le 2\phi - O_\sS.
\end{eqnarray}}
}
Thus the worths of the coalitions corresponding to Horizontal partition $\CHC = \{ \M, \S\}$ are given by: 
$$
\nu_\sS^\pCHC = U_\sS^* \mbox{ and }  \nu_\sM^\pCHC = U_\sM^* . 
$$
Thus we have,
 $\{f_\sS^{\pCHC}, f_\sM^{\pCHC} \} = \{0,0\}$ and 
  $\{g_\sS^{\pCHC}, g_\sM^{\pCHC} \} = \{\frac{\dbar_\sM^2}{8\tilde\alpha},\frac{\dbar_\sM^2}{16\tilde\alpha}\}$ 
  }

\vspace{-2mm}
 \subsection{Worth-Limits}\label{subsec_worth_limit}
\vspace{-2mm}

For further analysis, one needs to study the ALC and VC partitions. However   the expressions for these two partitions are significantly complex  and hence we begin with a  specific yet an important asymptotic case study in this conference paper -- while the complete generality would be considered in future.  We consider an \textit{asymptotic regime near  $(\varepsilon, \gamma) \approx (1,1)$; as mentioned previously, here the customers are willing to switch the loyalties towards their manufacturers and hence we call such  a regime as Essential and Substitutable-Manufacturer (ESM) regime.} We also consider manufacturers of equal reputation, \textit{i.e., with $\tilde\alpha_\sMo = \tilde\alpha_\sMt$}.
Towards obtaining the  asymptotic study we consider the following procedure. 

\noindent{\bf ESM  Regime:} For any  partition-coalition    $(\P,  \C)$, consider the function  $(\gamma, \varepsilon) \mapsto  (1-\gamma)(1-\varepsilon) \nu_\sC^\P$.  From all the expressions derived in this paper,  i.e., for all $(\P,  \C)$, these functions are continuous. Hence the following limits exist (for each $(\P,  \C)$) and can be rewritten as below: 
%If  we establish that the above function is jointly continuous in    $(\gamma, \varepsilon)$,  then  following limits exist for each $(\C, \P)$:

\vspace{-4mm}
{\small\begin{eqnarray}
\label{eqn_worth_lim}
f_\sC^{\P} &:=& \hspace{-3mm}  \lim_{ (\gamma, \varepsilon)  \to (1,1) } \hspace{-2mm}  (1-\gamma) (1-\varepsilon)  \nu_\sC^\P \nonumber \\
&= & \lim_{  \varepsilon  \to 1}  \lim_{ \gamma   \to 1 } (1-\gamma) (1-\varepsilon)  \nu_\sC^\P.
\end{eqnarray}}
We refer to the above as worth-limits, with slight abuse of notation. Similarly define    $\{f_\sC^{pa}\}$ using anticipated worths $\{\nu^{pa}_\sC\}$.   We will also require the limits of the following derivatives

\vspace{-3mm}
{\small\begin{eqnarray}
\label{Eqn_lim_derivatives}
f_\sC^{(1), \P} \hspace{-1mm} :=  \hspace{-1mm}   \lim_{\varepsilon \to 1}   %
  \frac{ d{\tilde  \nu}_\sC^\P } {d \varepsilon} \mbox{ with }   {\tilde  \nu}_\sC^\P := (1-\varepsilon) \lim_{\gamma \to 1} (1-\gamma) \nu_\sC^\P, 
\end{eqnarray}}}%
and also that of the  anticipated worths,  $\{f_\sC^{(1), pa}\}$.
\textit{The idea is to derive  the stability results by comparing the worth-limits   instead of the actual worths $\{\nu_\sC^\P\}$, and further using   the   derivative limits  \eqref{Eqn_lim_derivatives}  when the worth-limits are equal.}  We claim that such  stability results are applicable for all $(\varepsilon,\gamma)$ in a neighbourhood of $(1,1)$ because of the following reasons and procedure.

\ignore{: i)  one needs to prove strict inequality in \eqref{defn_blocking} to establish blocking; on the other hand  to show that a particular coalition does not block a particular partition we require $\le$, however  sufficient if we instead show strict inequality $<$;  ii) thus in all, one can work with strict inequalities to establish required stability results; iii) any strict inequality at limit also implies the same in the neighbourhood by continuity; 
iv) if and when certain  quantities are equal at limit, we consider derivatives to show strict inequalities in the neighbourhood; and v) finally one can find a common neighbourhood where all the 
required inequalities hold at limit because we only require finitely many inequalities to establish stability results. }

\ignore{

the results  are established using some inequalities, and the 
ii) in case of

Then if we prove results using  $\{f_\sC^{\P}\}_{\sC, \P}$, because of finitely many components across $(\sC, \P)$ and because of strict inequalities, there exists a neighborhood of $(\gamma, \varepsilon)$ around $(1,1)$ in which the result is true

In a similar way, we define similar limits  
$
f_\sC^{pa}
$ corresponding to  anticipatory worths $
\nu_\sC^{pa},$ to derived required analysis. }

From \eqref{defn_blocking} a configuration $(\P, \x) $
 is stable 
 \ignore{if there    exists no $\C \notin \P$ such that the anticipatory worth
 $$
 \nu^{pa}_\sC >  \sum_{i \in \sC }  x_i. 
 $$
 In other words the configuration is stable} 
 if the following  set of  inequalities are satisfied:
\begin{eqnarray}\label{eqn_worth_pa}
 \sum_{i \in \sC }   x_i   \ge   \nu^{pa}_\sC   \mbox{ for all }  \C \notin \P. 
\end{eqnarray}
  (we identify only the configurations that satisfy the above with strict inequality, a more complete study is again a part of the future work, and the reasons for this omission is evident in the immediate next).

If the  inequalities in \eqref{eqn_worth_pa} are satisfied in a strict manner  by some vector $\y$ and for some partition $\P$ using limits   $\{f_\sC^{pa}\}$ in place of $\nu_\sC^{pa}$, then by continuity there exists $\bar \gamma$ and $\bar \varepsilon$ such that the above inequalities  (finitely many) are satisfied for all $\gamma > \bar \gamma$ and $\varepsilon > \bar\varepsilon$ --- this implies that for all those $(\gamma, \varepsilon)$, the  configuration 
$$\left (\P, \beta \y    \right ), \mbox { with, } \beta := \frac {1} {(1-\varepsilon)(1-\gamma)},$$ 
is stable; thus one can obtain stability results near ESM regime  
using the   worth-limits  $\{f_\sC^\P, f_\sC^{pa}\}$ (when strict inequalities are considered in \eqref{eqn_worth_pa}). 

During blocking by mergers, i.e., say when blocking coalition $\C = \C_1 \cup \C_2$, then  recall by consistency in   \eqref{eqn_consistency},  the  inequality \eqref{eqn_worth_pa} modifies to
\begin{eqnarray}\label{eqn_block_merger}
\sum_{i \in \sC }   x_i  =  \nu_{\sC_1}^P + \nu_{\sC_2}^\P  \ge  \nu_\sC^{pa}.  
\end{eqnarray}
And if now  the worth-limits of both right and left hand sides are equal, then the comparison in neighbourhood is possible only by considering the derivatives. This is because for  such limits, by Taylors series expansion,  near $\varepsilon \approx 1$ we have (see \eqref{Eqn_lim_derivatives}):

\vspace{-2mm}
{\small\begin{align}
{\tilde \nu}_{\sC_1}^P &+ {\tilde \nu}_{\sC_2}^\P  -  {\tilde \nu}_\sC^{pa} \label{eqn_taylor_series} \\ 
 &= (\varepsilon-1)   
\left . \frac{ d \left ( {\tilde \nu}_{\sC_1}^P + {\tilde \nu}_{\sC_2}^\P  -  {\tilde \nu}_\sC^{pa}  \right )  }{ d\varepsilon} \right |_{\varepsilon \to 1} \hspace{-6mm}+  o ((1-\varepsilon)^2 ) \nonumber\\ \nonumber
%&& +  o ((1-\varepsilon)^2 ) \\ 
& = (\varepsilon-1)  \left( f_{\sC_1}^{(1), \P} + f_{\sC_2}^{(1),\P} - f_\sC^{(1), pa}  \right )+  o ((1-\varepsilon)^2 ),  
\ignore{
\\
\mbox{ where,}  
{\tilde \nu}_\sC^\P  &=& (1-\varepsilon) \hat \nu_\sC^P  \\
\hat \nu_\sC^P &:=& \lim_{\gamma \to 1} (1-\gamma) \nu_\sC^\P 
\\
f_\sC^{(1), \P} &:= &   \lim_{\varepsilon \to 1}   %
  \frac{ d{\tilde  \nu}_\sC^\P } {d \varepsilon}  .}
\end{align}}%
where the limits $\{f_\sC^{(1), pa}, f_\sC^{(1), \P} \}$
are defined in \eqref{Eqn_lim_derivatives}. Thus the required stability results can be established if now the derivative limits satisfy the required   inequalities -- and then there exists an $\bar \varepsilon  < 1$ such that the stability results are true in a neighbourhood as below: 
\begin{eqnarray}
  \label{Eqn_gam_eps_nhd}
\left \{ (\varepsilon, \gamma) :  \varepsilon \ge  \bar \varepsilon \mbox{ and }  \gamma \ge  \bar \gamma_\varepsilon  \right \},
\end{eqnarray}where   $\bar \gamma_\varepsilon < 1 $ is a  lower bound depending upon  $\varepsilon $.

\ignore {
$$
f(a+(1-\varepsilon)) = f(a) + (1-\varepsilon) f'(a) +  o ( (1-\varepsilon)^2 )
$$
because of continuity,
$$f(a) = \lim_{\varepsilon \to 1} f(a+(1-\varepsilon))
$$

\newpage
In the nhd of $\varepsilon \approx 1$ we have:
\begin{eqnarray*}
{\tilde \nu}_\sC^\P (\varepsilon) &=& f_\sC^\P +  (1-\varepsilon)
  \frac{ d{\tilde \nu}_\sC^\P } {d \varepsilon} + o((1-\varepsilon)^2) \\
  && \hspace{-18mm} = f_\sC^\P -  (1-\varepsilon) \hat \nu_\sC^P  +    (1-\varepsilon) \left .
  \frac{ d{\hat  \nu}_\sC^\P } {d \varepsilon} \right |_ {\varepsilon \to 1} + o((1-\varepsilon)^2)
\end{eqnarray*}
and recall $f_\sC^\P = \lim_{\varepsilon\to 1}  (1-\varepsilon) \hat \nu_\sC^P $. Now consider
\begin{eqnarray*}
\lim_{\varepsilon \to 1} (1-\varepsilon) {\tilde \nu}_\sC^\P (\varepsilon)
  =  \lim_{\varepsilon \to 1}  (1-\varepsilon)^2 \left .
  \frac{ d{\hat  \nu}_\sC^\P } {d \varepsilon} \right |_ {\varepsilon \to 1} \ne 0
\end{eqnarray*}
}

Further to ensure that the coalitions under consideration are operating, one would require conditions like that in {\bf A}.2. However these conditions are trivially satisfied in the limits $\gamma \to 1$, once $\dbar_\sMi > 0$ for all $i$; furthermore, the conditions will also be satisfied in the neighbourhood of $(1,1)$ (if required by shrinking the neighbourhood further) due to similar reasons.  

\ENMremoved{
\noindent{\bf ENM (essential, non-substitutable) regime: }
When $\varepsilon \approx 0$, as seen from \eqref{Eqn_Demand_Function}, the customers are `loyal' to their manufacturer,  the amount of demand $\varepsilon p_{i} \alpha_i$ that shifts from $M_i$ to the other manufacturer is negligible.  
We thus  consider \textit{a second regime in which the manufacturers are non-substitutable, but the products are essential, by considering systems near $(\varepsilon, \gamma) \approx (0, 1)$.} Towards this, in a similar manner, we consider the  limits (we will not require the derivatives in this case): 

\vspace{-4mm}
{\small$$
g_\sC^{\P} := \hspace{-3mm}  \lim_{ (\gamma, \varepsilon)  \to (1,0) } \hspace{-2mm}  (1-\gamma)\nu_\sC^\P  = \lim_{  \varepsilon  \to 0}  \lim_{ \gamma   \to 1 } (1-\gamma) \nu_\sC^\P.
$$}
}
\ignore{
Even  to that a configuration $(\P, \x) $ is unstable by blocking by a coalition $\C$ we actually prove strict inequality for \eqref{defn_blocking} which facilitates similar limit analysis.

Observe that the assumption {\bf{A.2}} modifies for two manufacturers in a way that 
$\dbar_\sM = \dbar_\sMo + \dbar_\sMt$, $ \alpha_\sM = \min\{\alpha_\sMo,\alpha_\sMt\}$ and $C_\sG = C_\sS + \min\{C_\sMo,C_\sMt\}$ and $O_\sG = O_\sS + \min \{O_\sMo, O_\sMt\}.$ but the similar conditions hold. Also observe that we are intereted in the regimes $(\varepsilon,\gamma) \to (1,1)$ and $(\varepsilon,\gamma) \to (0,1)$. Thus these assumptions on operating become trivially true at these regimes.}

\begin{table*}[ht]
 \addtolength{\tabcolsep}{-2pt} 
\begin{subtable}{.55\textwidth}
\centering    
\begin{tabular}{ |c | c |} \hline  \vspace{-3mm}
&\\ \vspace{-3mm}
$f^\GC_\sG = 0$ &  $\left (f^\CHC_\sS, f^\CHC_\sM \right ) = \left (0,0 \right)$ \\
&\\
 \hline  \vspace{-3mm}
&\\ \vspace{-3mm}
 $\left(f^{\VC_{i}}_{\sV_i}, f^{\VC_{i}}_\sMminusi \right) = \left( \frac{\dbar_\sM^2}{8\tilde\alpha}, 0 \right) $ \   & 
  $\left( f^\ALC_\sS, f^\ALC_\sMo, f^\ALC_\sMt \right) = \left(\frac{\dbar_\sM^2}{8\tilde\alpha},0, 0\right) $  \\
 &\\
\hline
\end{tabular}
\caption{Worth-limits}
\label{tab:Worth_limits}
\end{subtable}
%\end{minipage}
%\hspace{2mm}
 %\begin{minipage}{.35\textwidth}
\begin{subtable}{.35\textwidth}
\begin{tabular}{|l|}\hline \vspace{-3mm} \\ 

   $f^{(1),\VC_i}_{\sV_i}   =     \frac{2\dbar_\sMi\dbar_\sMminusi + \dbar_\sMminusi^2 -\dbar_\sMi^2}{16\tilde\alpha},  \  f^{(1),\VC_i}_{\sMminusi} = -\frac{\dbar_\sMminusi^2}{16\tilde\alpha} 
 $ 
 
\\
\vspace{-3mm}
 \\ \hline
 \vspace{-3mm}
 \\ 
 $f^{(1),\ALC}_{\sMi}   = \frac{-\left(5\dbar_\sMi + \dbar_\sMminusi \right)^2}{144\tilde\alpha}  \ \forall i  $, \  $f^{(1),\ALC}_{\sS}   =    \frac{\dbar_\sM^2}{8\tilde\alpha}  $
  \vspace{-3mm} \\ 
 \\ \hline
\end{tabular}
\caption{Derivative-limits \label{tab:derive_ESM}}
\end{subtable}
%\end{minipage}
\vspace{2mm}
\caption{\footnotesize ESM regime near $(\gamma, \varepsilon) \approx (1, 1)$ \label{tab:ESM}}

%\end{table*}

\vspace{5mm}

%\begin{table*}[ht]
 \addtolength{\tabcolsep}{-2pt}

\ENMremoved{
\begin{tabular}{lllll}
\cline{1-3}  \multicolumn{1}{|l|}{ } & \multicolumn{2}{l|}{ }   &  &  \vspace{-3mm} \\ 
\multicolumn{1}{|l|}{$g^\GC_\sG = \frac{\dbar_\sM^2}{4\tilde \alpha}$} & \multicolumn{2}{l|}{$\left (g^\CHC_\sS, g^\CHC_\sM \right ) = \left (\frac{\dbar_\sM^2}{8\tilde \alpha}, \frac{\dbar_\sM^2}{16\tilde \alpha} \right)$}                                                &  & \vspace{-3mm} \\ 
\multicolumn{1}{|l|}{ } & \multicolumn{2}{l|}{ }   &  &  \\  \cline{1-3} 
\multicolumn{1}{|l|}{ } & \multicolumn{1}{l|}{ }   & \multicolumn{1}{l|}{ } &  \vspace{-3mm} \\  
\multicolumn{1}{|l|}{$\left(g^{\VC_i}_{\sV_{i}}, g^{\VC_{i}}_\sMminusi \right) = \left( \frac{\dbar_\sMi^2}{4\tilde\alpha} + \frac{\dbar_\sMminusi^2}{8\tilde\alpha}, \frac{\dbar_\sMminusi^2}{16\tilde\alpha}\right) $} & \multicolumn{1}{l|}{\small$\left (g^\ALC_\sS, g^\ALC_\sMo ,g^\ALC_\sMt\right ) = \left (\frac{\dbar_\sMo^2}{8\tilde \alpha}, \frac{\dbar_\sMo^2}{16\tilde \alpha}, \zero \right)$} & 
\multicolumn{1}{l|}{$\left (g^\ALC_\sS, g^\ALC_\sMi \right) = \left (\frac{\dbar_\sM^2}{16\tilde \alpha}, \frac{(3\dbar_\sMi -\dbar_\sMminusi)^2}{64\tilde \alpha} \right)$}   &     \\ 
\multicolumn{1}{|l|}{for all $\dbar_\sMo,\dbar_\sMt$} & \multicolumn{1}{l|}{$\dbar_\sMo > \frac{1}{\sqrt{2}-1}\dbar_\sMt$} & \multicolumn{1}{l|}{$\dbar_\sMt < \dbar_\sMo \le \frac{1}{\sqrt{2}-1}\dbar_\sMt$} & 
 \vspace{-3mm} \\ 
\multicolumn{1}{|l|}{ } & \multicolumn{1}{l|}{ }   & \multicolumn{1}{l|}{ } &  \\  
\cline{1-3}
\end{tabular} 
\caption{ENM-regime: The worth-limits near $(\gamma, \varepsilon) \approx (1, 0)$  \label{table_EIP}}

\vspace{4mm}
The GC-core in ENM limit when $\dbar_\sMo > \frac{1}{\sqrt{2} -1}\dbar_\sMt$ 
\Details{
\begin{eqnarray}
     \left \{\beta\x: x_\sS > \frac{\dbar_\sMo^2}{8\tilde\alpha} ,  \ \  x_\sMo > \frac{\dbar_\sMo^2}{16\tilde \alpha} , \ \   x_\sMt > 0, \ \  
 x_\sS + x_\sMt > \frac{\dbar_\sMt^2}{4\tilde\alpha} + \frac{\dbar_\sMo^2}{8\tilde\alpha},   \ \ 
 x_\sS + x_\sMo >  \frac{\dbar_\sMo^2}{4\tilde\alpha} + \frac{\dbar_\sMt^2}  {8\tilde\alpha}, \ \ 
 x_\sMo + x_\sMt > \frac{\dbar_\sM^2}{16\tilde\alpha},   x_\sMo + x_\sMt + x_\sS = \frac{\dbar_\sM^2}{4\tilde\alpha} \right \}.
\end{eqnarray}

{\small \begin{eqnarray*}
{\cal C}  = \left \{ \beta \x : 
  \frac{\dbar_\sMo^2}{16\tilde\alpha}<   x_\sMo  \le \frac{\dbar_\sMo^2+ 4 \dbar_\sMt \dbar_\sMo}{8\tilde\alpha} ,  \
 \frac{(3\dbar_\sMt -\dbar_\sMo)^2}{64\tilde\alpha} <  x_\sMt \le \frac{\dbar_\sMt^2+ 4 \dbar_\sMt \dbar_\sMo}{8\tilde\alpha}, \ 
\frac{\dbar_\sMo^2}{8\tilde\alpha} <  x_\sS \le \frac{3 \dbar_\sM^2 }{16\tilde\alpha}, \    x_\sMo + x_\sMt + x_\sS = \frac{\dbar_\sM^2}{4\tilde\alpha} \right \} 
\end{eqnarray*}
}
 We thus have
\begin{eqnarray*}
0 <  x_\sMt &=& \nu_\sG^\GC - (x_\sMo + x_\sS) \le \frac{\dbar_\sM^2}{4 \tilde \alpha} - \frac{\dbar_\sMo^2}{4\tilde\alpha}  - \frac{\dbar_\sMt^2}{8\tilde \alpha} = \frac{\dbar_\sMt^2+ 4\dbar_\sMt \dbar_\sMo}{8\tilde\alpha} \\ 
\frac{\dbar_\sMo^2}{16\tilde\alpha}< x_\sMo &\le& \frac{\dbar_\sMo^2+ 4 \dbar_\sMt \dbar_\sMo}{8\tilde\alpha} \\
\frac{\dbar_\sMo^2}{8\tilde\alpha}< x_\sS &\le &  \frac{3 \dbar_\sM^2 }{16\tilde\alpha}
\end{eqnarray*}
}
{\small\begin{equation}
{\cal C}_> = \left \{ \beta \y : 
 \frac{\dbar_\sMo^2}{16\tilde\alpha} <   y_\sMo  \le  \Gamma_1 ,  \
 0 <  y_\sMt \le  \Gamma_2, \ 
\frac{\dbar_\sMo^2}{8\tilde\alpha} <  y_\sS \le \frac{3 \dbar_\sM^2 }{16\tilde\alpha}, \    y_\sMo + y_\sMt + y_\sS = \frac{\dbar_\sM^2}{4\tilde\alpha} \right \}  \mbox{ with } \Gamma_i = \frac{\dbar_\sMi^2+ 4 \dbar_\sMt \dbar_\sMo}{8\tilde\alpha}.
\label{Eqn_GC-core_ENMone}
\end{equation}}
 \Details{
Now the core in limit $\dbar_\sMt \to 0$ we have
\begin{eqnarray*}
{\cal C}_>& =& \left \{ \beta \x : 
 \frac{\dbar_\sM^2}{16\tilde\alpha} \le  x_\sMo  \le \frac{\dbar_\sM^2}{8\tilde\alpha} ,  \  \frac{\dbar_\sM^2}{8\tilde\alpha} \le  x_\sS \le \frac{3\dbar_\sM^2}{16\tilde\alpha}, \  0< x_\sMt , \   x_\sS + x_\sMo = \frac{\dbar_\sM^2}{4\tilde\alpha} \right \} \\
 &= &
 \left \{ \beta \x : x_\sMo =  \frac{\dbar_\sM^2}{8\tilde\alpha}  - \delta, \  x_\sS =  \frac{\dbar_\sM^2}{8\tilde\alpha}  + \delta, \ x_\sMt = 0 , \  \  0 \le  \delta \le  \frac{\dbar_\sM^2}{16\tilde\alpha}   \right \} .
\end{eqnarray*}

The GC - core in ENM limit in other case is:
\begin{eqnarray}
     \left \{\beta\x:  x_\sS > \frac{\dbar_\sM^2}{16\tilde\alpha} ,  \ \  x_\sMo > \frac{\dbar_\sMo^2}{16\tilde \alpha} , \ \   x_\sMt > \frac{(3\dbar_\sMt - \dbar_\sMo)^2}{64\tilde\alpha}, \ \  
 x_\sS + x_\sMt > \frac{\dbar_\sMt^2}{4\tilde\alpha} + \frac{\dbar_\sMo^2}{8\tilde\alpha},   
 x_\sS + x_\sMo >  \frac{\dbar_\sMo^2}{4\tilde\alpha} + \frac{\dbar_\sMt^2}  {8\tilde\alpha}, \ \ 
 x_\sMo + x_\sMt > \frac{\dbar_\sM^2}{16\tilde\alpha} \right \}.
\end{eqnarray}

We thus have 
\begin{eqnarray*}
\frac{(3\dbar_\sMt -\dbar_\sMo)^2}{64\tilde\alpha}<  x_\sMt &=& \nu_\sG^\GC - (x_\sMo + x_\sS) \le \frac{\dbar_\sM^2}{4 \tilde \alpha} - \frac{\dbar_\sMo^2}{4\tilde\alpha}  - \frac{\dbar_\sMt^2}{8\tilde \alpha} = \frac{\dbar_\sMt^2+ 4\dbar_\sMt \dbar_\sMo}{8\tilde\alpha} \\ 
\frac{\dbar_\sMo^2}{16\tilde\alpha}< x_\sMo &\le& \frac{\dbar_\sMo^2+ 4 \dbar_\sMt \dbar_\sMo}{8\tilde\alpha} \\
\frac{\dbar_\sM^2}{16\tilde\alpha}< x_\sS &\le &  \frac{3 \dbar_\sM^2 }{16\tilde\alpha}   
\end{eqnarray*}
}
The GC-core in ENM limit when $\dbar_\sMt \le \dbar_\sMo \le \frac{1}{\sqrt{2} - 1}\dbar_\sMt$
{ \begin{equation}\label{eqn_core_two}
{\cal C}_{eq}  = \left \{ \beta \y : 
  \frac{\dbar_\sMo^2}{16\tilde\alpha}<   y_\sMo  \le \Gamma_1,  \
 \frac{(3\dbar_\sMt -\dbar_\sMo)^2}{64\tilde\alpha} <  y_\sMt \le \Gamma_2, \ 
\frac{\dbar_\sMo^2}{8\tilde\alpha} <  y_\sS \le \frac{3 \dbar_\sM^2 }{16\tilde\alpha}, \    y_\sMo + y_\sMt + y_\sS = \frac{\dbar_\sM^2}{4\tilde\alpha} \right \}.
\end{equation}
\Details{Now the core when $\dbar_\sMo \approx \dbar_\sMt$
\begin{eqnarray*}
{\cal C}& =& \left \{ \beta \x : 
 \frac{\dbar_\sM^2}{64\tilde\alpha} \le  x_\sMi  \le \frac{5\dbar_\sM^2}{32\tilde\alpha} \ \forall i,    \  \frac{\dbar_\sM^2}{16\tilde\alpha} \le  x_\sS \le \frac{3\dbar_\sM^2}{16\tilde\alpha}, \    x_\sMt +  x_\sS + x_\sMo = \frac{\dbar_\sM^2}{4\tilde\alpha} \right \} 
 %
% &= &
% \left \{ \beta \x : x_\sMo =  \frac{\dbar_\sM^2}{8\tilde\alpha}  - \delta, \  x_\sS =  \frac{\dbar_\sM^2}{8\tilde\alpha}  + \delta, \ x_\sMt = 0 , \  \  0 \le  \delta \le  \frac{\dbar_\sM^2}{16\tilde\alpha}   \right \} .
 \end{eqnarray*}}}
}

\end{table*}

\vspace{-2mm}
\subsection {VC Partition}  
\vspace{-2mm}
Recall in   the partition with vertical cooperation,   $\VC_i$ the supplier collaborates with one of the manufacturers $M_i$ and competes with the other.
 The Stackelberg game is between the coalition $\V_i$ as leader and the   manufacturer  $M_{-i}$ as follower. The   
 manufacturer  $M_{-i}$ (when it operates) obtains raw material from $\V_i$, quotes $p_{-i}$ and the demand $D_\sMminusi$ attracted by $M_{-i}$ also contributes towards the revenue of $\V_i$;  the VC coalition $\V_i$ also derives utility due to its own demand $D_\sMi $ (recall here a direct price $p_i$ is quoted to the customers). Thus the utilities of the two coalitions are:

 \vspace{-3mm}
{\small\begin{align}\label{eqn_util_vc_v}
U_{\sV_i} &= \big [ D_\sMi(\a_\sM)(p_i - C_\sMi - C_\sS) + \F_\sMminusi D_\sMminusi (\a_\sM)  (q-C_\sS)   \nonumber  \\
& \hspace{45pt} 
 - O_\sS - O_\sMi \big ] \F_{\sV_i}  ,  \\
\label{eqn_util_vc_manu}
 U_\sMminusi &= \big (D_\sMminusi(\a_\sM)(p_{-i}-q-C_\sMminusi)\F_{\sV_i}   -  O_\sMminusi\big )\F_\sMminusi  .
\end{align}}
The SBE $(\a_{\sV_i}^{*}, a_\sMminusi^{*})$ (when exists) satisfies:% $a_\sMminusi^{*} = a_\sMminusi^{*}(\a_{\sV_i}^*)$

\vspace{-3mm}
{\small\begin{align*}
a_\sMminusi^{*} &= a_\sMminusi^{*}(\a_{\sV_i}^*),    \ \
a_\sMminusi^{*}(\a_{\sV_i}) := \arg \max_{a_\sMminusi} U_\sMminusi (a_\sMminusi, \a_\sV), 
\mbox{ and } \\
\a_{\sV_i}^{*} &\in \arg\max_{\a_\sV \in \Aset_{\sV_{i}}}U_{\sV_i} (\a_\sV, a_\sMminusi^{*}(\a_\sV)), 
\end{align*}}%
and defines the  worths,
{\small$
\nu_{\sV_{i}}^\VC =  U_{\sV_{i}} (\a^*) $} and {\small$    \nu_\sMminusi^\VC =  U_\sMminusi (\a^*).
$}
As already mentioned, it is complicated to analyze this game theoretically,  we instead obtain the ESM limits in the following:

\begin{lemma}\label{lem_vc_esm_worth_limits}
Assume {\small$\alpha_\sMo = \alpha_\sMt = \alpha$.}  The worth-limits {\small$(f^{\VC_{i}}_{\sV_i}, f^{\VC_{i}}_\sMminusi )$} and the derivative limits {\small$(f_{\sV_{i}}^{(1),\VC_i},f_{\sMminusi}^{(1),\VC_i})$} for ESM regime are respectively in Tables \ref{tab:Worth_limits}  and  \ref{tab:derive_ESM}. \ENMremoved{he worth-limits {\small$(g_{\sV_i}^{\VC_i} , g_{\sMminusi}^{\VC_i})$} for ENM regime are in   Table~\ref{table_EIP}.}
\ignore{
\begin{itemize}
    \item The worth-limits \eqref{eqn_worth_lim} and the required derivative limits \eqref{Eqn_lim_derivatives}   under VC partition  in the ESM regime are given by:
 \begin{eqnarray*}
f_{\sV_i}^{\pVC}  &=&
\frac{\dbar_\sM^2}{8\tilde\alpha}, \  \ 
f_\sMminusi^{\pVC}   \ =  \  0 \mbox{ for all } i, \mbox{ and }\\
f_{\sV_i}^{(1),\VC_i}  &=&  
\frac{2\dbar_\sMi\dbar_\sMminusi + \dbar_\sMminusi^2 -\dbar_\sMi^2}{16\tilde\alpha} .
\end{eqnarray*}
\end{itemize}
\begin{itemize}
    \item The worth-limits   in the ENM regime equal:
\begin{eqnarray*}
g_{\sV_i}^{\pVC}  =
\frac{\dbar_\sMi^2}{4\tilde\alpha} + \frac{\dbar_\sMminusi^2}{8\tilde\alpha},  \mbox{ and }
g_{\sMminusi}^{\pVC}  =  \frac{\dbar_\sMminusi^2}{16\tilde\alpha}.
\end{eqnarray*}
\end{itemize}}
\end{lemma}
\Rev{\begin{proof}
Refer to \TR{\cite{TR}}{Appendix \ref{sec_Appendix_VC}} for the proof.  
\end{proof}
}

\ignore{\ignore
{\color{red} 

Fix a partion  $\P$ and then a coalition $\C$ and consider $\nu_\sC^\P$. Now consider the following function of $(\gamma, \varepsilon)$, $f_\sC^\P (\gamma, \varepsilon) := (1-\gamma)(1-\varepsilon) \nu_\sC^\P$; sufficient to work with $\{f_\sC^\P\}_{\sC, \P}$ as we compare for a given pair $(\gamma, \varepsilon)$. If we can establish that the function $(\gamma, \varepsilon) \mapsto f_\sC^\P$ is continuous, then  we have by $R^2$ topology that 
$$
f_\sC^{\P, *} :=  \lim_{ (\gamma, \varepsilon)  \to (1,1) }  f_\sC^\P  = \lim_{  \varepsilon  \to 1}  \lim_{ \gamma   \to 1 }  f_\sC^\P
$$
}
}

%%%%%%%%%%%%%%%%%%%%%%%%%%%%%%%%%%%%%%%%%%%%%%%%%%%%%%%%%
\vspace{-2mm}
\subsection{ALC Partition}
\vspace{-2mm}
Recall the partition, $\ALC = \{ \S, \Mo , \Mt \}$, where all the agents operate alone and compete with each other. In this  partition, we have a SB game, where supplier ~$\S$ is the leader, quoting its price via the action~$a_\sS$; the competing manufacturers~$\{ \Mo , \Mt\}$ are followers in the lower echelon, who respond to~$a_\sS$ via a non-cooperative strategic form game (inner game between the manudacturers), parametrized by fixed action of supplier~$a_\sS$. As a leader of SB game, the supplier is aware of the Nash Equilibrium (NE)~$\a_\sM^* (a_\sS)$ of the inner game~$\a_\sM^* (a_\sS)$ for every~$a_\sS$. The utility of each of these agents are given in~\eqref{Eqn_UtilityofManufacturer}-\eqref{UtilityofSupplier}. The utility of the supplier $\S$ and a manufacturer $\M_i$ is ,  

\vspace{-3mm}
{\small
\begin{align}
    \label{eqn_supplier_monopoly_supplier_utility}
U_{\sS}(a_\sS) &= \left( \left (\sum_{i=1}^{2}D_{\sMi}(\a^*_\sM (a_\sS))\F_\sMi \right )(q-C_{\sS})-O_{\sS}\right)\F_\sS ,\\
\label{suppliermonopolymanufacturerutility}
U_{\sMi}(\a_\sM; a_\sS) & = \left((D_{\sMi}(\a_\sM(a_{\sS}))(p_{i}-C_{\sMi}-q)\F_\sS-O_{\sMi}\right)\F_\sM .  
\end{align}}
The equilibrium of the SB game  satisfies

\vspace{-3mm}
{\small \begin{align*}%\label{bestresponsesupplierj}
 a_\sS^*  & = \arg \max_{a_{\sS}} U_\sS \left (a_{\sS},  \a_\sM^* (a_\sS))  \right ),  \\
 \a_\sMi^* & = \arg \max_{a_{\sMi}} U_{\sMi}(a_{\sMi}, a_{\sMminusi }; a_\sS^*) \text{ for all } i.
\end{align*}}
\noindent The worth of supplier $\S$ is, {\small$ \nu_\sS^\ALC =  U_\sS (a_\sS^*) $}, and the worth of manufacturer $\Mi$ is, {\small$  \  \nu_\sMi^\ALC =  U_\sMi (\a_\sM^*).$} As already mentioned, it is complicated to analyze this game theoretically,  we instead obtain the ESM limits in the following:
\ignore{In SB game, the supplier begins with quoting its  prices, via the action~$a_\sS$.  The manufacturers respond to this, via a non-cooperative strategic form game, parametrized by fixed~$a_\sS$. This game (representing the response against~$a_\sS$) is between the two manufacturers and is described by~$\left < \{ \Mo, \  \Mt \}, \{ \Aset_\sMo, \Aset_\sMt\}, \{W_\sMo , W_\sMt \} \right >$, where the utilities of the manufacturers $\M_i$ is,
$$
W^{(a_\sS)}_{\sMi }(\a_\sM)  := U_\sMi (\a_\sM, a_\sS), \ \forall \ \a_\sM  \in \Aset_\sMo \times \Aset_\sMt  ,\ \forall \ i. 
$$
\begin{align}
\label{Eqn_NE_inner_game}
W^{(a_\sS)}_\sMi (a_\sMi^* (a_\sS), a^*_{\sM_{-i}}(a_\sS),a_\sS) \ge W_\sMi (a_\sMi , a^*_{\sM_{-i}}(a_\sS), a_\sS) \nonumber  \\ 
\mbox{ for all }  a_\sMi \in \Aset_\sMi \mbox{ and  for all }  \Mi. 
\end{align}

The optimal action of the supplier against this NE is: 

\vspace{-3mm}
{\small \begin{eqnarray}\label{bestresponsesupplierj}
 a_\sS^*   = \arg \max_{a_{\sS}} U_\sS \left (a_{\sS},  \a_\sM^* (a_\sS))  \right ),   
\end{eqnarray}}
where the utility of the supplier is given by:  

\vspace{-3mm}
{\small
\begin{equation}
    \label{eqn_supplier_monopoly_supplier_utility}
U_{\sS}(a_\sS)= \left( \left (\sum_{i=1}^{2}D_{\sMi}(\a^*_\sM (a_\sS))\F_\sMi \right )(q-C_{\sS})-O_{\sS}\right)\F_\sS  . 
\end{equation}}
Basically the supplier  maximizes the above utility   to find the optimal price ($q^{*}$) at which it would supply the raw material to both the manufacturers (when it is optimal to operate, i.e, when $\BR_{\sS}  \ne \{ n_o \}$).
To compute the equilibrium of the Stackelberg game, we begin with analysis of inner game between the manufacturers, whose solution is given by \eqref{Eqn_NE_inner_game}. 
To this end, recall the utility of the manufacturers (for each~$i \in \{1, 2\}$), for any given $a_{\sS} $ is,
\begin{equation}\label{suppliermonopolymanufacturerutility}
U_{\sMi}=\left((D_{\sMi}(\a_\sM(a_{\sS}))(p_{i}-C_{\sMi}-q)\F_\sS-O_{\sMi}\right)\F_\sM.    
\end{equation}
}
\begin{lemma} \label{lem_ALC_worth_limits}
Assume {\small$\alpha_\sMo = \alpha_\sMt = \alpha$.}  The worth-limits {\small$(f^{\ALC}_{\sS}, f^{\ALC}_{\sMi}, f^{\ALC}_\sMminusi)$} and the derivative limits {\small 
$(f^{(1),\ALC}_{\sS},$ \ $f^{(1),\ALC}_{\sMi} \ f^{(1),\ALC}_{\sMminusi})$
} for ESM regime are respectively in Tables \ref{tab:Worth_limits}  and  \ref{tab:derive_ESM}. 
\ENMremoved{
The worth-limits {\small$(g_{\sV_i}^{\VC_i} , g_{\sMminusi}^{\VC_i})$} .}

\ignore{
   Assume $\tilde\alpha_\sMo = \tilde\alpha_\sMt = \tus{ \tilde\alpha}$ and $\dbar_\sMo \ge \dbar_\sMt$, then we have the following:
   \begin{itemize}
       \item The worth limits of the agents in the ALC case in ESM regime are given by:
   \begin{eqnarray*}
 f_\sS^{\pALC}  =\frac{\dbar_\sM^2}{8\tilde\alpha}, 
f_\sMi^{\pALC}  = 0   \mbox{for all }i, \mbox{ and }
\end{eqnarray*}

   \item The worth-limits of the agents in the ALC case in ENM regime are given by:
   
If  $\dbar_\sMo \ge \frac{1}{\sqrt{2} - 1}\dbar_\sMt$ then
   \begin{eqnarray*}
  g_\sS^{\pALC} = \frac{\dbar_\sMo^2}{8\tilde\alpha}, \ 
  g_\sMo^{\pALC} =  \frac{\dbar_\sMo^2}{16\tilde\alpha}, \mbox{ and }
    g_\sMo^{\pALC} =  0 .   
   \end{eqnarray*} 
  In the other case, we have that
\begin{eqnarray*}
g_\sS^{\pALC} =   \frac{\dbar_\sM^2}{16\tilde\alpha}, \mbox{ and }
g_\sMo^{\pALC} =   \frac{(3\dbar_\sMi-\dbar_\sMminusi)^2}{64\tilde\alpha} \forall i.
\end{eqnarray*}
\end{itemize}
}
\end{lemma}
\Rev{\begin{proof}
Refer to \TR{\cite{TR}}{Appendix \ref{sec_Appendix_AA}} for the proof.  
\end{proof}
}

\vspace{-5mm}
\section{Stability Results -- ESM regime} \label{sec:stability_results}
\vspace{-2mm}
We have derived the worths $\{\nu_\sC^\P\}$ and the worth-limits $\{f_\sC^\P\}$ in the previous section and now aim to identify the partitions and configurations that are stable in ESM regime. % 
\ENMremoved{
We begin with 
  the ESM regime.}% 
We  have worth-limits only for VC and ALC partitions, and thus for the comparison purposes (see \eqref{defn_blocking},~\eqref{eqn_worth_pa} and~\eqref{eqn_block_merger}),  compute the same for GC and HC partitions. By Corollaries~\ref{cor_gc}-\ref{cor_hc}, $f_\sG^\GC =  
f_\sM^\CHC = f_\sS^\CHC = 0$; these are also tabulated in Table \ref{tab:ESM}.  

From Table~\ref{tab:ESM}, the immediate result is that,  the GC and HC partitions (irrespective of the pay-off vectors) are both  blocked   by  coalition $\V_i$. Thus none among these two    partitions are  stable. 
As discussed at  the end of Subsection~\ref{sec_market_seg}, the customers may no longer feel the product is essential   in the absence of choices, and this may be the reason for  non-stability of the GC and HC partitions.\footnote{Observe when a worth-limit is non-zero,the corresponding worth/optimal-revenue increases to infinity as $(\varepsilon, \gamma) \to (1,1)$ and this is true only for worths related  to VC and ALC partitions in ESM regime; by Corollaries \ref{cor_gc}- \ref{cor_hc}  the worths for GC and HC partitions also  increase to infinity  with $\gamma \to 1$, but not with $\varepsilon \to 1$ and hence the corresponding worth-limits are zero.}

We now identify the stable configurations. Towards this,
the  pessimal anticipatory worth-limits   (see equation \ref{defn_blocking}), using Table \ref{tab:ESM}, are:
\begin{eqnarray} \label{eqn_PAC_utilities}
  f^{pa}_\sS &=& \min \{ f^\CHC_\sS, f^\ALC_\sS \} = 0\nonumber  \\ f^{pa}_\sM &=& f^\CHC_\sM = 0  \label{Eqn_horizontal_pessimal_limit} \\ 
  f^{pa}_\sMi &=& \min \{ f^{\VC_{-i}}_\sMi, f^{\ALC}_\sMi \}  = 0 ,  \ f^{pa}_\sG = 0 , \mbox{ and }  \nonumber \\
  f^{pa}_{\sV_{i}} &=& f^{\VC_{i}}_{\sV_{i}} = \frac{\dbar_\sM^2}{8\tilde\alpha} . 
  \label{Eqn_pessimal_limit}
\end{eqnarray}
\ignore{Thus the pessimal utilities of all single agents is zero.  Hence, \textit{any configuration   $(\VC_i, \nicefrac{\x}{((1-\varepsilon)(1-\gamma) )})$ involving vertical cooperation and with $x_\sS \ne 0, x_\sMo \ne 0$ is stable} -- merger GC gets zero (which is obviously lesser) and any split of $\V_i$, either $\S$ or $\M_i$,   also anticipates to get zero; }
In the following we prove  that:  i)  the  partitions $\ALC$  and $\VC_2$ are not stable, while   $\VC_1$ is stable  when 
\begin{eqnarray}\label{eqn_cond_stability}
    \dbar_\sMo >  
    (\sqrt{2} +1) \dbar_\sMt  ;
 \end{eqnarray}%
and ii) none of the partitions are stable when the above inequality \eqref{eqn_cond_stability}  is negated   strictly. 
 Towards this we establish few (strict) inequalities at ESM limit  for each configuration -- the set of inequalities  are strict and demonstrate either that the configuration is stable or that a coalition blocks it. 
Then the said configuration remains stable (or is blocked by the said coalition) for all $(\gamma, \varepsilon)$ around $(1,1)$ and in a set as in \eqref{Eqn_gam_eps_nhd} of  subsection \ref{subsec_worth_limit}.  We begin with ALC partition. 

\medskip
 \noindent
\textbf{ALC partition is not stable:}
Note that there is only one scaled configuration  at limit involving ALC $(\ALC, \y)$ with $y_\sS = f^\ALC_\sS$, $y_\sMo = f^\ALC_\sMo$ and $y_\sMt = f^\ALC_\sMt$, because of  consistency. The   coalitions that can possibly block ALC partition are the merger coalitions such as $\G$, $\M$, $\V_{i}$.
%Clearly, from Table~\ref{tab:Worth_limits} and \eqref{Eqn_pessimal_limit}, the mergers $\G $ and $\M$  cannot block the ALC partition (strict inequalities are satisfied).
Consider a merger $\V_{i}$  which gets the same utility as  $y_\sS +y_\sMi$ of ALC at  limit. Thus we compare  using  the derivative limits of Table \ref{tab:derive_ESM} , 

 \vspace{-3mm}
 {\small\begin{align}
 f^{(1),\ALC}_{\sS} + f^{(1),\ALC}_{\sMi} &= - \frac{\dbar_\sM^2}{8\tilde\alpha} +  \frac{\left(5\dbar_\sMi + \dbar_\sMminusi \right)^2}{144\tilde\alpha} \nonumber\\
& <  \frac{\dbar_\sMi^2 - 2\dbar_\sMi\dbar_\sMminusi - \dbar_\sMminusi^2 }{16\tilde\alpha} = f^{(1),\VC}_{\sV_{i}},\label{eqn_derv_lim_aa}    
 \end{align}}%
and hence the merger $\V_i$ blocks $\ALC$ (see \eqref{eqn_taylor_series} and observe $(\varepsilon - 1)$ is negative) and this is true with a strict inequality at limit. Therefore, $\ALC$ is not a stable partition in the ESM regime.

Continuing in this manner we obtain the following result (the remaining proof is in the Appendix):

\ignore{
\noindent{\bf Stable VC based configurations:}
Any scaled configuration  at ESM limit involving vertical cooperation,  $(\VC_i,  \y)$ with  $y_\sMi > 0$ is not stable -- this is because with non-zero  $y_\sMi$  we have $y_\sS <  \nicefrac{\dbar_\sM^2}{8\tilde\alpha}$  and so the coalition $\V_{-i} = \{S, M_{-i}\}$   blocks it (see Table~\ref{tab:Worth_limits} and observe  $y_\sMminusi =f_\sMminusi^{\VC_i} = 0$), as:
\begin{equation*}
  y_\sS + y_\sMminusi <  \frac{\dbar_\sM^2}{8\tilde\alpha} + 0 =  f^{pa}_{\sV_{-i}}    .    
\end{equation*}
On the other hand, when $y_\sMi = 0$,  we have that: $$y_\sS = f^{\VC_i}_{\sV_i} =\frac{\dbar_\sM^2}{8\tilde\alpha}=f^{\VC_{-i}}_{\sV_{-i}} = f^{pa}_{\sV_{-i}} , $$ and hence
one needs to compare the corresponding derivative limits of Table \ref{tab:derive_ESM}. Comparing the same, \textit{ when $\dbar_\sMo > \dbar_\sMt$,  the only stable scaled configuration at limit involving vertical cooperation is} 
\begin{eqnarray}\label{eqn_stable_config}
 \left (\VC_2, \  \left (\frac{\dbar_\sM^2}{8\tilde\alpha}, 0, 0\right ) \right )   
\end{eqnarray}
and this is    because:  
$$
(\varepsilon-1) \left ( f^{(1), \VC_i}_{\sV_i} - f^{(1), pa}_{\sV_{-i}} \right ) = 
\frac{\dbar_\sMminusi^2 -\dbar_\sMi^2}{8\tilde\alpha} > 0 \mbox{   when } i = 2 $$
and is strictly negative for  $i = 1. 
$ 
\textit{When $\dbar_\sMo = \dbar_\sMt$,} then by symmetry the two worths are equal and hence $\nu_{\sV_i}^{\VC_i} = \nu_{\sV_{-i}}^{pa}$ for both $i$; further from \eqref{defn_blocking} blocking is possible only with strict improvement,  and hence \textit{we have two stable scaled configurations at limit involving vertical cooperation}:
{\small
\begin{equation*}
\left (\VC_1, \  \left (\frac{\dbar_\sM^2}{8\tilde\alpha}, 0, 0\right ) \right ) \mbox{ and }  \left (\VC_2, \ \left (\frac{\dbar_\sM^2}{8\tilde\alpha}, 0, 0\right ) \right ) . 
\end{equation*}}
}

\Canticipate{

Thus the pessimal core associated with $\VC_i$ is 
$$
{\cal C}^{pa} ( {\VC}_i )= 
\left 
\{ \x : x_\sS > 0, x_\sMi > 0 , x_\sMminusi = 0, x_\sS + x_\sMi = \frac{ \dbar_{\sM}^2 }{8 \tilde \alpha} \right \} 
$$

On the other hand if agents anticipate according Cournot's anticipation rule (where they assume others to remain in their positions), the corresponding anticipatory utilities after a split from VC are:
\begin{equation} \label{eqn_CAC_utilities}
 f^{ca}_\sS =  f^\ALC_\sS  =  \nicefrac{\dbar_\sM^2}{8\tilde \alpha},   \mbox{ and }    f^{ca}_\sMo =   f^\ALC_\sMo = 0    
\end{equation}
Hence under Cournot's anticipation rule 
  any configuration $(\VC, \x)$ involving vertical cooperation is not stable if $x_\sMo > 0$ due to following reason:  as  (scaled) $\x$ is consistent with $\VC$, we have that $x_\sS + x_\sMo = f^\VC_\sV$;  hence  $x_\sS < f^\VC_\sV$ when  
$x_\sMo > 0$ and then   the configuration $(\VC, \x)$ is blocked by  coalition $\S$ (from Table \ref{table_ECL} and \eqref{eqn_CAC_utilities}). Thus the only configuration stable with VC is $(\VC, (\nicefrac{\dbar_\sM^2} {8 \tilde \alpha}, 0, 0) )$, i.e., 
$$
{\cal C}^{ca} ( {\VC}_i )= 
\left 
\{ \x :  x_\sMi =  0 , x_\sMminusi = 0, x_\sS  = \frac{ \dbar_{\sM}^2 }{8 \tilde \alpha} \right \} 
$$}

\Canticipate{
When a partition is blocked under  pessimal anticipation it would also be blocked under Cournot's anticipation.}

\begin{theorem}{\bf[ESM regime]}\label{thm_esm_regime} 
Consider any $\tilde \alpha, \dbar_\sMo, \dbar_\sMt, \{C_\sC\}$ and $\{O_\sC\}$ with $\dbar_\sMo\ge \dbar_\sMt$. 
   There exists a $\bar \varepsilon<1$ such that for every $  \varepsilon \in ( \bar \varepsilon, 1)$, there exists a $\bar \gamma_\varepsilon< 1$ and for any system with above parameters and
 with  $(\varepsilon, \gamma) \in \{ \varepsilon \ge \bar \varepsilon, \gamma \ge \bar \gamma_\varepsilon \}$,
the following are true:
\begin{enumerate}[$i)$]
    \item When $\dbar_\sMt \le \dbar_\sMo <  
    (\sqrt{2} +1) \dbar_\sMt$, none of the partitions are stable. 
     
    \item When $ \dbar_\sMo >  
    (\sqrt{2} +1) \dbar_\sMt$, then only   
$\VC_1$ partition is stable. Further  the configuration $(\VC_1, \x)$ is stable if
\begin{eqnarray}\label{eqn_share_of_manu}
 x_\sMo  
 \in \left (
\nu_\sM^{pa} - \nu_\sMt^{\VC_1}     , \       \nu_{\sV_1}^{\VC_1}  - \nu_{\sV_2}^{\VC_2} + 
      \nu_\sMt^{\VC_1} \right ), 
\end{eqnarray}and the above interval is non-empty.   \eop

\RB{     \item Partition $\VC_i$ is stable against restricted blocking    for each $i$, further, the configuration $(\VC_i, \x)$  is restricted-stable if 
     $
x_\sMi \ge \nu_\sM^{\CHC} - \nu_\sMminusi^{\VC_i}.
$}   
\end{enumerate}

\end{theorem}

\ignore{
({\color{red} wrong result, we dont have such a good result, need to write what we have properly}):

\begin{theorem}{\bf[ESM regime]} 
Consider any $\tilde \alpha, \dbar_\sMo, \dbar_\sMt, \{C_\sC\}$ and $\{O_\sC\}$. 
   There exists a $\bar \varepsilon<1$ such that for every $  \varepsilon \in ( \bar \varepsilon, 1)$, there exists a $\bar \gamma_\varepsilon< 1$ and for any system with above parameters and with  $(\varepsilon, \gamma) \in \{ \varepsilon \ge \bar \varepsilon, \gamma \ge \bar \gamma_\varepsilon \}$,
the following are true:
\begin{enumerate}[$i)$]
    \item when $\dbar_\sMo > \dbar_\sMt$,
only two   configurations are stable one with 
   $\VC_1$ partition and the other with $\ALC$ partition. 
   \item when $\dbar_\sMo = \dbar_\sMt$. there are exactly three stable configurations one each with $\VC_1$, $\VC_2$ and $\ALC$ partitions.
\end{enumerate}
In all the cases, the pay-off vectors $\x$ in the stable configuration are approximately $\nicefrac{\x}{\beta} \approx  ( \nicefrac{\dbar_\sM^2}{8\tilde\alpha} , 0, 0). $  \eop
\end{theorem}}

\noindent {\textbf{Remarks:}}  When there is a huge disparity between the two manufacturers in terms of the dedicated market demands,
the SC has stable configurations. The supplier prefers to collaborate with stronger manufacturer ($\VC_1$ is stable) and the weaker manufacturer has no choice (no partner finds it beneficial to oppose $\VC_1$ by collaborating with weaker $M_2$).

However the share of the manufacturer that collaborates with supplier is negligible in comparison to that of the supplier  (the upper bound on scaled pay-off $(1-\varepsilon)(1-\gamma)x_\sMo$ to 0 as seen from \eqref{eqn_share_of_manu} and Table \ref{tab:ESM}, while the lower bound on that of the supplier  converges to $\nicefrac{\dbar_\sM^2}{8\tilde\alpha}$);  in fact the scaled share of the non-collaborating manufacturer $M_2$ also converges to~0 (from Table \ref{tab:ESM} $f_\sMt^{\VC_1} = 0$). Thus in the essential and substitutable manufacturer regime, the supplier has even higher advantage than that in the single manufacturer SC -- the competition at the lower echelon added with substitutability significantly improved the benefits and the position of the supplier.

Another interesting aspect is that the supplier prefers to collaborate with stronger manufacturer ($\VC_1$ is stable, but not $\VC_2$) -- in the prelimit, the supplier derives better  coalitional utility when it collaborates with stronger $M_1$. 

On the other hand, when the manufacturers are of comparable strengths, as in part i of Theorem \ref{thm_esm_regime}, the SC has  no stable configuration. The system probably keeps switching configurations (any operating configuration is blocked by one or the other coalition).  \RB{ We however have stability if the SC is only opposed by mergers and splits of the existing coalitions. }

\vspace{-5mm}
\section{Conclusions}
\vspace{-2mm}
\Rev{
The main takeaway of this work is that it establishes the possible stability of vertical mergers (the collaboration between the supplier and a manufacturer) and instability of centralised supply chain (all the members of SC); this is true for the SC that supplies essential goods and that caters to not-so loyal customers. This is in contrast to the current literature which usually establishes the stability of centralized SC. This contrast is the consequence of the realistic consideration of the partition-form aspects -- where the worth of a coalition depends upon the arrangement of the agents outside the coalition.  When the manufacturers are significantly different in terms of market power, the vertical cooperation (or merger) between the supplier and the stronger manufacturer is stable, while the weaker manufacturer is left-out to compete with the collaborating pair. Surprisingly, no collaboration is stable when the manufacturers are of comparable market strengths.}

\ignore{
We examine a two-manufacturer (lower echelon) and one supplier (higher echelon) supply chain to analyze stable operating configurations. considers that the centralised supply chain (grand coalition) is stable, contrary to the belief in the existing literature that states the stability of the completely centralised supply chain (grand coalition), when the manufacturers are significantly different in terms of market power and the supply chain deals with the product that is essential to customers (necessities, not luxury items), the vertical merger between the stronger manufacturer and the supplier is stable (i.e. the supplier prefers to merge with the manufacturer that dominates the market). Observe that the manufacturer with less market power is left out of the merger, who in turn will compete with the merger for the market share.

Additionally/More interestingly/Surprisingly, when the manufacturers are of similar market powers and the supply chain deals with the essential product, the customers are majorly indifferent to the manufacturers. Surprisingly, in this scenario, no configuration is stable, implying that no agent finds it beneficial to merge with other agents, nor does it find it beneficial to stay alone, leading to the instability of the system. 
Contrary to the common belief in the literature, collaboration among the supply chain agents may not always be beneficial. 

We observe that when the product is essential to customers (necessities, not luxury items) (which means that it is not a luxurious item and is important for survival) and if the manufacturers are of similar reputations in the market, the customers are majorly indifferent to the manufacturers. Additionally, when the manufacturers are significantly different in terms of market power, the vertical cooperation (or merger) between the supplier and the stronger manufacturer is stable (i.e. the supplier prefers to merge with the manufacturer that dominates the market); the weaker manufacturer is left-out to compete with the collaborating pair. Surprisingly, no collaboration is stable when the manufacturers are of comparable market strength.

We consider a   two-manufacturer (lower echelon) and one supplier (higher echelon) supply chain to study the operating configurations that are stable. As opposed to the common belief in the literature, the grand coalition of all agents  is not stable -- when the product is essential, the customers are majorly indifferent to the manufacturers and the manufacturers are of similar reputations. Additionally, if the manufactures are significantly  different in terms of market power, the vertical cooperation between the supplier and the stronger manufacturer is stable; the weaker manufacturer is left-out to compete with the collaborating pair. Surprisingly, no collaboration is stable when the manufacturers are  of comparable market strength.}

In reality, the worth of any coalition depends upon the arrangement of opponents in the market space; for example,  the revenue generated by the supplier with the  two manufacturers operating together is different from that when the two manufacturers operate independently. This realistic aspect is captured by studying the SC using partition-form games, which paved way to the above mentioned contrasting results.  
%These games are examples of  what are known as partition form games, where the worth of any coalition depends upon the arrangement of opponents.
%, our results are drastically different from the known results in literature, as we captured a more realistic framework for supply chains via partition form games. 

Further, the competition at the lower echelon significantly favors the higher level supplier -- 
 the supplier enjoys a huge fraction of the revenue generated, while  the manufacturers draw a negligible fraction irrespective of their market powers and irrespective of whether they collaborate with the supplier or not.
 %If a single manufacturer operates at lower echelon, it derives a better revenue, however the supplier still derives a larger share. 

\Rev{This research has opened up many new questions -- which configurations are stable in an SC that supplies luxury goods (non-essential) or in a SC with loyal customers?  More interesting questions are about the stable configurations with competition at both the echelons along with vertical competition.}

%happens if some customers are loyal to a   manufacturer? what happens if the product is not significantly essential? More interesting questions are about the stable configurations with competition at both the echelons along with vertical competition.

\vspace{-4mm}

\ENMremoved{
\vspace{-2mm}
\subsection{ENM Regime}
\vspace{-2mm}

\subsubsection{When Manufacturer 1 has significantly large customer base}
The pessimal anticipation for the case when $\dbar_\sMo > \frac{1}{\sqrt{2}-1}\dbar_\sMt$, i.e., when   $\dbar_\sMo > 2.41 \dbar_\sMt$
$$
g^{pa}_{\sMt} = 0, \  g^{pa}_{\sS} =  \frac{\dbar_\sMo^2}{8\tilde\alpha} 
\mbox{ and }
g^{pa}_\sMo = \frac{\dbar_\sMo^2}{16\tilde \alpha}  \mbox{, further,  }
$$
$$
g^{pa}_{\sV_{i}} = \frac{\dbar_\sMi^2}{4\tilde\alpha} + \frac{\dbar_\sMminusi^2}{8\tilde\alpha}\mbox{ , and } g^{pa}_{\sM} = \frac{\dbar_\sM^2}{16\tilde\alpha}
 $$
The configuration $(\G, \x)$ with
 payoff vector $\x = (x_\sS, x_\sMo, x_\sMt)$  is stable when:
 $$
 x_\sS > \frac{\dbar_\sMo^2}{8\tilde\alpha} ,  \ \  x_\sMo > \frac{\dbar_\sMo^2}{16\tilde \alpha} , \ \   x_\sMt > 0, \ \  
 x_\sS + x_\sMt > \frac{\dbar_\sMt^2}{4\tilde\alpha} + \frac{\dbar_\sMo^2}{8\tilde\alpha},   
 $$
 $$
 x_\sS + x_\sMo >  \frac{\dbar_\sMo^2}{4\tilde\alpha} + \frac{\dbar_\sMt^2}  {8\tilde\alpha}, \ \ 
 x_\sMo + x_\sMt > \frac{\dbar_\sM^2}{16\tilde\alpha}.
 $$
 Thus the core would be empty and one would not have a single pay-off vector satisfying the above if any of the following
the impossibility conditions  are true:
$$
%
%x_\sS + x_\sMo +x_\sMt 
%
 \dbar_\sM^{2} 
\le  \dbar_\sMt^2  + \frac{3\dbar_\sMo^2}{4}
\mbox{, 
or }
 \dbar_\sM^{2} \le  \dbar_\sMo^2 + \frac{\dbar_\sMt^2}{2} \mbox{ or }
\dbar_\sM^{2} \le  \frac{\dbar_\sMo^2}{2} + \frac{\dbar_\sM^2}{4}
$$
Clearly none of the above conditions are satisfied and hence we have a non-empty core. 
 One such example payoff vector is $x_\sS = \frac{\dbar_\sMo^2}{8\tilde\alpha} + \frac{\dbar_\sMt^2}{4\tilde\alpha}$, $x_\sMo = \frac{\dbar_\sMo^{2}}{8\tilde\alpha}$ and $x_\sMt = \frac{\dbar_\sMo\dbar_\sMt}{2\tilde\alpha}$.

\subsubsection{When the two manufacturers are not too different in terms of customer base}
 The pessimal anticipation for the case when $\dbar_\sMt < \dbar_\sMo \le \frac{1}{\sqrt{2}-1}\dbar_\sMt$
{\color{red}
\begin{eqnarray*}
 g^{pa}_{\sS} =   \frac{\dbar_\sM^2}{16\tilde \alpha} 
, \quad 
g^{pa}_\sMo = \frac{\dbar_\sMo^2}{16\tilde\alpha} , \quad  g^{pa}_{\sMt} =  \frac{(3\dbar_\sMt -\dbar_\sMo)^2}{64\tilde \alpha} , \quad \\
g^{pa}_{\sV_1} = \frac{\dbar_\sMo^2}{4\tilde\alpha} + \frac{\dbar_\sMt^2}{8\tilde\alpha}  , \quad 
 g^{pa}_{\sV_2} = \frac{\dbar_\sMt^2}{4\tilde\alpha} + \frac{\dbar_\sMo^2}{8\tilde\alpha},\text{ and } g^{pa}_{\sM} = \frac{\dbar_\sM^2}{16\tilde\alpha}.
 \end{eqnarray*}
 }
One of the impossibility conditions are:
$
d_1^2+ d_2^2 - 38 d_1 d_2  \stackrel{?}{ >} 0 
 $
But the above is never true, as using the bounds of this case we have
 $
d_1^2+ d_2^2 - 38 d_1 d_2   < 2.4^2 d_2^2 + d_2^2 - 38 d_2^2  =  (5.4-38) d_2^2 < 0
 $

\begin{theorem}
In the ENM regime, GC completely blocks all the other partitions.    
\end{theorem}

}

\bibliographystyle{apalike}

\bibliography{example}

\vspace{-5mm}
\section*{APPENDIX}
\vspace{-2mm}
In this appendix, we consider the following  optimization problem and derive its solution:
\begin{eqnarray}
\label{Eqn_Sing_sing}
U(a) = \left( (\bar{d} - \alpha p)^+(p - c) - O_{c}\right) \indc{a \neq n_o}.
\end{eqnarray}
%We immediately have:
\begin{lemma} 
\label{lem_basic_unconstrained_optimization}
Define 
$
\Delta = \bar{d} - \alpha c - 2 \sqrt{\alpha O_c}. 
$
The maximizer and the maximum value of \eqref{Eqn_Sing_sing} is given by:

\vspace{-3mm}
{\small\begin{align*}
a^* &= p^*\indc{\Delta > 0} + n_{o}\indc{\Delta < 0}
 , \mbox{ where } \ p^* =   \frac{\bar{d} + c \alpha }{2 \alpha} , 
 \\U(a^*) &= \left( \frac{(\bar{d} - \alpha c)^2}{4 \alpha} -O_c\right)\indc{\Delta > 0}.
  \end{align*}}% 
When $\Delta = 0$,  we have two optimizers, $p^*$ and $ n_o$.  
\end{lemma}
\begin{proof}
Towards solving \eqref{Eqn_Sing_sing}, we first consider optimizing the interior objective, more precisely,  $w (p) = (\bar{d} - \alpha p)(p - c)$ only w.r.t. $p$.  The solution to this optimization problem (using derivative techniques) is,
\begin{equation} \label{eqn_given_q_single_M_optimizer}
p^* = \frac{\bar{d}}{2 \alpha} + \frac{c}{2}, \text{ and } w^* =  w(p^*) = \frac{(\bar{d} - \alpha c)^2}{4 \alpha}.   
\end{equation}
Returning to the original problem \eqref{Eqn_Sing_sing}, if $\Delta > 0$, then,
$$
\alpha p^* = \frac{\bar d}{2} + \frac{\alpha c} {2} <   
 {\bar d}  - \sqrt{\alpha O_c} < {\bar d}, 
 $$and hence,
   $(\dbar - \alpha p^*)^+ = \dbar - \alpha p^*$, $w^* - O_c >0$. Thus, $p^*$ is also the maximizer of  \eqref{Eqn_Sing_sing} with $U^* = w^* - O_c$.  If $\Delta< 0$ then  $n_o$ is the optimizer, as  $U^* = U(n_o) = 0 > w^*-O_c$. 
   The last sentence now follows trivially.
 \end{proof}

%\section*{APPENDIX B}
\label{sec_VC_stable}

\noindent{\bf Proof continued, Theorem \ref{thm_esm_regime}:}
 Without loss of generality, consider stability of $\VC_1$.  
To ensure $(\VC_1, \x)$ is stable, it should not be blocked by HC coalition $\M$, as well as VC coalition $\V_2$. In other words, we require,
%
%
%
%One of the inequalities required to ensure $(\VC_2, \x)$ is stable is given below -- thisx
\begin{eqnarray}\label{eqn_v2_stable_hc_v1}
\nu_\sM^{pa} - \nu_\sMt^{\VC_1}     \le     x_\sMo \le   \nu_{\sV_1}^{\VC_1}  - \nu_{\sV_2}^{\VC_2} + 
      \nu_\sMt^{\VC_1},      
\end{eqnarray}
and this is because for any pay-off vector consistent  with $\VC_1$, we have $x_\sMt = \nu_\sMt^{\VC_1}$ and $x_\sMo + x_\sS = \nu_{\sV_1}^{\VC_1}$ 
and $\nu^{pa}_{\sV_2} =\nu^{\VC_2}_{\sV_2} . $
Towards this, consider the limit\footnote{It is easy to observe that the derivative limit in case of HC from Corollary \ref{cor_hc} is equal to $f^{1,\CHC}_\sM = -\nicefrac{\dbar_\sM^2}{16\tilde\alpha}$.}

\vspace{-2mm}
{\small\begin{align}
    \frac{\tilde\nu_{\sV_1}^{\VC_1}  - \tilde\nu_{\sV_2}^{\VC_2} + 
      \tilde\nu_\sMt^{\VC_1}   -  (\tilde\nu_\sM^{pa} - \tilde\nu_\sMt^{\VC_1}   )}{1-\varepsilon} \hspace{-49mm}\nonumber \\
      &= \left(\frac{-2\dbar_\sMo\dbar_\sMt + \dbar_\sMo^2 -\dbar_\sMt^2}{16\tilde\alpha}\right) - \left(\frac{-2\dbar_\sMo\dbar_\sMt + \dbar_\sMt^2 -\dbar_\sMo^2}{16\tilde\alpha}\right)\nonumber \\
      & \hspace{20mm} + o ( (1-\varepsilon) ) +2\frac{\dbar_\sMt^2}{16\tilde\alpha }-\frac{\dbar_\sM^2}{16\tilde\alpha}\nonumber \\
      &= 
      \frac{\dbar_\sMo^2 -\dbar_\sMt^2}{8\tilde\alpha} + \frac{\dbar_\sMt^2}{8 \tilde\alpha} - \frac{\dbar_\sM^2}{16\tilde\alpha} + o ( (1-\varepsilon) ) \nonumber\\
      &= \frac{\dbar_\sMo^2-\dbar_\sMt^2-2\dbar_\sMo \dbar_\sMt} {16\tilde\alpha} + o ( (1-\varepsilon) )\hspace{-49mm} \label{eqn_third_inequality}
\end{align}}
We get the above as $\tilde\nu_\sM^{pa} = \tilde\nu_\sM^\CHC$ and then refer to Table~\ref{tab:derive_ESM} and equation \eqref{eqn_taylor_series}.
Similarly, to ensure the configuration $(\VC_1, \x)$ is not blocked by singletons  $\S$ and $\M_2,$ we require
\begin{eqnarray}\label{eqn_stable_ag_singleton}
  \nu_\sMo^{pa}  \le   x_\sMo  \le  \nu_{\sV_1}^{\VC_1} - \nu_\sS^{pa}.
\end{eqnarray}
Finally, for stability against blocking by GC, we require
\begin{eqnarray}\label{eqn_stable_ag_gc}
      \nu_{\sV_1}^{\VC_1} +   \nu_\sMt^{\VC_1}  \ge \nu_\sG^\GC. 
\end{eqnarray}
In view of the above three  required  inequalities~\eqref{eqn_v2_stable_hc_v1},~\eqref{eqn_stable_ag_singleton} and ~\eqref{eqn_stable_ag_gc}, we require  (if possible) 
an $\bar \varepsilon < 1$ such that the following inequalities are satisfied for each $\varepsilon \ge
\bar \varepsilon$:
\begin{eqnarray}
\label{Eqn_Ineq_VC_stable}
  \tilde\nu_{\sV_1}^{\VC_1}  - \tilde\nu_{\sV_2}^{\VC_2} + 
      \tilde\nu_\sMt^{\VC_1}   -  (\tilde\nu_\sM^{pa} - \tilde\nu_\sMt^{\VC_1}   ) %\hspace{-50mm}\\
     % &=&  f_{\sV_2}^{\VC_2}  - f_{\sV_1}^{\VC_1} + 
   %   f_\sMo^{\VC_2}   -  (f_\sM^{pa} - f_\sMo^{\VC_2}   )  
  % 
  & >&  0, \nonumber\\
   \tilde\nu_{\sV_1}^{\VC_1} +  \tilde\nu_\sMt^{\VC_1} - \tilde \nu_\sG^\pGC  &>& 0, \mbox{ and } 
   %(1-\varepsilon) \lim_{\gamma \to 1} (1-\gamma) \nu_\sG^\GC \nonumber 
 \nonumber   \\
 \tilde \nu_{\sV_1}^{\VC_1} - \tilde \nu_\sS^{pa} -  \tilde \nu_\sMo^{pa}  &>& 0
.\end{eqnarray}
If the above inequalities are satisfied, then one can choose by continuity a $\bar \gamma_\varepsilon  < 1$ for each $\varepsilon
 \ge \bar \varepsilon$   such that the  strict inequalities in~\eqref{Eqn_Ineq_VC_stable} are now satisfied with  $\{\nu_\sC^\P\}$ in place of $\{\tilde \nu_\sC^\P\}$, for all   $\gamma \ge \bar \gamma_\varepsilon$ and for each~$\varepsilon \ge \bar \varepsilon$. Clearly, for such $(\varepsilon, \gamma)$, the configuration $(\VC_1, \x)$ is stable when:
\begin{eqnarray}\label{eqn_cond_stability_1}
 x_\sMo  
 \in \left (
\nu_\sM^{pa} - \nu_\sMt^{\VC_1}     , \       \nu_{\sV_1}^{\VC_1}  - \nu_{\sV_2}^{\VC_2} + 
      \nu_\sMt^{\VC_1} \right ),    
\end{eqnarray}and the above interval is non-empty. 

There exists an $\bar \varepsilon<1$ such that the last two inequalities of~\eqref{Eqn_Ineq_VC_stable}  are satisfied  (see Table \ref{tab:ESM} and \TR{\cite{TR}}{Appendices \ref{sec_Appendix_VC}  and \ref{sec_Appendix_AA}} for the proof of the values in the table, which follows in similar lines as~\eqref{eqn_derv_lim_aa}).

 Thus from \eqref{eqn_third_inequality}, all three  strict inequalities   of \eqref{Eqn_Ineq_VC_stable}   are definitely satisfied (if required for a larger $\bar \varepsilon$),    when \eqref{eqn_cond_stability} is satisfied.

\noindent{\bf Instability:}
On the other hand, say \eqref{eqn_cond_stability} is negated with strict inequality.
%, i.e., say:
%$$
%\dbar_\sMo < (\sqrt{2} + 1)\dbar_\sMt.
% $$
 Then from \eqref{eqn_third_inequality}
 there exists an $\bar \varepsilon<1$ such that for all $\varepsilon \ge \bar \varepsilon$ the following is satisfied:
\begin{eqnarray*}
     \tilde\nu_{\sV_1}^{\VC_1}  - \tilde\nu_{\sV_2}^{\VC_2} + 
      \tilde\nu_\sMt^{\VC_1}   <  (\tilde\nu_\sM^{pa} - \tilde\nu_\sMt^{\VC_1}   )  
\end{eqnarray*}
As before there exists $\bar \gamma_\varepsilon$ for each  $\varepsilon \ge \bar \varepsilon$, and then  for any  $\gamma \ge \bar \gamma_\varepsilon$ and $\varepsilon$ we have: 
\begin{eqnarray*}
     \nu_{\sV_1}^{\VC_1}  -  \nu_{\sV_2}^{\VC_2} + 
       \nu_\sMt^{\VC_1}   <  ( \nu_\sM^{pa} -  \nu_\sMt^{\VC_1}   )  .
\end{eqnarray*}
For all such $(\varepsilon, \gamma)$
there exists no pay-off division, to be more precise, no $x_\sMo$ such that  \eqref{eqn_v2_stable_hc_v1}  is satisfied. 
Thus $\VC_1$ is blocked for any configuration either by $\V_2$ or by HC. 

Now assume without loss of generality, $\dbar_\sMo \ge \dbar_\sMt$. Then \eqref{eqn_cond_stability} can never be satisfied when roles of $\Mt$ and $\Mo$   interchanged and thus 
  the partition $\VC_2$ is never stable in ESM regime.  However, as proved above,  $\VC_1$ is stable with payoff vectors additionally satisfying \eqref{eqn_cond_stability_1}   when \eqref{eqn_cond_stability} is satisfied and unstable when \eqref{eqn_cond_stability} is negated with strict inequality. 

\RB{
\noindent{\textbf{Restricted Blocking:}}
From  Corollary~\ref{cor_hc}  and using the estimates in \cite{TR}
 we have:
$$
 \tilde \nu_\sM^{pa} - \tilde \nu_\sMminusi^{\VC_i} = (1-\varepsilon) \left (\frac{\dbar _\sM^2 -\dbar_\sMminusi}{16\tilde\alpha} \right ) > 0
$$
Thus when one considers stability only against mergers and splits (termed as restricted blocking in \cite{singhal2021coalition}), one can always find stable configurations (in the ESM regime) with $\VC_i$ (for any $i$) and this is true when 
$$
x_\sMi \ge \nu_\sM^{\CHC} - \nu_\sMminusi^{\VC_i}.
$$}
Thus we have proved the theorem. \eop

\TR{}{
\onecolumn
\section{Appendix -- Generic Game}

In this appendix, we solve the game between the manufacturers, where the supplier announces a price $q$ and the manufacturers respond via the game $\left < \{M_1, M_2\},  (W_1, W_2), (\Aset_\sMo, \Aset_\sMt) \right >$. The result of this game will be used extensively in the further proofs. Here, the utility of manufacturer $i$ is, %that have similar structure as that of the inner game between the manufacturers in All Alone Case. The results from this generic game will help us to interpret the results of inner game between the manufacturer thus it will be useful in analysis of All Alone Case Configuration.

%\noindent{\bf Generic Game:} In a similar way, consider a generic strategic form game that resembles  the one that exists between the two manufacturers  against a single supplier. Consider the game $\left < \{M_1, M_2\},  (W_1, W_2), (\Aset_\sMo, \Aset_\sMt) \right >$, where the utility of manufacturer $i$ is,
{\small
\begin{equation}\label{Eqn_Gen_game}
W_{i} (a_i, a_{-i}) =\left(\left(\dbar_i-\alpha_{i}p_{i}+ \varepsilon \alpha_{-i}p_{-i}\right)^+\left(p_{i}-c_{i}\right)-O_{c_i}\right)\F_{a_{i}} .
\end{equation}
}
We define the following function~$\Delta_i$, which is a bound on demand, and will be 
\begin{eqnarray}\label{def_Delta}
  \Delta_i({p}) :=  \dbar_{i} + \varepsilon\alpha_{-i}p - \alpha_{i}c_{i} -2\sqrt{\alpha_{i} O_{c_{i}}} \mbox{ as in Lemma \ref{lem_basic_unconstrained_optimization}.}  
\end{eqnarray}

\begin{lemma}\label{lem_onesupplier_twomanu}
 
If $a_{-i} = n_{o} $, then set $p_{-i} = 0$. The best response (BR) of the agent $i$ in Game \eqref{Eqn_Gen_game} is unique  when   $\Delta_{i}(p_{-i}) \ne 0$ and is given by:
\begin{equation}\label{eqn_onesuppliertwomanubestresponse}
 a_{i}^{*}(a_{-i}) = \left(p_{i}^{*}(p_{-i})
 \indc{\Delta_{i}(p_{-i}) > 0}
 + n_{o}
 \indc{ \Delta_{i}(p_{-i}) < 0} \right).
\end{equation}
where,
\begin{eqnarray}\label{best_response}
 p_{i}^{*}(p_{-i}) = p^b_i  + 
\frac{\varepsilon \alpha_{-i}p_{-i}}{2\alpha_{i}} \mbox{ with }  p^b_i = \frac{\dbar_{i}+c_{i} \alpha_i }{2\alpha_{i}}.  
\end{eqnarray}
Otherwise, $a_i^*(a_{-i}) = \{ p_i^* (p_{-i}), n_o\}$, i.e., BR has two elements.
\end{lemma}
\begin{proof}
The utility function in~\eqref{Eqn_Gen_game} has a similar structure as the utility function given in~\eqref{Eqn_Sing_sing}, and thus the proof follows from Lemma~\ref{lem_basic_unconstrained_optimization}.   
\end{proof}

\begin{definition}[Operating NE and Fully Operating NE]
\begin{enumerate}[$(i)$]
    \item A NE is called an Operating NE if the actions of at least one of the agents in the NE differ from the non-operating action~$n_o$.
    \item A NE is called a Fully Operating NE if the actions of both the agents in the NE differ from~$n_o$.
\end{enumerate}
\end{definition}

\begin{lemma} {\bf [Inner Game between Manufacturers]}\label{lem_game_btw_manufacturers}
We define the following action:
$$
p_i^* := \frac{\varepsilon\left(\dbar_{-i} + c_{-i}\alpha_{-i}\right) +2\left(\dbar_{i}+c_{i}\alpha_{i}\right)}{(4-\varepsilon^{2})\alpha_{i}} \text{ for any manufacturer } M_i.
$$
Without loss of generality, consider 
$\Delta_{2} (p_{1}^{*})  \le  \Delta_{1} (p_{2}^{*}).$ Then the following is true  
for the  game defined in~\eqref{Eqn_Gen_game}.
\begin{enumerate}[$(i)$]
    \item  If $\Delta_2 ( (p_{1}^{*}) \ge 0 $, then 
$(p_{1}^{*}, p_{2}^{*})$ is the unique fully operating NE. In fact, this is the unique NE if $\Delta_{2} ( p_1^*) > 0$.

\item  If $\Delta_2 (p_{1}^{*}) < 0 \leq \Delta_{1} ( 0) $, then $(p_{1}^{b},n_{o})$ is  the unique operating  NE. 
In fact, this is the unique NE if $\Delta_{1} ( 0) > 0$.

\item If $\Delta_2(p_{1}^{*} )  < 0$ and $\Delta_{1} ( 0)  < 0$  then there is no operating NE. Here $(n_{o},n_{o})$ is the unique NE.

\eop
\end{enumerate}
\end{lemma}
\begin{proof}
If there exists a fully operating NE $({\tilde p}^*_1, {\tilde p}_2^*)$,  then by definition, the NE should simultaneously satisfy the following two equations:
$$\tilde p_{i}^{*}({\tilde p}_{-i}^{*}) = {\tilde p}_{i}^{*} \mbox{ for both }i=1,2.$$
That is,   by  Lemma \ref{lem_onesupplier_twomanu},
 any such NE must satisfy 
\begin{eqnarray*}
 \frac{\dbar_{i} + \varepsilon\alpha_{-i} {\tilde p}^*_{-i} +c_{i}\alpha_{i}}{2\alpha_{i}}  = {\tilde p}^*_{i} \mbox{ for both } i=1,2.
\end{eqnarray*}
By linearity and by full rank condition, 
the above pair of equations have an unique solution  given by $(p_{1}^{*},p_{2}^{*})$ (easy to verify by substitution). 
Thus  there exists at maximum one fully operating NE --
if there exists a fully operating NE, then it has to be $(p_1^*, p_2^*)$ --  no other fully operating pair of actions can form an NE.

\noindent {\bf Case 1:} Consider the case, where $\Delta_{i}(p_{-i}^{*})\geq 0$ for both $i =1,2$. Using the above arguments and Lemma \eqref{lem_onesupplier_twomanu},  observe that 
$(p_1^*, p_2^*)$ is the unique fully operating NE. 

%From Lemma \eqref{lem_onesupplier_twomanu}, we have that
%\begin{eqnarray}
%\BR_{i}(p_{-i}) =
%\begin{cases}
 %\{p_{i}^{*}(p_{-i}) \mbox{ if }  \Delta_{i}(p_{-i}) \geq 0\\
 %\{n_{o}\} \mbox{ else }.
%\end{cases}
%\end{eqnarray}

\noindent {\bf Case 2:} Consider the case, where   $\Delta_{2}(p_{1}^{*}) < 0 \le \Delta_{1}(0)$. Thus by Lemma $\eqref{lem_onesupplier_twomanu}$,   the best response against $p_1^{*}$ equals, $\BR_{2}(p_{1}^{*}) = \{n_{o}\}$. Thus,  $(p_{1}^{*},p_{2}^{*})$ is not an NE. Again, by initial arguments of this Lemma, there exists no fully operating NE.

By \eqref{def_Delta}, the mapping  $p_1 \to \Delta_{2}(p_{1})$ is strictly increasing. Thus, using~$\Delta_{2}(p_{1}^{*}) < 0$ and Lemma~\ref{lem_onesupplier_twomanu}, 
\begin{eqnarray}\label{Eqn_br_player_2}
  \BR_{2}(p_1) =\{ n_{o}\}  \  \forall p_{1} \leq p_{1}^{*}.  
\end{eqnarray}
Similarly, as  $\Delta_{1}(0) \ge 0$ and the mapping $p_2 \mapsto \Delta_{1}(p_{2})$ is strictly increasing,   we have $\Delta_{1}(p_{2}) >  0$ for all $p_2 > 0$. Thus, again by Lemma~\ref{lem_onesupplier_twomanu}, we have
{
\begin{equation*}
 n_o \notin \BR_{1}(p_{2}),  \mbox{ and further, }
\BR_{1}(p_{2}) =  \{ p_{1}^{*}(p_{2}) \} \ \forall \ p_2  >0.   
\end{equation*}}
Further, using similar logic and \eqref{best_response}, for $p_{2} > p_{2}^{*}$, we have $p_1^* (p_{2}) > p_{1}^{*} = p_1^*(p_2^*)$. Thus, by initial arguments of this lemma, any pair $(p_1, p_2) \ne (p_1^*, p_2^*)$ with $p_1 > 0$, $p_2 > 0$ is not a NE. % Hence such a $p_2$ can't be a part of any  NE, as any pair $(p_1, p_2) \ne (p_1^*, p_2^*)$ and with $p_1 > 0$, $p_2 > 0$ can't be an NE. 
Now, consider the remaining case, where $p_2 < p_2^*$. For all such $p_2$, by strict monotonicity of  $p_2 \mapsto p^*_{1}(p_{2})$, we have $p_1^*(p_2) <  p_1^*$, and thus by equation~\eqref{Eqn_br_player_2}, we have
$$
\BR_{2}(p_{1}^{*}(p_{2})) = \{n_{o}\}.
$$
Thus, in all, any $p_2 >0 $ i.e., any $a_2\ne n_o$ is not NE strategy for manufacturer~2.

As $\Delta_{1}(0) \geq 0$, we have
$$
p_1^b \in \BR_{1}(n_{o}) \mbox{ and } \BR_2 (p_1^b) = \{n_o\} \mbox{ as } p_1^b < p_1^*.
$$
Thus we have that $(p_{1}^{b},n_{o})$ is an NE and this the only operating NE. Further, observe that when $\Delta_1 (0) > 0$, this is the unique NE.

\noindent {\bf Case 3:} Consider the case where $\Delta_{2}(p_{1}^{*}) < 0$ and $\Delta_{1}(0) < 0$. 
By direct substitution (substituting $p = 0$ in \eqref{def_Delta})  (by monotonicity, we also have $\Delta_{2}(0) < 0$) one can verify that $(n_o, n_o)$ is NE. 

Similar to case~2 above, for all $p_1 \le p_1^*$, we have $\Delta_2 (p_1) <0$,  and hence the corresponding $\BR_2 (p_1) = \{n_o\}$. Further, again similar to case~2 above, $p_1 > p_1^*$ cannot be a NE. 
However, observe that $\BR_1 (n_o) = \{n_o\}$, thus there is no operating NE.
\end{proof}

\section{Appendix -- Vertical Cooperation Partition}\label{sec_Appendix_VC}

\begin{proof}[Proof of Lemma \ref{lem_vc_esm_worth_limits}]
Without loss of generality, we will prove the lemma for the partition $\pVC_1 = \{\V_1,\M_2 \}$, the case with $\pVC_2 = \{\V_2,\M_2 \}$ follows because of similar arguments. 
Firstly, we will simplify the utilities of different agents/coalition in~\eqref{eqn_util_vc_v},~\eqref{eqn_util_vc_manu}, and then obtain the best response of $M_2$ (i.e., $p_2^{*}(p_1,q)$) given the actions $p_1,q$ of coalition $\V_1$. Then we obtain the optimizers, $p_1^{*}$ and $q^*$ assuming $\alpha_\sMo = \alpha_\sMt = \alpha$ using derivative based argumets. After obtaining these relevant quantities, and checking the operating condition for $\M_2$ and the coalition $\V_1$, we find the worth limits and derivative limits of coalition $\V_1$ and the agent $\M_2$ for the ESM regime. Now, we begin with the detailed proof.

The utility of the vertical coalition $\V_1$, as defined in~\eqref{eqn_util_vc_v}, is modified as follows, by substituting the demand $D_{\sMi}(\a_\sM)$ from~\eqref{Eqn_Demand_Function},
\begin{align}\label{eqn_util_V_Vertical}
U_{\sV_1}(\a_{\sV_1}; a_\sMt)  
&= \bigg (  D_{\sMo}(\a_\sM)(p_1 -C_{\sMo}- C_{\sS}) + D_{\sMt}(\a_\sM)\F_\sMt(q-C_\sS)  - O_{\sS} - O_\sMo \bigg )\F_{\sV_1}, \\
&=   \bigg(( \dbar_\sMo -\alpha_\sMo p_1 + \varepsilon\alpha_\sMt p_2 ) (p_1 -C_{\sMo}- C_{\sS}) + 
 ( \dbar_\sMt -\alpha_\sMt p_2 + \varepsilon\alpha_\sMo p_1 ) \F_\sMt(q- C_\sS) - O_\sS - O_\sMo \bigg)\F_{\sV_1} . 
 \ignore{
 \\
 &=& 
 ( \dbar_\sMo -\alpha_\sMo p_1 + \varepsilon\alpha_\sMt p_2 ) (p_1 -C_{\sMo}- C_{\sS} - \varepsilon   (q-C_s) ) + ( \dbar_\sMo -\alpha_\sMo p_1 + \varepsilon\alpha_\sMt p_2 )  \varepsilon   (q-C_s) \nonumber  \\
 && + 
 ( \dbar_\sMt -\alpha_\sMt p_2 + \varepsilon\alpha_\sMo p_1 ) (q- C_\sS) - O_\sS - O_\sMo \nonumber \\
  &=&
\bigg  ( \dbar_\sMo -\alpha_\sMo p_1 + \varepsilon\alpha_\sMt p_2  \bigg ) \bigg (p_1 -C_{\sMo}- C_{\sS} - \varepsilon   (q-C_s) \bigg )   \nonumber  \\
 && + 
 \bigg ( \dbar_\sMt -\alpha_\sMt p_2 +  \varepsilon ( \dbar_\sMo   + \varepsilon\alpha_\sMt p_2 )  \bigg ) (q- C_\sS) - O_\sS - O_\sMo }
\end{align}
Using \eqref{eqn_util_vc_manu}, the utility of the manufacturer $M_2$ for a given $ \a_{\sV}$ is,
\begin{eqnarray}\label{eqn_util_m2_Vertical}
U_{\sMt }(a_\sMt; \a_{\sV}) = \bigg( D_{\sMt}(\a_\sM)(p_{2}-C_{\sMt}-q)\F_{\sV_{1}}  - O_\sMt\bigg )\F_{\sMt}.
\end{eqnarray}
Since $M_2$ is follower, for any given $(q, p_1)$ the optimizer of $U_{\sMt }$ using Lemma \ref{lem_basic_unconstrained_optimization} is,
\begin{equation} \label{eqn_p_2_star}
a_\sMt^* (p_1, q) = p_2^{*}(p_1,q) = \indc{q \le \theta_2(p_1) } \left ( \frac{ \dbar_\sMt + \varepsilon \alpha_\sMo p_1  }{ 2 \alpha_\sMt } + \frac{C_\sMt + q}{2} \right )  + n_o \indc{ q >  \theta_2 (p_1) },  
\end{equation}
where 
\begin{equation} \label{eqn_theta_2_p_1}
\theta_2 (p_1) = \frac{\dbar_\sMt + \varepsilon\alpha_\sMo p_1 -\alpha_
\sMt C_\sMt -2\sqrt{\alpha_\sMt O_\sMt}}{ \alpha_\sMt}  .  
\end{equation}
\ignore{\color{red}
For a given $q$, and $p_1$, the BR of $\Mt$ and BR of $\V$ are respectively given using Lemma \ref{lem_basic_unconstrained_optimization}:
\begin{eqnarray*}
 % \max_{a_\sMt}  U_\sMt (a_\sMt; p_1, q) &=&   \\
  a_\sMt^* (p_1, q) &=&  \indc{q \le \theta_2(p_1) } \left ( \frac{ \dbar_\sMt + \varepsilon \alpha_\sMo p_1  }{ 2 \alpha_\sMt } + \frac{C_\sMt + q}{2} \right )  + n_o \indc{ q >  \theta_2 (p_1) } \\
  a_\sMo^* (p_2, q) &=& \left(\frac{\dbar_\sMo + \varepsilon\alpha_\sMt p_2}{2\alpha_\sMo} + \frac{C_\sMo + C_\sS}{2 } + \frac{\varepsilon(q-C_\sS)}{2}\right) \indc{q \ge  \theta_1(p_2)} + n_o \indc{q < \theta_1(p_2) } \\
  a_\sMo^* (n_o, q) &=& \left(\frac{\dbar_\sMo  }{2\alpha_\sMo} + \frac{C_\sMo + C_\sS}{2 } \right)   \mbox{ under A.4, it is always opreating}
\end{eqnarray*}}
\ignore{
{\color{red} 
Thee derivative a $q < \theta_2 (p)$
\begin{eqnarray*}
\frac{d U_{\sV_1}}{d p_1} 
&=& - \alpha_\sMo ( p_1-C_\sMo - C_\sS ) + 
 ( \dbar_\sMo -\alpha_\sMo p_1 + \varepsilon\alpha_\sMt p_2 ) + \varepsilon \alpha_\sMo (q-C_\sS) 
+ \left  ( \varepsilon\alpha_\sMt ( p_1-C_\sMo - C_\sS ) - \alpha_\sMt (q-C_\sS) 
\right )  \frac{d p_2}{ d p_1}
\end{eqnarray*}

Solve using gradient method:
\begin{eqnarray}
    p_{1, k+1} &=& p_{1, k} + \epsilon_k \left . \frac{d U_{\sV_1}}{d p_1} \right |_{p_{1,k}, q_{k}} \\
        q_{k+1} &=& q_{k} + \epsilon_k \left . \frac{d U_{\sV_1}}{d q} \right |_{p_{1,k}, q_{k}}
\end{eqnarray}
}
}
For all the following calculations in this section, we assume $\alpha_\sMo = \alpha_\sMt = \alpha $. Define 
\begin{align*}
 e_1 &:= \frac{\dbar_\sMt}{2\alpha} + \frac{C_\sMt}{2}, \\
 e_2 &:= \alpha \left (1 - \frac{\varepsilon^2}{2} \right)(C_\sMo + C_\sS) + \dbar_\sMo +\varepsilon\alpha \left( \frac{\dbar_\sMt}{2\alpha} + \frac{C_\sMt}{2} \right) -\frac{C_\sS \varepsilon \alpha}{2},\\
 e_3 &:= \frac{\dbar_\sMt}{2} -\alpha\frac{C_\sMt}{2}- (C_\sM + C_\sS)\frac{\varepsilon\alpha}{2} + C_\sS\frac{\alpha}{2} .
\end{align*}
If $q < \theta_2(p_1)$ then from Lemma~\ref{lem_basic_unconstrained_optimization}, we have $p_2^*(p_1,q) = \nicefrac{q}{2} + \nicefrac{p_1\varepsilon}{2} + e_1$. Since coalition $\V_1$ is the leader, and $M_2$ is the follower, as in proof of Theorem~\ref{thm_all_alone_two_}, substituting $a_\sMt = p_2^*(p_1,q)$ into \eqref{eqn_util_V_Vertical} (by neglecting the operating conditions), consider
\begin{eqnarray*}
  U_{\sV_1}(p_1, q) :=  U_{\sV_1}(p_1, q; p_2(p_1, q)) %&=&   
  %( \dbar_\sMo -\alpha p_1 + \varepsilon\alpha  (e_1 + \varepsilon p_1/2 + q/2)  ) (p_1 -C_{\sMo}- C_{\sS})   \\
  %  && +
 %( \dbar_\sMt -\alpha  (e_1 + \varepsilon p_1/2 + q/2) + \varepsilon\alpha p_1 ) (q- C_\sS) - O_\sS - O_\sMo \nonumber \\
 &=&   \left( \dbar_\sMo -\alpha p_1 \left(  1- \frac{\varepsilon^2}{2} \right)+ \varepsilon\alpha  e_1 + \frac{\varepsilon\alpha  q}{2}  \right) (p_1 -C_{\sMo}- C_{\sS})   \\
    && \hspace{20pt}+
 \left( \dbar_\sMt -\alpha  e_1 + \frac{\alpha p_1  \varepsilon - \alpha q}{2}  \right) (q- C_\sS) - O_\sS - O_\sMo.
\end{eqnarray*}
The first step of this analysis is to derive the optimizers of the function~$U_\sV$. Later we show that the operating conditions are satisfied at these optimizers (as in proof of Theorem \ref{thm_all_alone_two_}).  Now, differentiating $U_{\sV_1}$ with respect to $p_1$ and $q$, we get
\begin{eqnarray*}
   \frac{d  U_{\sV_1}} { d p_1 }  &=& - \alpha
\left( 1-  \frac{\varepsilon^2}{2}  \right)  (p_1 - C_\sMo - C_\sS) + \dbar_\sMo -\alpha p_1 \left(  1- \frac{\varepsilon^2}{2}\right) + \varepsilon\alpha  e_1 + \frac{\varepsilon \alpha q}{2} +  \frac{(q-C_\sS)  \varepsilon \alpha}{2}, \\
&=&  -\alpha  (2 - \varepsilon^2)p_1  + \alpha \varepsilon q + e_2, \\
 \frac{d  U_{\sV_1}} { d q }  &=&   \frac{(p_1 - C_\sMo - C_\sS) \varepsilon  \alpha  } {2}  + \dbar_\sMt -\alpha  e_1 + \frac{\alpha p_1 \varepsilon - \alpha q}{2}  - \frac{(q-C_\sS) \alpha}{2}  \  =  \  \alpha \varepsilon p_1 - \alpha q + e_3 .
\end{eqnarray*}
Observe that the Hessian matrix $H(p_1, q)$ of second order partial derivatives of $U_{\sV_1}$ is always positive for any $(p_1, q)$, and the diagonal entries of the Hessian matrix are always negative for any $(p_1, q)$. Solving the simultaneous equations $\nicefrac{d  U_{\sV_1}}{d p_1} = 0$ and $\nicefrac{d  U_{\sV_1}}{d q} = 0$, we get the unique extreme $p_1^{*}$ and $q^{*}$, where
%Thus the stakelberg equilibrium is 
\begin{eqnarray} \label{eqn_VC_p_1_q_star}
p_1^* =  \frac{e_2 + e_3 \varepsilon}{\alpha( 2 -2 \varepsilon^2 )  } \mbox{ and }
    q^* = \frac{e_3 + \alpha\varepsilon p_1^*}{\alpha}  .\ignore{\mbox{ this means } 
    p_1^*\varepsilon - q^* = - \frac{e_3}{\alpha} \mbox{ and } q^*+p_1^* \varepsilon = 2p_1^* \varepsilon + \frac{e_3}{\alpha}  = \frac{e_2 \varepsilon+e_3}{\alpha (1-\varepsilon^2)}  }
\end{eqnarray}
Since these optimizers are unique, using the second derivative test, we conclude that $p_1^{*}$ and $q^{*}$ are the global optimizers.

Now consider the ESM regime, where $(\varepsilon,\gamma) \to (1,1)$. Now consider the following limits, which will be used in the further calculations,
\begin{eqnarray} \label{eqn_e_is_limits}
\lim_{\varepsilon \to 1}\lim_{\gamma \to 1}\alpha e_1 = \frac{\dbar_\sMt}{2}, \ 
 \ 
\lim_{\varepsilon \to 1}\lim_{\gamma \to 1}e_2  = \frac{\dbar_\sMt}{2} + \dbar_\sMo,
\mbox{ and } 
\lim_{\varepsilon \to 1}\lim_{\gamma \to 1}e_3 = \frac{\dbar_\sMt}{2} 
\end{eqnarray}
The demand $D_\sMo (\a_\sM)$ of coalition $\V_1$ in the limit at SBE   is (see \eqref{eqn_util_V_Vertical}):
\begin{align}\label{eqn_d_v}
 \lim_{\varepsilon \to 1}\lim_{\gamma \to 1} D_\sMo (\a_\sM) &= \lim_{\varepsilon \to 1}\lim_{\gamma \to 1} \left(\dbar_\sMo -\alpha p_{1}^{*} + \varepsilon\alpha p_{2}^{*}(p_1^{*}, q^{*})\right) \nonumber \\
 &= \lim_{\varepsilon \to 1}\lim_{\gamma \to 1} \left(\dbar_\sMo -\alpha p_1^{*} (1 - \varepsilon^2) + \varepsilon \alpha e_1 +\frac{\varepsilon e_3}{2}\right)
   \ignore{ \dbar_\sMo - \frac{e_2 + \varepsilon e_3}{2} + \varepsilon \alpha e_1 +\frac{\varepsilon e_3}{2} =   \dbar_\sMo - \frac{e_2  }{2} + \varepsilon \alpha e_1}
    = \frac{\dbar_\sMo}{2} + \frac{\dbar_\sMt}{4} .
\end{align}
\ignore{
\begin{eqnarray*}
 \left(\dbar_\sMo -\alpha(\frac{e_2 + e_3\varepsilon}{(2-2\varepsilon^2)\alpha} (1-\frac{\varepsilon^2}{2})) + \varepsilon\alpha e_1+ \varepsilon\alpha\frac{e_3 + \alpha\varepsilon(\frac{e_2+ e_3\varepsilon}{(2-2\varepsilon^2)\alpha})}{2\alpha}\right) \hspace{-70mm} \\
 &=& \frac{1}{2 (2-2\varepsilon^2)} \left( 2 \dbar_\sMo  (2-2\varepsilon^2) - ( e_2 + e_3\varepsilon ) (2-\varepsilon^2)  +2 \varepsilon\alpha e_1(2-2\varepsilon^2)+ \varepsilon  e_3  (2-2\varepsilon^2)  +  \varepsilon^2 (e_2+ e_3\varepsilon)  \right)   
 \\
 &=& \frac{1}{2 (2-2\varepsilon^2)} \left( 2 \dbar_\sMo  (2-2\varepsilon^2) - ( e_2 + e_3\varepsilon ) (2-2\varepsilon^2)  +2 \varepsilon\alpha e_1(2-2\varepsilon^2)+ \varepsilon  e_3  (2-2\varepsilon^2)    \right) \\
 &=&
  \frac{1}{2 } \left( 2 \dbar_\sMo    - ( e_2 + e_3\varepsilon )    +2 \varepsilon\alpha e_1 + \varepsilon  e_3      \right) \  = \  \frac{1}{2 } \left( 2 \dbar_\sMo    -  e_2     +2 \varepsilon\alpha e_1        \right) 
\end{eqnarray*}}
The demand $D_\sMt(\a_\sM)$ of $M_{2}$ in the limit at SBE is (see \eqref{eqn_util_m2_Vertical}):
\begin{align}
\lim_{\varepsilon \to 1}\lim_{\gamma \to 1} D_\sMt(\a_\sM) &= \lim_{\varepsilon \to 1}\lim_{\gamma \to 1} \left(\dbar_\sMo -\alpha p_{2}^{*}(p_1^{*}, q^{*}) + \varepsilon\alpha p_{1}^{*}\right) \nonumber \\   &= \lim_{\varepsilon \to 1}\lim_{\gamma \to 1} \left(\dbar_\sMt -\alpha e_1 + \alpha p_1^{*} \frac{\varepsilon}{2} -\alpha \frac{q^{*}}{2}\right)  =
\frac{\dbar_\sMt}{4} .
\end{align}
To ensure that at limit $q^* \le \theta_2(p_1^*)$, using \eqref{eqn_theta_2_p_1}, \eqref{eqn_VC_p_1_q_star}, and \eqref{eqn_e_is_limits},
\begin{eqnarray*}
\lim_{\varepsilon \to 1}\lim_{\gamma \to 1} \theta_2(p_1^*) - q^* =  \dbar_\sMt + \varepsilon\alpha \lim_{\varepsilon\to 1}\lim_{\gamma \to 1}p_1^{*} -\alpha \lim_{\varepsilon \to 1}\lim_{\gamma \to 1} q^{*} =  \dbar_\sMt  - e_3  = \frac{\dbar_\sMt}{2} > 0.
\end{eqnarray*}
Therefore, there exists a neighbourhood of $(\gamma,\varepsilon)$ near $(1,1)$ such that $q^* < \theta_2(p_1^*)$.
Now, using \eqref{eqn_p_2_star}, \eqref{eqn_VC_p_1_q_star}, \eqref{eqn_e_is_limits}, we get (recall that $\alpha = \tilde\alpha (1-\gamma)$)
\begin{align*}
 \lim_{\varepsilon \to 1}\lim_{\gamma \to 1} (1-\gamma) (1-\varepsilon) ( p_2^*(p_1^*, q^*) - q^*) = \lim_{\varepsilon \to 1}\lim_{\gamma \to 1}\frac{(1-\gamma) (1-\varepsilon)  (-q^*+p_1^*\varepsilon + 2e_1)}{2}  &=  0 ,
\mbox{ and} \\
\lim_{\varepsilon \to 1}\lim_{\gamma \to 1}(1-\gamma) (1-\varepsilon) p_2^*(p_1^*, q^*)   = \lim_{\varepsilon \to 1}\lim_{\gamma \to 1}\frac{(1-\gamma) (1-\varepsilon) (q^*+p_1^*\varepsilon + 2e_1)}{2} &=
\frac{1}{4 \tilde \alpha} (\dbar_\sMo + \dbar_\sMt) .
\end{align*}

Finally, substituting the values of $p_1^{*}, q^*, p_2^*(p_1^{*},q^{*})$ from \eqref{eqn_VC_p_1_q_star}, \eqref{eqn_p_2_star} into the utility of a coalition $\V_1$ (see \eqref{eqn_util_V_Vertical}) and manufacturer  $M_2$  (see \eqref{eqn_util_m2_Vertical}) at SBE in limits are:
\begin{eqnarray*}
    f_{\sV_1}^{\VC_1}  &=& \lim_{ (\varepsilon, \gamma) \to (1,1)} (1-\gamma) (1-\varepsilon)\nu_\sV^\pVC= \frac{1}{ 8\tilde \alpha}  \left ( \dbar_\sMo + \dbar_\sMt  \right )^2  = \frac  {\dbar_\sM^2}{8\tilde\alpha},\\
     f_\sMt^{\VC_1}  &=& \lim_{ (\varepsilon, \gamma) \to (1,1)} (1-\gamma) (1-\varepsilon)\nu_\sMt^\pVC  = 0 .
\end{eqnarray*}
Again, by substituting the values of  $p_1^{*}, q^*, p_2^*(p_1^{*},q^{*})$ from \eqref{eqn_VC_p_1_q_star}, \eqref{eqn_p_2_star} into the utility of a coalition $\V_1$ (see \eqref{eqn_util_V_Vertical}) and manufacturer  $M_2$  (see \eqref{eqn_util_m2_Vertical}) at $\gamma \to 1$, we get,
\begin{eqnarray}\label{eqn_v_gamma}
\lim_{\gamma \to 1}(1-\gamma)\nu_{\sV_1}^{\pVC_1}
& =& \left(\frac{\dbar_\sMo}{2} + \frac{\varepsilon\dbar_\sMt}{4}\right)\left(\frac{\dbar_\sMo + \dbar_\sMt\varepsilon}{2\tilde\alpha(1-\varepsilon^{2})}\right) + \left( \frac{\dbar_\sMt}{4}\right)\left(\frac{\dbar_\sMt + \varepsilon\dbar_\sMo}{2(1-\varepsilon^2)\tilde\alpha}\right)\\
 \lim_{\gamma \to 1}(1-\gamma)\nu_\sMt^\pVC
& =& \frac{\dbar_\sMt^2}{16\tilde\alpha}\label{eqn_manu_gamma}
\end{eqnarray}
Thus after scaling the equations \eqref{eqn_v_gamma} and \eqref{eqn_manu_gamma} by $(1-\varepsilon)$ and differentiating the scaled equations with respect to $\varepsilon$ and then taking the limit $\varepsilon \to 1$, we get,
\begin{eqnarray*}
f_{\sV_1}^{(1),\VC_1} &=&  \frac{2\dbar_\sMo\dbar_\sMt + \dbar_\sMt^2 -\dbar_\sMo^2}{16\tilde\alpha}.  \\
f_{\sMt}^{(1),\VC_1} &=& -\frac{\dbar_\sMt^2}{16\tilde\alpha}.
\end{eqnarray*}
Along the similar lines, we can derive the expressions of worth-limits and the derivative-limits for the other partition $\pVC_2$ which completes the proof.
\end{proof}

\ENMremoved{
\vspace{-2mm}
\subsection*{ENSM regime} 
\vspace{-2mm}
We now derive the limits with   $(\gamma, \varepsilon) \to (1,0)$
 in a similar way. To begin with the  limits of
$(\alpha e_1, e_2, e_3)$ are
\begin{eqnarray}
\lim_{\varepsilon \to 0}\lim_{\gamma \to 1}\alpha e_1 = \frac{\dbar_\sMt}{2}, \ 
 \ 
\lim_{\varepsilon \to 0}\lim_{\gamma \to 1}e_2  = \dbar_\sMo ,
\mbox{ and } 
\lim_{\varepsilon \to 0}\lim_{\gamma \to 1}e_3 = \frac{\dbar_\sMt}{2} 
\end{eqnarray}
The demand $D_\sMo(\a_\sM)$ of coalition $\V$ in the limit at SBE is: 
\begin{align}
\lim_{\varepsilon \to 0}\lim_{\gamma \to 1} D_\sMo (\a_\sM) &= \lim_{\varepsilon \to 0}\lim_{\gamma \to 1} \left(\dbar_\sMo -\alpha p_{1}^{*} + \varepsilon\alpha p_{2}^{*}(p_1^{*}, q^{*})\right)\nonumber \\ &= 
   \lim_{\varepsilon \to 0}\lim_{\gamma \to 1} \left( \dbar_\sMo -\alpha p_1^{*} (1 - \varepsilon^2) + \varepsilon \alpha e_1 +\frac{\varepsilon e_3}{2}\right)
      =   \frac{\dbar_\sMo}{2} 
\end{align}
The demand $D_{\sMt}(\a_\sM)$ of $M_2$ in limit at SBE is:
\begin{align}
\lim_{\varepsilon \to 0}\lim_{\gamma \to 1} D_\sMo (\a_\sM) &= \lim_{\varepsilon \to 0}\lim_{\gamma \to 1} \left(\dbar_\sMo -\alpha p_{2}^{*}(p_{1}^{*},q^{*}) + \varepsilon\alpha p_{1}^{*}\right)\nonumber \\ &=
   \lim_{\varepsilon \to 0}\lim_{\gamma \to 1} \left(\dbar_\sMt -\alpha e_1 + \alpha p_1^{} \frac{\varepsilon}{2} -\alpha \frac{q}{2}\right)  = \frac{\dbar_\sMt}{4}
\end{align}
To ensure that at limit $q^* \le \theta_2(p_1^*)$, using \eqref{eqn_theta_2_p_1}, \eqref{eqn_VC_p_1_q_star}, and \eqref{eqn_e_is_limits},
\begin{eqnarray*}
\lim_{\varepsilon \to 0}\lim_{\gamma \to 1} \theta_2(p_1^*) - q^* =  \dbar_\sMt + \varepsilon\alpha \lim_{\varepsilon\to 0}\lim_{\gamma \to 1}p_1^{*} -\alpha \lim_{\varepsilon \to 1}\lim_{\gamma \to 1} q^{*} =  \dbar_\sMt  - e_3  = \frac{\dbar_\sMt}{2} > 0.
\end{eqnarray*}
Therefore, there exists a neighbourhood of $(\gamma,\varepsilon)$ near $(0,1)$ such that $q^* < \theta_2(p_1^*)$.

Finally substituting the values of $p_1^*, p_2^*, q^*$ from \eqref{eqn_p_2_star}, \eqref{eqn_VC_p_1_q_star} into the utility of a coalition $\V$ (see \eqref{eqn_util_V_Vertical}) and manufacturer  $M_2$  (see \eqref{eqn_util_m2_Vertical}) at SBE in limits are:
\begin{eqnarray}
    g_\sV^\VC  &=& \lim_{ (\varepsilon, \gamma) \to (0,1)} (1-\gamma) U_\sV^*  = \frac{\dbar_\sMo^2}{4\tilde\alpha} + \frac{\dbar_\sMt^2}{8\tilde\alpha}\\
     g_\sMt^\VC  &=& \lim_{ (\varepsilon, \gamma) \to (0,1)} (1-\gamma)U_\sMt^*  = \frac{\dbar_\sMt^2}{16\tilde\alpha}
\end{eqnarray}
}

\section{Appendix -- All Alone Case Partition }\label{sec_Appendix_AA}
\begin{proof}[ Proof of Lemma \ref{lem_ALC_worth_limits} ]
In this partition, the supplier quotes a price $q,$ and the manufacturers respond to it via an inner game between them. Initially, we find the inner game equilirium for given supplier price $q,$ and then we find the optimal action of the supplier by optimizing it's utility. %To begin with we compute the NE of the inner game. Towards this observe that the utilities of the inner game \eqref{suppliermonopolymanufacturerutility} exactly resemble the utilities of the game considered in \eqref{Eqn_Gen_game} and hence Lemma  \ref{lem_game_btw_manufacturers} is applicable. 
Towards this, we define the following quantities, which will be extensively used later in the proof.
\begin{align}
\sigma_{i}^{a} & := \frac{\dbar_{\sMi}-\alpha_{\sMi}C_{\sMi} - 2\sqrt{\alpha_{\sMi}O_{\sMi}}}{\alpha_{\sMi}} , \label{eqn_sigma_ia}\\
\sigma_{i}^{b} & := \frac{4\dbar_{\sMi} + 2\varepsilon\dbar_{\sMminusi}+2\varepsilon\alpha_{\sMminusi}C_{\sMminusi} + 2(\varepsilon^{2}-2)\alpha_{\sMi}C_{\sMi}-2(4-\varepsilon^{2})\sqrt{\alpha_{\sMi}O_{\sMi}}}{2((2-\varepsilon^{2})\alpha_{\sMi} -\varepsilon\alpha_{\sMminusi})} ,\label{eqn_sigma_ib}\\
p_{i}^{b}(q) &:= \frac{\dbar_{\sMi}}{2\alpha_{\sMi}} + \frac{C_{\sMi} + q}{2} , \label{eqn_pib}\\
p_{i}^{*}(q) &:= \frac{\varepsilon\left(\dbar_{\sMminusi} + (C_{\sMminusi} + q)\alpha_{\sMminusi}\right) +2\left(\dbar_{\sMi}+(C_{\sMi} + q)\alpha_{\sMi}\right)}{(4-\varepsilon^{2})\alpha_{\sMi}} ,\label{eqn_pstar_q}\\
\ignore{p_i^*(q) - q
 & = \frac{\varepsilon \dbar_{\sMminusi}  +2\dbar_{\sMi}}{(4-\varepsilon^{2})\alpha_{\sMi}} + \frac{\varepsilon C_{\sMminusi}  +2C_{\sMi} }{(4-\varepsilon^{2}) } - \frac{q (1-\varepsilon)}{2-\varepsilon} , \nonumber \\}
\Delta_{\sMi}(p;q) & :=  \dbar_{\sMi} + \varepsilon\alpha_{\sMminusi}p - \alpha_{\sMi}(C_{\sMi} + q)-2\sqrt{\alpha_{\sMi} O_{\sMi}} \label{eqn_delta_aa}.
\end{align}
Here, $\sigma_{i}^{a}$ and $\sigma_{i}^{b}$ are different thresholds on action of supplier, %depending on which supplier will decide the price to be quoted to the manufacturers
based on which manufacturers will decide whether to operate or not. Observe that the utilities of the inner game between the manufacturers, given by~\eqref{suppliermonopolymanufacturerutility}, are of similar form as utilities of the generic game considered in~\eqref{Eqn_Gen_game}. Therefore, Lemma~\ref{lem_game_btw_manufacturers} is applicable.
Observe from \eqref{eqn_sigma_ia}, \eqref{eqn_delta_aa}  that $ \Delta_{\sMi}(0;q) < 0$ is equivalent to $q >\sigma_i^a$; and from \eqref{eqn_sigma_ib}, \eqref{eqn_delta_aa} that $ \Delta_{\sMi}(p_i^*(q);q) < 0$ is equivalent to $q > \sigma_i^b$.
\ignore{Thus by Lemma \ref{lem_game_btw_manufacturers}, the NE of the inner-game  for any $a_\sS \ne n_o$  and then the utility of supplier \eqref{eqn_supplier_monopoly_supplier_utility} after substituting the NE,
is given by (assume  w.l.o.g that $\sigma_{1}^{b} \ge
\sigma_{2}^{b}$):}
{Thus  by Lemma \ref{lem_game_btw_manufacturers}, one can derive the NE of the inner-game $(p_{1}^{*}(q), p_{2}^{*}(q))$ for any supplier action $a_\sS = q$ (which is not equal to $ n_o$) when both the manufacturers operate and $(p_i^b(q), n_o)$ when only manufacturer $i$ operates; after substituting this NE into~\eqref{eqn_supplier_monopoly_supplier_utility}, we get the utility of supplier as below (assume  w.l.o.g that $\sigma_{1}^{b} \ge
\sigma_{2}^{b}$):}
%Thus because of the utilities of the inner game as defined in \eqref{suppliermonopolymanufacturerutility}, one can use Lemma \ref{lem_game_btw_manufacturers} 
%to  derive the NE $\a_\sM^* (a_\sS)$ (when $a_\sS \ne n_o$) and further substituting the NE of inner-game the utility of supplier $U_{\sS}(a_\sS)$  becomes:
%
  \begin{eqnarray}\label{util_supp_All_alone_simp}
  U_{\sS}(a_\sS)= \left \{ \begin{array}{cl}
  0  & \mbox{ if }  a_\sS = n_o \\
     -O_\sS  &   \mbox{ if } a_\sS \ne n_o \mbox{ and }  q > \max\{\sigma_{1}^{a},\sigma_{2}^{b}\}  \ \\\
     
   \left( \frac{\dbar_{\sMo} - \alpha_{\sMo}(C_{\sMo} + q)}{2}\right)(q-C_{\sS}) - O_{\sS} & \mbox{ if } a_\sS \ne n_o \mbox{ and }  \sigma_{2}^{b} < q  \le \sigma_{1}^{a}   \\
 \left(\frac{(\dbar_{\sMo}+\dbar_{\sMt}+(\varepsilon-1)(\alpha_{\sMo} C_{\sMo}+ \alpha_{\sMt}C_{\sMt})-(1-\varepsilon)(\alpha_{\sMt}+ 
 \alpha_{\sMo})q}{(2-\varepsilon)}\right)\left(q-C_{\sS} \right) -O_{\sS}  & \mbox{ if } a_\sS \ne n_{o} \mbox{ and } q \le \sigma_{2}^{b}.
  \end{array}\right .   
\end{eqnarray}
Observe here that the case when only the first manufacturer operates is given by third row of \eqref{util_supp_All_alone_simp} and the case when both manufacturers operate is given by the fourth row of \eqref{util_supp_All_alone_simp}.
\ignore{
{\color{red}
If $\sigma_2^b <   \sigma_1^a$, the third row is relevant  and the optimizer in that sub-regime is $$q_1^* = \min\{ \sigma_1^a, \max\{q^{*},\sigma_2^b\} \}, \mbox{ with } q^* = \frac{\dbar_\sMo - \alpha_\sMo C_\sMo}{2\alpha_\sMo}  + \frac{C_\sS}{2}, $$ and the optimizer of the fourth line is
$$
q_2^* =  \frac{\dbar_\sMo + \dbar_\sMt}{2(1-\varepsilon)(\alpha_\sMo + \alpha_\sMt)} + \frac{C_\sS}{2}.
$$
The optimizer $q^*$ is one among $q_1^*$ and $q_2^*$.}
}

Initially, we need to find the optimizer $q^*$ of the utility function $U_{\sS}$, and then the optimal utilities of the supplier and the manufacturers to compare with other cases. Specifically, we find these quantities in the ESM regime, as $(\gamma,\varepsilon) \to (1,1)$. As before, we assume that $\alpha_\sMo = \alpha_\sMt = \alpha$ and also recall $\alpha = \tilde\alpha (1-\gamma)$.
Towards optimizing \eqref{util_supp_All_alone_simp}, we first need to analyse  $\{\sigma_i^b, \sigma_i^a\}$; by \eqref{eqn_sigma_ia} and \eqref{eqn_sigma_ib}, the corresponding limits in the ESM regime are:
\begin{alignat}{3}
\lim_{(\varepsilon,\gamma) \to (1,1)}(1-\gamma)(1-\varepsilon)
\sigma_{i}^a &=
0, \hspace{3mm} &&
\ignore{
\lim_{(\varepsilon,\gamma) \to (0,1)}
(1-\gamma)\sigma_{i}^a  &&= \frac{\dbar_\sMi}{\tilde\alpha} },\label{eqn_lim_sigma_ia} \\
\lim_{(\varepsilon,\gamma) \to (1,1)}(1-\gamma)(1-\varepsilon)\sigma_{i}^{b}
  &= \frac{2\dbar_\sMi + \dbar_\sMminusi}{3\tilde\alpha}. \hspace{3mm} && 
  \ignore{
\lim_{(\varepsilon,\gamma) \to (0,1)}(1-\gamma)
\sigma_{i}^b &&= \frac{\dbar_\sMi}{\tilde\alpha}. }  \label{eqn_lim_sigma_ib}
\end{alignat}
Observe that in the ESM regime, at limit, when $\dbar_\sMo \ge \dbar_\sMt$, we have at the limit, $0=\sigma_1^a  < \sigma_2^b < \sigma_1^b $.  Now, using this strict inequality and continuity of functions $\sigma_1^a, \sigma_1^b,$ and $ \sigma_2^b$ with respect to ($\varepsilon, \gamma$), the same strict inequality is satisfied for all $(\varepsilon,\gamma)$ in a neighbourhood of $(1,1)$. For all such $(\varepsilon,\gamma)$, after substituting the inner-game NE $(p_{1}^{*}(q),p_2^{*}(q))$ from \eqref{eqn_pstar_q}, the utility of supplier as given in~\eqref{util_supp_All_alone_simp} modifies as follows:
\begin{eqnarray}\label{eqn_supp_util_case2}
  U_{\sS}(a_\sS)= \begin{cases}
  0  & \mbox{if }  a_\sS = n_o \\
  -O_\sS  &   \mbox{if } a_\sS \ne n_o \mbox{ and }  q >  \sigma_{2}^{b}  \\
 \left(\frac{\dbar_{\sMo}+\dbar_{\sMt}+(\varepsilon-1)\alpha (C_{\sMo}+C_{\sMt})-(1-\varepsilon)2\alpha q}{(2-\varepsilon)}\right)\left(q-C_{\sS} \right) -O_{\sS}  & \mbox{if } a_\sS \ne n_{o} \mbox{ and } q \le \sigma_{2}^{b}.
  \end{cases}   
\end{eqnarray}
For any such $(\gamma, \varepsilon)$, using Lemma \ref{lem_basic_unconstrained_optimization}, we get the optimizer of $U_{\sS}$ as given below:
\begin{eqnarray*}
  q^{*} =\min \left \{ \frac{\left(\dbar_\sMo+\dbar_\sMt+(\varepsilon-1)\alpha(C_{\sMo} + C_{\sMt})\right)}{4\alpha(1-\varepsilon)}+\frac{C_{\sS}}{2} ,\sigma_{2}^{b}  \right \} 
 \end{eqnarray*}
 Using simple algebra, and~\eqref{eqn_lim_sigma_ib}, observe that
 \begin{align*}
\lim_{(\varepsilon,\gamma) \to 1}(1-\gamma)(1-\varepsilon)\left( \frac{\left(\dbar_\sMo+\dbar_\sMt+(\varepsilon-1)\alpha(C_{\sMo} + C_{\sMt})\right)}{4\alpha(1-\varepsilon)}+\frac{C_{\sS}}{2} \right) &= \frac{\dbar_\sMo + \dbar_\sMt}{4\tilde\alpha}
, \text{ and } \\
\lim_{(\varepsilon,\gamma) \to (1,1)}(1-\gamma)(1-\varepsilon)\sigma_{2}^{b} &= \frac{2\dbar_\sMt + \dbar_\sMo}{3\tilde\alpha} . 
 \end{align*}
Thus as $(\nicefrac{\dbar_\sMo + \dbar_\sMt)}{4\tilde\alpha} < (\nicefrac{2\dbar_\sMt + \dbar_\sMo)}{3\tilde\alpha} $, we have that in the neighbhourhood of $(\varepsilon,\gamma) \to (1,1)$, the optimizer of function in \eqref{eqn_supp_util_case2} is,
\begin{eqnarray}\label{eqn_opt_q_star}
q^{*} =  \frac{\left(\dbar_\sMo+\dbar_\sMt+(\varepsilon-1)\alpha(C_{\sMo} + C_{\sMt})\right)}{4\alpha(1-\varepsilon)}+\frac{C_{\sS}}{2} 
\end{eqnarray}
Thus after substituting this $q^{*}$ in \eqref{eqn_supp_util_case2} and then taking the limits after scaling, we get
\begin{eqnarray}\label{eqn_worth_lim_supp_alc_conf}
\lim_{(\varepsilon,\gamma) \to (1,1)}(1-\gamma)(1-\varepsilon) U_\sS(a_\sS^{*}) = \frac{\dbar_\sM^2}{8\tilde\alpha}
\end{eqnarray}
From \eqref{suppliermonopolymanufacturerutility}, that the optimal utility of the manufacturer $i$ is given by:
\begin{eqnarray}\label{eqn_opt_util_manuu_alc_case}
U_\sMi(\a_\sM^{*},a_\sS^{*}) =   \left((\dbar_\sMi + \varepsilon \alpha_\sMminusi p_{-i}^{*}(q^{*})-\alpha_\sMi p_{i}^{*}(q^{*}))(p_{i}^{*}(q^{*}) - q^{*} - C_\sMi)\F_\sS - O_\sMi\right)\F_\sMi
\end{eqnarray}
\ignore{
{\color{blue}
$$
\lim_{\gamma \to 1}(1-\gamma)(p_{i}^{*}(q) -q) = \frac{\varepsilon\dbar_\sMminusi + 2\dbar_\sMi + (\varepsilon^2 + \varepsilon -2)\tilde\alpha(\lim_{\gamma \to 1}(1-\gamma)q^{*})}{(4-\varepsilon^2)\tilde\alpha}
= \frac{\varepsilon\dbar_\sMminusi + 2\dbar_\sMi +(\varepsilon - 1)(\varepsilon+2) \frac{\dbar_\sMi +\dbar_\sMminusi}{4\tilde\alpha (1-\gamma)(1-\varepsilon)}(1-\gamma)\tilde\alpha}{(4-\varepsilon^2)\tilde\alpha}
$$
$$
= \frac{(6-\varepsilon)\dbar_\sMi + (3\varepsilon - 2)\dbar_\sMminusi}{(4-\varepsilon^2)\tilde\alpha}
$$
 This is always positive as $\varepsilon \to 1$

}
}
Also observe that after substituting $q^{*}$ from \eqref{eqn_opt_q_star} in \eqref{eqn_pstar_q}, and then taking the limits we get that 
\begin{eqnarray*}
\lim_{(\varepsilon,\gamma) \to (1,1)}(1-\gamma)(1-\varepsilon)(p_{i}^{*}(q^{*})- q^{*}) \to 0 \quad \forall i
\end{eqnarray*}
Thus we have,
\begin{eqnarray}\label{eqn_Worth_lim_manu_alc_config}
\lim_{(\varepsilon,\gamma) \to (1,1)}(1-\gamma)(1-\varepsilon) U_\sMi(\a_\sM^{*},a_\sS^{*})  = 0 
  \quad \forall i.  
\end{eqnarray}
Thus from \eqref{eqn_worth_lim_supp_alc_conf}and \eqref{eqn_Worth_lim_manu_alc_config}, the ESM limit of  worths of coalition ALC are,
\begin{eqnarray*}
  f_\sS^{\pALC} =\lim_{(\varepsilon,\gamma) \to (1,1)}(1-\gamma)(1-\varepsilon)\nu_\sS^\pALC = \frac{\dbar_\sM^2}{8\tilde\alpha}\\
 f_\sMi^{\pALC} = \lim_{(\varepsilon,\gamma) \to (1,1)}(1-\gamma)(1-\varepsilon)\nu_\sMi^\pALC = 0  \ \  \forall i. 
\end{eqnarray*}

Observe that after substituting  $q^*$ from m \eqref{eqn_opt_q_star} in \eqref{eqn_pstar_q}, and then substituting $q^*$ and the obtained $p_i^{*}(q^*)$ in \eqref{eqn_supp_util_case2} and \eqref{eqn_opt_util_manuu_alc_case} we get
\begin{eqnarray}\label{eqn_worth_epsilon_alc}
\lim_{\gamma\to 1}(1-\gamma)\nu_\sS^\pALC = \frac{\dbar_\sM^2}{8\tilde\alpha(1-\varepsilon)(2-\varepsilon)}
, \text{ and } \quad 
\lim_{\gamma \to 1}(1-\gamma)\nu_\sMi^\pALC = \frac{((6-\varepsilon)\dbar_\sMi + (3\varepsilon -2)\dbar_\sMminusi)^2}{16(4-\varepsilon^2)^2\tilde\alpha} \mbox{ for all $i$ }.
\end{eqnarray}
Thus after scaling the worths in equation \eqref{eqn_worth_epsilon_alc} by $(1-\varepsilon)$ and differentiating the scaled worths with respect to $\varepsilon$ and then taking the limit $\varepsilon \to 1$, we get,
\begin{eqnarray*}
f_{\sS}^{(1),\pALC} &=&  \frac{\dbar_\sM^2}{8\tilde\alpha}.  \\
f_{\sMi}^{(1),\pALC} &=& \frac{-(5\dbar_\sMi + \dbar_\sMminusi)^2}{144\tilde\alpha}.
\end{eqnarray*}
This completes the proof.
\end{proof}

\ENMremoved{
\vspace{-2mm}
\subsection*{ESNM Regime}
\vspace{-2mm}
From equations \eqref{eqn_lim_sigma_ia} and \eqref{eqn_lim_sigma_ib}, observe that in the ESNM regime with $\dbar_\sMo \ge \dbar_\sMt$, at the limit, we have $\sigma_2^{b} < \min\{ \sigma_1^{a}, \sigma_1^{b}\}$. Now, using this strict inequality and continuity of functions $\sigma_1^a, \sigma_1^b,$ and $ \sigma_2^b$ with respect to ($\varepsilon, \gamma$), the same strict inequality is satisfied for all $(\varepsilon,\gamma)$ in a neighbourhood of $(0,1)$. For all such $(\varepsilon,\gamma)$, the supplier utility \eqref{util_supp_All_alone_simp} after substituting the NE of inner game, modifies to 
\begin{eqnarray}\label{util_supp_eip}
  U_{\sS}(a_\sS)= \left \{ \begin{array}{cl}
  0  & \mbox{ if }  a_\sS = n_o \\
     -O_\sS  &   \mbox{ if } a_\sS \ne n_o \mbox{ and }  q > \sigma_1^{a}  \\
     
   \left( \frac{\dbar_{\sMo} - \tilde\alpha (1-\gamma)(C_{\sMo} + q)}{2}\right)(q-C_{\sS}) - O_{\sS} & \mbox{ if } a_\sS \ne n_o \mbox{ and } \sigma_{2}^{b} < q \le  \sigma_{1}^{a} \\
 \left(\frac{\dbar_{\sMo}+\dbar_{\sMt}+(\varepsilon-1)(1-\gamma)\tilde\alpha (C_{\sMo}+C_{\sMt})-(1-\varepsilon)(1-\gamma)2\tilde\alpha q}{(2-\varepsilon)}\right)\left(q-C_{\sS} \right) -O_{\sS}  & \mbox{ if } a_\sS \ne n_{o} \mbox{ and } q \le \sigma_{2}^{b}
  \end{array}\right .   
\end{eqnarray}
Now, we define the unconstrained optimizers of the above utilities, where
\begin{align*}
m_1^{*} &:= \arg\max_{q} \left( \frac{\dbar_{\sMo} - \tilde\alpha (1-\gamma)(C_{\sMo} + q)}{2}\right)(q-C_{\sS}) - O_{\sS} =  \frac{\dbar_\sMo - \tilde\alpha(1-\gamma) C_\sM}{2\tilde\alpha(1-\gamma)} + \frac{C_\sS}{2},\\
m_2^{*} &:= \arg\max_{q}\left(\frac{\dbar_{\sMo}+\dbar_{\sMt}+(\varepsilon-1)(1-\gamma)\tilde\alpha (C_{\sMo}+C_{\sMt})-(1-\varepsilon)(1-\gamma)2\tilde\alpha q}{(2-\varepsilon)}\right)\left(q-C_{\sS} \right) -O_{\sS} \\
&= \frac{(\dbar_\sMo+\dbar_\sMt)+(\varepsilon-1)(1-\gamma)\tilde\alpha(C_{\sMo} + C_{\sMt})}{4\tilde\alpha(1-\gamma)(1-\varepsilon)}+\frac{C_{\sS}}{2} .
\end{align*}
Now we find the scaled limits of the above optimizes, 
\begin{eqnarray}\label{eqn_m1_star_m2_star}
\lim_{(\varepsilon,\gamma) \to (0,1)}(1-\gamma)  m_{1}^{*} = \frac{\dbar_\sMo}{2\tilde\alpha} , \text{ and }
  \lim_{(\varepsilon,\gamma) \to (0,1)}(1-\gamma)  m_{2}^{*}  = \frac{\dbar_\sMo + \dbar_\sMt}{4\tilde\alpha}
\end{eqnarray}
In this subsection, whenever we talk about the limit of $m_i^*$, $\sigma_i^a$, $\sigma_i^b$, we refer to the scaled limit. Now from \eqref{eqn_m1_star_m2_star} and \eqref{eqn_lim_sigma_ia}, in the ESNM regime, at limit, we get $m_1^{*} \le \sigma_1^{a}$.
From \eqref{eqn_m1_star_m2_star} and \eqref{eqn_lim_sigma_ib},
if $\dbar_\sMo > 2\dbar_\sMt$ then in the ESNM regime, at limit, we have $\sigma_2^{b}< m_1^{*} \le \sigma_{1}^{a}$; else, we have $ m_1^{*} \le \sigma_2^{b}$. Similarly, from \eqref{eqn_lim_sigma_ib} and \eqref{eqn_m1_star_m2_star}, if $\dbar_\sMo \le 3\dbar_\sMt$ in the ESNM regime, at limit we have $m_2^{*} \le \sigma_2^{b}$; otherwise we have $m_2^{*} > \sigma_2^{b}$.

{\textbf{Case 1 :} $ 2\dbar_\sMt <\dbar_\sMo \le 3\dbar_\sMt $}. From the above explaination, in the ESM regime, at limit, we have $ \sigma_2^{b} < m_1^{*} \le \sigma_1^{a}$ and $m_2^{*} \le \sigma_2^{b}$. To find the optimizer $q^*$ of $U_{\sS}$ as given in~\eqref{util_supp_eip}, we compare the scaled limit of the utility at $m_1^*$ and $m_2^*$ below: 
\begin{equation}\label{eqn_util_at_m}
\lim_{(\varepsilon,\gamma) \to (0,1)}(1-\gamma)U_{\sS}(m_{1}^{*})= \frac{\dbar_\sMo^2}{8\tilde\alpha}
, \text{ and }
\lim_{(\varepsilon,\gamma) \to (0,1)}(1-\gamma)U_{\sS}(m_2^{*}) = \frac{\dbar_\sM^2}{16\tilde\alpha} .
\end{equation}
Compairing above utilities, we get that if $\left(\sqrt{2} - 1\right)\dbar_\sMo > \dbar_\sMt$
then  $q^{*} = m_{1}^{*}$, else,  $q^{*} = m_2^{*}$. 
\begin{enumerate}[(i)]
    \item $\left(\sqrt{2} - 1\right)\dbar_\sMo > \dbar_\sMt$\\ \noindent
In this case observe from \eqref{util_supp_eip} which is obtained by substituting the NE of inner game of the manufacturers that as $q^{*} = m_{1}^{*}$, only the first manufacturer operates . Thus from \eqref{eqn_pib} which is obtained by applying Lemma \ref{lem_game_btw_manufacturers},  $p_{1}^{*}(q^{*}) = p_{1}^{b}(q^{*})$ and the second manufacturer chooses not to operate. Thus the supplier being the leader of Stackelberg game quotes the raw material price in a way that second manufacturer does not operate as he gets more utility by doing so. Now by substituting obtained values of $q^{*}$ in \eqref{util_supp_eip} and and substituting $p_{i}^{*}(q^{*})$ in \eqref{??} and then taking the limits after scaling we get the following:
\begin{eqnarray*}
\lim_{(\varepsilon,\gamma) \to (0,1)}(1-\gamma)q^{*} =  \frac{\dbar_\sMo}{2\tilde\alpha}, \quad
\lim_{(\varepsilon,\gamma) \to (0,1)}(1-\gamma)U_\sS^{*} = \frac{\dbar_\sMo^{2}}{8\tilde\alpha}, \quad 
\lim_{(\varepsilon,\gamma) \to (0,1)}(1-\gamma)U_\sMo^{*} = \frac{\dbar_\sMo^2}{16\tilde\alpha}, \quad
\lim_{(\varepsilon,\gamma) \to (0,1)}(1-\gamma)U_\sMt^{*} = 0.
\end{eqnarray*}
Thus, the ENSM limits of the worths are,
\begin{eqnarray*}
 g_\sS^{\pALC}   = \frac{\dbar_\sMo^{2}}{8\tilde\alpha}
, \quad 
g_\sMo^{\pALC} =  \frac{\dbar_\sMo^{2}}{16\tilde\alpha}
, \quad
 g_\sMt^{\pALC} = 0 .
\end{eqnarray*}

\item $\left(\sqrt{2} - 1\right)\dbar_\sMo \leq \dbar_\sMt$\\ \noindent
\begin{eqnarray*}
\lim_{(\varepsilon,\gamma) \to (0,1)}(1-\gamma)q^{*} =  \frac{\dbar_\sMo +\dbar_\sMt}{4\tilde\alpha},  \quad
\lim_{(\varepsilon,\gamma) \to (0,1)}(1-\gamma)U_{\sS}^{*} = \frac{\dbar_\sM^2}{16\tilde\alpha}, \quad
\lim_{(\varepsilon,\gamma) \to (0,1)}(1-\gamma)U_{\sMi}^{*} = \frac{(3\dbar_\sMi -\dbar_\sMminusi)^2}{64\tilde\alpha}.
\end{eqnarray*}
Thus, the EIP limits of the worths are,
\begin{eqnarray*}
 g_\sS^{\pALC}   = \frac{\dbar_\sM^{2}}{16\tilde\alpha}
, \quad 
g_\sMo^{\pALC} = \frac{(3\dbar_\sMo -\dbar_\sMt)^2}{64\tilde\alpha}, \quad
 g_\sMt^{\pALC}  = \frac{(3\dbar_\sMt -\dbar_\sMo)^2}{64\tilde\alpha} .
\end{eqnarray*}
\end{enumerate}

\ignore{
In this case we have the optimizer of the function \eqref{util_supp_eip} in the limit as
\begin{eqnarray*}
\lim_{(\varepsilon,\gamma) \to (0,1)}(1-\gamma)q^{*} =    \lim_{(\varepsilon,\gamma) \to (0,1)}(1-\gamma)m_1^{*}  = \frac{\dbar_\sMo}{2\tilde\alpha}
\end{eqnarray*}
Recall that from \eqref{suppliermonopolymanufacturerutility}, that the optimal utility of the manufacturer is given by
\begin{eqnarray*}
U_\sMi^{*} =   \left((\dbar_\sMi + \varepsilon \alpha_\sMminusi p_{i}^{*}(q^{*})-\alpha_\sMi p_{-i}^{*}(q^{*}))(p_{i}^{*}(q^{*}) - q^{*} - C_\sMi)\indc{a_\sS \ne n_{o}} - O_\sMi\right)\indc{a_\sMi \ne n_{o}} 
\end{eqnarray*}
But as here the optimizer $q^{*} = m_{1}^{*}$, 
Thus we have 
\begin{eqnarray*}
\lim_{(\varepsilon,\gamma) \to (0,1)}(1-\gamma)U_\sS^{*} = \frac{\dbar_\sMo^{2}}{8\tilde\alpha}
\end{eqnarray*}
Then we have that
\begin{eqnarray*}
\lim_{(\varepsilon,\gamma) \to (0,1)}(1-\gamma)U_\sMo^{*} = \frac{\dbar_\sMo^2}{16\tilde\alpha}
\end{eqnarray*}
Also
\begin{eqnarray*}
\lim_{(\varepsilon,\gamma) \to (0,1)}(1-\gamma)U_\sMt^{*} = 0
\end{eqnarray*}
Thus we have that in this case that the EIP limits of the worths of the ALC case are given by:
\begin{eqnarray*}
 g_\sS^{\pALC}   = \frac{\dbar_\sMo^{2}}{8\tilde\alpha}
\end{eqnarray*}
\begin{eqnarray*}
g_\sMo^{\pALC} =  \frac{\dbar_\sMo^{2}}{16\tilde\alpha}
\end{eqnarray*}
and
\begin{eqnarray*}
 g_\sMt^{\pALC} = 0 
\end{eqnarray*}

Now if $\dbar_\sMo^{2} -\dbar_\sMt^2 -2\dbar_\sMo\dbar_\sMt \le 0$, then $\lim_{(\varepsilon,\gamma) \to (0,1)}(1-\gamma)U_{\sS}(m_2^{*})   >  \lim_{(\varepsilon,\gamma) \to (0,1)}(1-\gamma)U_{\sS}(m_{1}^{*}) $. In this case we have the optimizer of the function \eqref{util_supp_eip} in the limit as
\begin{eqnarray*}
\lim_{(\varepsilon,\gamma) \to (0,1)}(1-\gamma)q^{*} =    \lim_{(\varepsilon,\gamma) \to (0,1)}(1-\gamma)m_1^{*}  = \frac{\dbar_\sMo +\dbar_\sMt}{4\tilde\alpha}
\end{eqnarray*}
Thus we have
$$
\lim_{(\varepsilon,\gamma) \to (0,1)}(1-\gamma)U_{\sS}^{*} = \frac{\dbar_\sM^2}{16\tilde\alpha}
$$
Also 
$$
\lim_{(\varepsilon,\gamma) \to (0,1)}(1-\gamma)U_{\sMi}^{*} = \frac{(3\dbar_\sMi -\dbar_\sMminusi)^2}{64\tilde\alpha}
$$
Thus we have that in this case that the EIP limits of the worths of the ALC case are given by:
\begin{eqnarray*}
 g_\sS^{\pALC}   = \frac{\dbar_\sM^{2}}{16\tilde\alpha}
\end{eqnarray*}
\begin{eqnarray*}
g_\sMo^{\pALC} = \frac{(3\dbar_\sMo -\dbar_\sMt)^2}{64\tilde\alpha}
\end{eqnarray*}
and
\begin{eqnarray*}
 g_\sMt^{\pALC}  = \frac{(3\dbar_\sMt -\dbar_\sMo)^2}{64\tilde\alpha}
\end{eqnarray*} }

{\textbf{Case 2:} $\dbar_\sMo \le 2\dbar_\sMt$} We have in the EIP regime that $m_1^{*} \le \sigma_2^{b}$ and $m_2^{*} \le \sigma_2^{b}$.
Observe  from \eqref{util_supp_eip} that when $q \in (\sigma_2^b, \sigma_1^a]$, the function $U_\sS$ is quadratic and concave with respect to $q$. Since in the scaled limit $m_1^* \leq \sigma_2^b$, for all $q \in (\sigma_2^b, \sigma_1^a]$, in the scaled limit, we have $ U_\sS(q) < U_\sS(\sigma_2^b)$.
Therefore, showing $U_\sS(\sigma_2^b) < U_\sS(m_2^*)$ in the scaled limit, suffices to conclude that $q^* = m_2^*$ is the optimizer of $U_\sS.$ Towards this, consider the following limits
\begin{eqnarray*}
\lim_{(\varepsilon,\gamma)\to (0,1)}(1-\gamma)U_\sS(\sigma_{2}^{b}) =    , \quad 
\lim_{(\varepsilon,\gamma)\to (0,1)}(1-\gamma)U_\sS(m_{2}^{*}) =  \frac{\dbar_\sM^2}{16\tilde\alpha}
\end{eqnarray*}

Thus we get that $q_{1}^{*} = \sigma_1^{a}$ and $q_2^{*} = m_2^{*}$ in the EIP limit regime. Then we get 

Thus it is easy to observe that $\lim_{(\varepsilon,\gamma)\to (0,1)}(1-\gamma)U_\sS(m_{2}^{*}) \ge \lim_{(\varepsilon,\gamma)\to (0,1)}(1-\gamma)U_\sS(\sigma_{1}^{a})$.
Hence we observe that the EIP limits of the worths of the ALC case are given by:
\begin{eqnarray*}
 g_\sS^{\pALC}   = \frac{\dbar_\sM^{2}}{16\tilde\alpha}
\end{eqnarray*}
\begin{eqnarray*}
g_\sMo^{\pALC} = \frac{(3\dbar_\sMo -\dbar_\sMt)^2}{64\tilde\alpha}
\end{eqnarray*}
and
\begin{eqnarray*}
 g_\sMt^{\pALC}  = \frac{(3\dbar_\sMt -\dbar_\sMo)^2}{64\tilde\alpha}
\end{eqnarray*}

{\textbf{Case 3}: $\dbar_\sMo \ge 3\dbar_\sMt$} We have in the EIP regime that  $\sigma_{1}^{a} \ge m_1^{*} > \sigma_2^{b}$ and $m_2^{*} >\sigma_2^{b}$. Thus we get that $q_{1}^{*} = m_1^{*}$ and $q_2^{*} = \sigma_2^{b}$

Also observe that $U_\sS(m_1^{*}) = \frac{\dbar_\sMo^2}{8\tilde\alpha}$ and $U_\sS(\sigma_2^{b}) = \frac{\dbar_\sMo\dbar_\sMt - \dbar_\sMt^2}{2\tilde\alpha}$
Also observe that here $U_\sS(m_1^{*}) > U_\sS(\sigma_2^{b}) $
Thus we have that in this case the EIP limits of the worths of ALC case are
\begin{eqnarray*}
 g_\sS^{\pALC}   = \frac{\dbar_\sMo^{2}}{8\tilde\alpha}
\end{eqnarray*}
\begin{eqnarray*}
g_\sMo^{\pALC} =  \frac{\dbar_\sMo^{2}}{16\tilde\alpha}
\end{eqnarray*}
and
\begin{eqnarray*}
 g_\sMt^{\pALC} = 0 
\end{eqnarray*}

\vspace{-2mm}
\subsection{Result of ALC case}
\vspace{-2mm}
\begin{itemize}
    \item If $2\dbar_\sMt < \dbar_\sMo \le 3\dbar_\sMt$ and $(\sqrt{2}-1)\dbar_\sMo \ge \dbar_\sMt$ hold or if  $\dbar_\sMo \ge 3\dbar_\sMt$ then the worths of ALC case in EIP regime are 
    $\{g_\sS^{\pALC},g_\sMo^{\pALC}, g_\sMt^{\pALC}  \} = \{\frac{\dbar_\sMo^{2}}{8\tilde\alpha},\frac{\dbar_\sMo^{2}}{16\tilde\alpha},0\}$
    \item  If $2\dbar_\sMt < \dbar_\sMo \le 3\dbar_\sMt$ and $(\sqrt{2}-1)\dbar_\sMo <\dbar_\sMt$ hold or if  $\dbar_\sMo \le 2\dbar_\sMt$ then the worths of ALC case in EIP regime are 
    $\{g_\sS^{\pALC},g_\sMo^{\pALC}, g_\sMt^{\pALC}  \} = \{\frac{\dbar_\sM^{2}}{16\tilde\alpha},\frac{(3\dbar_\sMo -\dbar_\sMt)^{2}}{64\tilde\alpha},\frac{(3\dbar_\sMt -\dbar_\sMo)^{2}}{64\tilde\alpha}\}$
    
\end{itemize}
}

\vspace{-2mm}
\ignore{
\subsection{Insights}
\vspace{-2mm}
i) Existence of two competitors ensures a much larger combined demand than in the case when the two operate together -- when they operate as opponents, a fraction customers of each tend to drift towards the another based on prices offered and essentialness of the product;  $\alpha_{-i} p_{-i} $  is the fraction that rejected manufacturer $\M_{-i}$, out of which  $\varepsilon\alpha_{-i}p_{-i}$  represents the fraction that blindly drifts towards $M_{i}$, depending on essentialness factor $\varepsilon$.
Thus the prices and the utility of ALC and VC are much larger than those at GC and HC when $\varepsilon \to 1$,

ii)  $0$ and $\zero$ are different type of zeros

iii)  In single manufacturer case supplier gets $2 \phi$, while manufacturer gets $\phi$ for some $\phi$.  
While in 2 manufacturer case the division depends upon the substituatbility of the manu.. 
If th manus ar substitutable then supplier gets $\phi$ and manus (0,0) in the othere case intrestingly we once again have $\phi, \phi$ and $2\phi$ for supplier 

iv) when supplier and manu are in nash then both of them will not operate

v) Constant thing : Sum total share of manu seems to be half of that of supplier

vi) When manufacturers are substitutible gc is not important and the supplier will  want to collaborate with at max one manu

vii) When manu are
\vspace{-2mm}
\subsection{ Calculating Derivative}
\vspace{-2mm}
Case 1: All Alone Case
$$
\tilde\nu_\sS^{\pALC} = 
(1-\varepsilon)\lim_{\gamma\to 1}(1-\gamma)\nu_\sS^\pALC = \frac{\dbar_\sM^2}{8\tilde\alpha(2-\varepsilon)} \Rightarrow  \text{ derivative is } \frac{\dbar_\sM^2}{8\tilde\alpha(2-\varepsilon)^2} \implies \frac{\dbar_\sM^2}{8\tilde\alpha}
$$
$$
\tilde\nu_\sMo^{\pALC} =
(1-\varepsilon)\lim_{\gamma \to 1}(1-\gamma)\nu_\sMi^\ALC=(1-\varepsilon) \frac{((6-\varepsilon)\dbar_\sMo + (3\varepsilon -2)\dbar_\sMt)^2}{16(4-\varepsilon^2)^2\tilde\alpha}  \mbox{ and derivative (at $\varepsilon=1$) = } -\frac{\left(5\dbar_\sMo + \dbar_\sMt \right)^2}{144\tilde\alpha}
$$

Case 2: VC Case
\begin{eqnarray*}
\tilde\nu_\sV^{\pVC} = (1-\varepsilon)\lim_{\gamma \to 1}(1-\gamma)\nu_\sV^{\pVC} 
& =& \left(\frac{\dbar_\sMo}{2} + \frac{\varepsilon\dbar_\sMt}{4}\right)\left(\frac{\dbar_\sMo + \dbar_\sMt\varepsilon}{2\tilde\alpha(1+\varepsilon)}\right) + \left( \frac{\dbar_\sMt}{4}\right)\left(\frac{\dbar_\sMt + \varepsilon\dbar_\sMi}{2(1+\varepsilon)\tilde\alpha}\right) \\ & &
 \mbox{ and derivative (at $\varepsilon=1$) = }
\left(\frac{2\dbar_\sMo\dbar_\sMt + \dbar_\sMt^2 -\dbar_\sMo^2}{16\tilde\alpha}\right)
\end{eqnarray*}
\begin{eqnarray*}
   \tilde\nu_\sMt^{\pVC} = (1-\varepsilon)\lim_{\gamma \to 1}(1-\gamma)\nu_\sMt^{\pVC} =  (1-\varepsilon)\frac{\dbar_\sMt^2}{16\tilde\alpha}  \mbox{ and derivative (at $\varepsilon=1$) = } -\frac{\dbar_\sMt^2}{16\tilde\alpha}  \\
    \tilde\nu_\sM^{\CHC} = (1-\varepsilon)\lim_{\gamma \to 1}(1-\gamma)\nu_\sM^{\CHC} =  (1-\varepsilon)\frac{\dbar_\sM^2}{16\tilde\alpha}  \mbox{ and derivative (at $\varepsilon=1$) = } -\frac{\dbar_\sM^2}{16\tilde\alpha}
\end{eqnarray*}

Further towards $\nu^{pa}_{\sMo}$
\begin{eqnarray*}
   \tilde\nu_{\sMo}^{\VC_2}  - \tilde \nu_\sMo^\ALC = (1-\varepsilon) \left ( \frac{\dbar_\sMo^2}{16\tilde\alpha}  - \frac{((6-\varepsilon)\dbar_\sMo + (3\varepsilon -2)\dbar_\sMt)^2}{16(4-\varepsilon^2)^2\tilde\alpha}    \right )  < 0  \mbox{ at limit }
\end{eqnarray*}
Thus in the regime,
$$
\nu^{pa}_{\sMo} = \nu_{\sMo}^{\VC_2}  \mbox{ and } \max \{ \nu_\sM^{pa} , \nu_\sMo^{pa} \} = \nu_\sM^{pa} = \max \{ \nu_\sM^{pa} , \nu_\sMt^{pa} \} 
$$
Similarly
$$
\nu^{pa}_{\sMt} = \nu_{\sMt}^{\VC_1} 
$$

Thus to ensure $V_2$ does not block when $x_\sMo = \max \{ \nu_\sM^{pa} , \nu_\sMo^{pa} \} = \nu_\sM^{pa}$ we need the below to be $> 0$
\begin{eqnarray*}
    \tilde \nu_\sMt^{\VC_1} + \tilde \nu_{\V_1}^{\VC_1} -  \max \{ \tilde \nu_\sM^{pa} , \tilde \nu_\sMo^{pa} \} - \tilde \nu_{\V_2}^{\VC_2} 
    &=&  (1-\varepsilon) \left  (  \frac{\dbar_\sMt^2}{16\tilde\alpha}  
     - \frac{\dbar_\sM^2}{16\tilde\alpha}  
    \right ) + \tilde \nu_{\V_1}^{\VC_1} - \tilde \nu_{\V_2}^{\VC_2}  
\end{eqnarray*}
%Using derivatives
%\begin{eqnarray*}
%  (\varepsilon-1) \left (  f_{\V_1}^{(1), \VC_1} - \tilde f_{\V_2}^{ (1), \VC_2} \right ) - \left  (  \frac{\dbar_\sMt^2}{16\tilde\alpha}  
  %   - \frac{\dbar_\sM^2}{16\tilde\alpha}  
%    \right ) > 0
%\end{eqnarray*}

for $\varepsilon \approx 1$, i.e.,  near 1,
$$
f(\varepsilon) = 
f( 1+  \varepsilon-1 ) = f(1)  + (\varepsilon-1) f' (1)  + o ( (1-\varepsilon)^2)
$$

Thus to ensure $V_1$ does not block $\VC_2$ when $x_\sMt = \max \{ \nu_\sM^{pa} , \nu_\sMt^{pa} \} = \nu_\sM^{pa}$ we need the below to be $> 0$
\begin{eqnarray*}
    \tilde \nu_\sMo^{\VC_2} + \tilde \nu_{\V_2}^{\VC_2} -  \max \{ \tilde \nu_\sM^{pa} , \tilde \nu_\sMt^{pa} \} - \tilde \nu_{\V_1}^{\VC_1} 
    &=&  (1-\varepsilon) \left  (  \frac{\dbar_\sMo^2}{16\tilde\alpha}  
     - \frac{\dbar_\sM^2}{16\tilde\alpha}  
    \right ) +  \tilde \nu_{\V_2}^{\VC_2}   - \tilde \nu_{\V_1}^{\VC_1}  \\
    &=&    (1-\varepsilon) \left  (  \frac{\dbar_\sMo^2}{16\tilde\alpha}  
     - \frac{\dbar_\sM^2}{16\tilde\alpha}  
    \right ) -  (1-\varepsilon)  \left ( f_{\V_2}^{(1), \VC_2}   - \tilde f_{\V_1}^{(1), \VC_1}  \right )
    \\
    &=& (1-\varepsilon)   \left  (  \frac{\dbar_\sMo^2}{16\tilde\alpha}  
     - \frac{\dbar_\sM^2}{16\tilde\alpha}  - (\frac{\dbar_\sMt^2 -\dbar_\sMo^2}{8\tilde\alpha})   
    \right ) \\
      &=& \frac{(1-\varepsilon) }{16 \tilde \alpha}  \left  (   \dbar_\sMo^2  
     - \dbar_\sM^2  -  2 \dbar_\sMt^2 + 2\dbar_\sMo^2    
    \right ) = \frac{(1-\varepsilon) }{16 \tilde \alpha} (  2\dbar_\sMo^2  -  3 \dbar_\sMt^2 -   2\dbar_\sMo \dbar_\sMt  ) 
\end{eqnarray*}

what is 
\begin{eqnarray*}
  \tilde\nu^{\VC_2}_\sMo +  \tilde \nu_\sMt^{pa}   -  \tilde \nu_\sM^{pa}  = (1-\varepsilon)   \left  (  \frac{\dbar_\sMo^2}{16\tilde\alpha}  +  \frac{\dbar_\sMt^2}{16\tilde\alpha}
     - \frac{\dbar_\sM^2}{16\tilde\alpha}      
    \right )
\end{eqnarray*}not good..

Thus to ensure $V_1$ does not block $\VC_2$ when $x_\sMt =   \nu_\sM^{pa} - \nu_\sMo^{\VC_2}$ we need the below to be $> 0$
\begin{eqnarray*}
    \tilde \nu_\sMo^{\VC_2} + \tilde \nu_{\V_2}^{\VC_2} -  {\tilde  x}_\sMt - \tilde \nu_{\V_1}^{\VC_1} 
    &=&  (1-\varepsilon) \left  (  \frac{\dbar_\sMo^2}{16\tilde\alpha}  
     - \frac{\dbar_\sM^2}{16\tilde\alpha} + \frac{\dbar_\sMo^2}{16\tilde\alpha} 
    \right ) +  \tilde \nu_{\V_2}^{\VC_2}   - \tilde \nu_{\V_1}^{\VC_1}  \\
    &=&    (1-\varepsilon) \left  (  \frac{\dbar_\sMo^2}{16\tilde\alpha}  
     - \frac{\dbar_\sM^2}{16\tilde\alpha} + 
   \frac{\dbar_\sMo^2}{16\tilde\alpha} \right ) -  (1-\varepsilon)  \left ( f_{\V_2}^{(1), \VC_2}   - \tilde f_{\V_1}^{(1), \VC_1}  \right )
    \\
    &=& (1-\varepsilon)   \left  (  \frac{\dbar_\sMo^2}{16\tilde\alpha}  
     - \frac{\dbar_\sM^2}{16\tilde\alpha} + 
   \frac{\dbar_\sMo^2}{16\tilde\alpha}  - (\frac{\dbar_\sMt^2 -\dbar_\sMo^2}{8\tilde\alpha})   
    \right ) \\
      &=& \frac{(1-\varepsilon) }{16 \tilde \alpha}  \left  (   \dbar_\sMo^2  
     - \dbar_\sM^2 +\dbar_\sMo^2 -  2 \dbar_\sMt^2 + 2\dbar_\sMo^2    
    \right ) = \frac{(1-\varepsilon) }{16 \tilde \alpha} (  3\dbar_\sMo^2  -  3 \dbar_\sMt^2 -   2\dbar_\sMo \dbar_\sMt  ) 
\end{eqnarray*}
\newpage
\vspace{-2mm}
\subsection{Quantities only with $\gamma \to 1$}
\vspace{-2mm}

GC: $$\lim_{\gamma \to 1}(1-\gamma) U^*_\sG = \frac{\dbar_\sM^2}{4\tilde\alpha}$$

ALC: 

\begin{eqnarray*}
    \lim_{\gamma\to 1}(1-\gamma)U_\sS^{*} &=& \frac{\dbar_\sM^2}{8\tilde\alpha(1-\varepsilon)(2-\varepsilon)} \\
\lim_{\gamma \to 1}(1-\gamma)U_\sMi^{*} &=& \frac{((6-\varepsilon)\dbar_\sMi + (3\varepsilon -2)\dbar_\sMminusi)^2}{16(4-\varepsilon^2)^2\tilde\alpha} 
\end{eqnarray*}

HC :

\begin{eqnarray*}
  \lim_{\gamma \to 1} (1-\gamma) U^*_\sS &=& \frac{\dbar_\sM^2}{8\tilde\alpha}\\
 \lim_{\gamma \to 1} (1-\gamma) U^*_\sM &=& \frac{\dbar_\sM^2}{16\tilde\alpha}    
\end{eqnarray*}

VC : 
\begin{eqnarray*}
\lim_{\gamma \to 1}(1-\gamma)U_\sV^{*} 
& =& \left(\frac{\dbar_\sMo}{2} + \frac{\varepsilon\dbar_\sMt}{4}\right)\left(\frac{\dbar_\sMo + \dbar_\sMt\varepsilon}{2\tilde\alpha(1-\varepsilon^{2})}\right) + \left( \frac{\dbar_\sMt}{4}\right)\left(\frac{\dbar_\sMt + \varepsilon\dbar_\sMi}{2(1-\varepsilon^2)\tilde\alpha}\right) \\
\lim_{\gamma \to 1}(1-\gamma)U_\sMminusi^{*} &=& \frac{\dbar_\sMminusi^2}{16\tilde\alpha}
\end{eqnarray*}}

`}
\end{document}

%% file: Introduction.tex
\ignore{
\section{Introduction}
\ignore{Supply chain has evolved very rapidly since 1990s showing an exponential growth
in papers in different journals of interest to academics and practitioners (Burgess
et al. 2006). The rise in papers on supply chain (SC) as well as the case studies in
different areas in different industries motivates to study SC issues further. Supply
chains are generally complex with numerous activities (logistics, inventory, purchasing and procurement, production planning, intra-and inter-organizational
relationships and performance measures) usually spread over multiple functions
or organizations and sometimes over lengthy time horizons. Supply chains tend to
increase in complexity and the involvement of numerous suppliers, service
providers, and end consumers in a network of relationships causes risks and
vulnerability for everyone (Pfohl et al. 2010)

In the last several years, the evolution of supply
chain management recognized that (i) a business
process consists of several decentralized firms
and (ii) operational decisions of these different
entities impact each others’ profit, and thus the
profit of the whole supply chain.
understanding came a great deal of interest in
modeling and understanding the impact of strategic operational decisions of the various players
in supply chains. To effectively model and analyse
decision making in such multi-person situation
where the outcome depends on the choice made
by every party, game theory is a natural choice.
Researchers in supply chain management now
use tools from game theory and economics to
understand, predict, and help managers to make
strategic operational decisions in complex multi - agent supply chain systems.
Loosely speaking, game theory models situations
where players make decisions to maximize their own
utility, while taking into account that other players
are doing the same and that decisions made by players impact each others utilities. There is a broad
division of game theory into two approaches: the
cooperative and the non-cooperative approach.
These approaches, though different in their theoretical content and the methodology used in their analysis, are really just two different ways of looking at
the same problem. To quote the words of Aumann:
‘‘the game is one ideal and the cooperative and non -cooperative approaches are two shadows’’
}

%It is well known that profitability of any firm depends upon the effectiveness of its Supply Chain (SC) which led the researchers to study SC coordination and information sharing among the members of SC 

Supply chains are complex systems that involve multiple agents at multiple echelons. These agents compete and/or collaborate with each other to acquire the maximum possible market share at `good' prices. The agents look for collaborative opportunities to provide better quality service thereby attracting more customers resulting in  enhanced individual performance  (for example, the collaboration of Exxon and Mobil (1999)), while others compete with each other if they find it beneficial  (Orange and SFR mobile networks).

In the handbooks in Operations Research,~\cite{chen2003information}, the authors discuss the importance of coordination and 
information sharing among the coordinating members on the effectiveness of the supply chain (SC). Cooperative game theoretic tools facilitate a systematic  study of the interactions among the agents of SC and in identifying the optimal strategies~(e.g., \cite{arshinder2011review}, \cite{thun2005potential}).

%We study the interplay between cooperation and competition, where the agents have an incentive to cooperate in order to enhance performance and attract more customers using tools of cooperative game theory. For example, they can collaborate by sharing resources, developing joint marketing campaigns, or negotiating with the agents as a single unit or compete with each other.

We examine the interplay between cooperation and competition in a two-echelon SC, with two manufacturers at the lower echelon and a single supplier at the upper echelon. Customers purchase the final product from the manufacturers, who in turn obtain the required raw materials from the supplier. Customers choose to buy the product from one of the manufacturers based on factors such as the quoted price, the reputation of the entities involved, the importance of the product (essentialness), etc. %Thus, the market of the customers is segmented between the two manufacturers. 
They may also decide not to buy the product from any of the manufacturers. The manufacturers compete with each other to attract `good' amount of customer base at `good' prices and rely on supplier for the raw material; the supplier at the upper echelon quotes a per-unit price, and the manufacturers respond to it by deciding whether or not to operate based on the price of raw materials,  production costs, demand response of the customers, etc. When a manufacturer decides to operate, it quotes a per-unit price of the final product to the customers.

Each agent can choose to collaborate with some or none of the other agents, not operate, select an agent for a particular service (such as supply of raw materials), or quote an appropriate price. The paper aims to find `optimal' pricing  and collaborative strategies of various agents using sophisticated game theoretic tools.

%The utility of the agents depends on the various factors such as the essentialness of the product, price quoted by the other agent, etc. 

When all the agents of the SC choose their actions simultaneously, we found that the agents do not operate at the resulting Nash equilibrium (NE). On the other hand, when the supplier leads by quoting a price for the raw material, and  subsequently the followers, the manufactures and the customers respond, the resultant Stacklberg (SB) game has an equilibrium at which all the agents operate. In fact, most of the papers in SC literature consider such a SB game~(e.g.,~\cite{zhao2023game} \cite{zheng2021willingness}). We consider the same to model the competition between lower and upper echelons.

\ignore{
To enhance the performance, any manufacturer can operate alone or can collaborate with the other manufacturer or, with the supplier, or with both of them. When any subset of the manufacturers operates without a supplier, the subset procures the raw material at the quoted price; when the subset collaborates with the supplier, they operate as a single unit, quote one price directly to the customers, and share the revenue generated amicably. When both manufacturers operate together, they choose the best among them for any aspect (e.g., reputation, production capacity). For example, if manufacturer 1  has a better reputation, the product would be projected as the product of manufacturer 1. 

Therefore, any agent can choose to collaborate with some or none of the other agents, not operate, select an agent for a particular service (such as supply of raw materials), or quote an appropriate price. As a result of the choices made by various agents in the system, each agent derives some revenue/share. The agents are selfish and aim to maximize their individual revenue/share, which drives their choices, including their collaboration attempts.
}

The cooperative game theoretic studies majorly focus on the stability of grand coalition and  further on scenarios where the worth of a coalition depends just upon its members (\cite{li2023pricing}, \cite{zheng2021willingness}). But the reality is far from such a study -- many a times grand coalition is not stable and further the worth/utility 
of the cooperating agents depends upon the arrangement of agents outside the coalition. 
The main contribution of this paper is to analyse SCs  considering  these realistic aspects -- to the best of our knowledge such a study is not considered in the existing SC based literature.  

%We analyse the concept of ``stable operating configurations" in two-echelon supply chains and study the pricing strategies and other decision choices made by these agents at equilibrium.

When the product is essential and the customers are (almost) indifferent to the manufacturers, the grand coalition is not stable. It is actually the  vertical cooperation between the supplier and one of the manufacturers  that results in stable configuration. More interestingly only the collaboration with  
the weaker manufacturer (weak in terms of market power) is 
stable; the stronger manufacturer is alone and has  to compete with the collaborating pair. No other attribute of the manufacturers contributes to this stable choice.  Even more interestingly, when the manufacturers are of  comparable market strengths, no collaboration is stable.

The model is described in Section~\ref{sec:model}, while the partition form game is discussed in Section~\ref{sec_coalition_formation_game}.
The two manufacturer SC is anlysed in Section~\ref{two_agent_supply_chain}. Some of the technical proofs are in Appendices and some are in technical report \cite{TR}.

%The remainder of this paper is organized as follows. A brief survey of the related literature is provided below, followed by the modeling details of the supply chain in Section~\ref{sec:model}. For the benchmarking purposes, we study the supply chain with the single manufacturer and supplier in Section~\ref{two_agent_supply_chain}. Next, we introduce some basic preliminaries about the partition form game, and different stability concepts in section~\ref{sec_coalition_formation_game}. We analyse the two manufacturer and single supplier SC in Section~\ref{sec_sing_Supp_two_manu}. We show the stability results of different coalitions in Section~\ref{sec:stability_results}. Throughout the paper, references to the appendix (primarily for proofs) point to the `additional material' document referred separately.

\input{Survey}

}
\vspace{-5mm}
\section{Introduction}
\vspace{-2mm}
Supply chains are complex systems that involve multiple agents at multiple echelons. These agents compete and/or collaborate with each other to acquire the maximum possible market share at `good' prices. The agents look for collaborative opportunities to provide better quality service, thereby attracting more customers, resulting in enhanced individual performance, while others compete with each other if they find it beneficial  (e.g., in 2016, Walmart teamed with \href{https://corporate.jd.com/aboutUs}{JD.com} to compete with Amazon and Alibaba in China). \ignore{Another example to Amazon and Walmart) motivate the co-existence of cooperation and competition in supply chains is (see \cite{li2023pricing}) that in China, Huawei has many suppliers of core components for its mobile phones and cooperation among these suppliers is incentivized to capture the competitive advantages of 
Apple and Samsung.}

Handbooks in Operations Research,~\cite{chen2003information}, discusses the importance of coordination on the effectiveness of the supply chain (SC). Cooperative game theory facilitates a systematic study of these interactions among the agents of SC~(e.g., \cite{arshinder2011review,thun2005potential,nagarajan2008game}). 

We examine the interplay between cooperation and competition in a two-echelon SC, with two manufacturers at the lower echelon directly facing the customers, and a single supplier at the upper echelon. %Customers purchase the final product from the manufacturers, who in turn obtain the required raw materials from the supplier.
Customers choose to buy (or not buy) the product from one of the two manufacturers based on factors like the quoted price, the reputation of the entities involved, the importance of the product (essentialness), etc. %Thus, the market of the customers is segmented between the two manufacturers. 
%hat 
The manufacturers compete with each other to attract `good' amount of customer base at `good' prices and rely on the supplier for the raw material. The supplier at the upper echelon quotes a per-unit price for raw materials to the manufacturers, and, the latter respond by either quoting a price of final product to the customers or by deciding not to operate; the choice of manufacturers also depends upon the production costs, demand response of the customers, etc. %, it quotes a per-unit price of the final product to the customers.

%Each agent can choose to collaborate with some or none of the other agents, not operate, select an agent for a particular service (such as supply of raw materials), or quote an appropriate price. 
This paper aims to find `optimal' pricing and collaborative strategies of the agents using sophisticated cooperative game theoretic tools. %similar to~\cite{nagarajan2007stable,nagarajan2009coalition}.
Majority of these games (e.g.,  in~\cite{li2023pricing,zheng2021willingness})  focus on the stability of the grand coalition and further on scenarios where the worth of a coalition depends just upon its members. But many times, the grand coalition may not be stable, and further, the worth of the cooperating agents may depend upon the arrangement of agents outside the coalition. Such games are referred as partition form games, and a recent thesis, \cite{shikshathesis},  provides a comprehensive summary of these games (see also~\cite{aumann1974cooperative}).

\Rev{
In any real-world SC, the revenue or the worth generated (for example) by a supplier, when all the manufacturers collaborate (i.e., operate as a single unit) will obviously be different from that in a scenario where the manufacturers also compete among themselves. Thus partition-form based study is essential to capture the frictions in SC. }

The first main contribution of this paper is to capture the above realistic aspects in an SC by modeling it as a partition form game and deriving the ingredients of the same -- to the best of our knowledge, none of the papers in SC literature consider this. We further consider that the agents in any coalition operate together as a single unit by pooling the best resources from each partner; furthermore, the possibilities of vertical and/or horizontal cooperation are also explored.
%resources

%-- to the best of our knowledge such a study incorporating alliance formation among the SC agents including both vertical and horizontal cooperation when the agents cooperate by operating together and by pooling their resources and then analysing the stable coalition structure is not considered in the existing SC based literature.  

The exhaustive partition-form game based study resulted in some interesting insights. When the product is essential, and when the customers are (almost) indifferent to the manufacturers, the grand coalition is not stable. It is actually the vertical cooperation between the supplier and one of the manufacturers that results in a stable configuration. More interestingly, only the collaboration with the stronger manufacturer (strong in terms of market power) is stable -- no other attribute of the manufacturers makes a difference (when their reputation among the customers is almost the same); the weaker manufacturer operates alone and competes with the collaborating pair. Even more interestingly, no collaboration is stable when the manufacturers are of comparable market strengths.

When the supplier leads by quoting a price, there exists a Stackelberg equilibrium (SBE) at which the agents operate, in contrast, a scenario where all the agents make a simultaneous move results in a  Nash equilibrium at which none of the agents operate. 
In fact, majority of the literature in SC seems to understand this at some level and considers the Stackelberg (SB) framework (e.g.,~\cite{li2023pricing,zheng2021willingness}). 
The SB framework  significantly favours the supplier -- the supplier enjoys a huge fraction of the revenue generated, which becomes even higher with the
 competition at the lower echelon. 
 
 %We consider the same to model the competition between lower and upper echelons.

%It is also observed that in all the configurations manufacturers get negligible share with respect to the supplier. This can be attributed to the supplier being the leader of the Stackelberg (SB) game in which the supplier leads by quoting a raw material price and the manufacturers which are the followers respond , the resultant game has an equilibrium at which all the agents operate and thus we consider such a SB game framework to model competition between supplier and manufacturers.  In fact, most of the literature in SC consider such a SB game~(e.g.,~\cite{zhao2023game,zheng2021willingness}). We consider the same to model the competition between lower and upper echelons.
%Further, when all the agents of the SC choose their actions simultaneously, we found that the agents do not operate at the resulting Nash equilibrium (NE).

%

The model is described in Section~\ref{sec:model},  the partition form games are in Section~\ref{sec_coalition_formation_game}, and 
the  SC is analysed in Section~\ref{sec_sing_Supp_two_manu}.  \TR{Some of the   proofs are in Appendices, and others are in technical report \cite{TR}.}{{ All the proofs are in Appendices.}}
\medskip

\noindent {\bf Literature Survey:}
There is a vast literature that studies the scope of SC coordination. Almost all the studies consider contract based cooperation (e.g.,~\cite{cachon2003supply} and subsequent papers). There are few strands of literature that study coalition formation ideas, where the agents are bound without any such  enforcement, because they find it beneficial to do so. Important and relevant papers in this category are \cite{nagarajan2009coalition}, \cite{zheng2021willingness} and \cite{li2023pricing} etc.

In \cite{zheng2021willingness}, authors study a two-echelon sustainable SC with two manufacturers and a single supplier; they neglect the partition-form aspects by defining the worth of a coalition to be the pessimal worth, the minimum  (anticipated worth) that the said coalition  can generate irrespective of the arrangement of the left-over agents.  However, if a coalition (not currently operating) has to block/oppose an operating configuration (the set of operating coalitions and revenues/shares of all the  agents of SC), the coalition should anticipate to derive a better revenue than the sum total revenue that its members are currently deriving. In other words, the anticipation is required only for estimating the worth of future (or blocking) coalition (as upfront it is not sure of the retaliatory actions of the others), and not for the worth currently derived (as considered in \cite{zheng2021willingness}).
We consider pessimal worth as the anticipated worth of blocking-coalition, while the (current) worth(s) in any operating configuration is derived by solving an appropriate game or optimization problem. 
%in the operating configuration. 
%There are many anticipatory rules including pessimal rule which are relevant for different scenarios (see a recent study in  \cite[Chapter 2]{shikshathesis}), and we consider the more widely used pessimal rule. 
%We omit the results using other anticipation rules due to lack of space. 

\ignore{  Partition form games capture the fact that the worth of a coalition depends upon the arrangement of the players outside the coalition and thus the members outside the coalition can arrange themselves in any possible way which will determine how mmuch worth the coalition will gain.}

Another recent study in \cite{li2023pricing} considers two assemblers (like manufacturers in our study) and many irreplaceable suppliers,  where the second assembler only competes for customer base and has its own  set of suppliers --  hence this study is not directly comparable to ours. However,  the study again neglects the partition form nature of the game -- the worth of any coalition is defined just based on its size.
%, but it actually also depends on how the remaining suppliers/assembler   arrange themselves (see results of \cite[section 4.1]{li2023pricing}). 
As already argued, when one neglects the inherent  partition form nature, the results could be misleading -- it would be interesting to analyze the SC of \cite{li2023pricing},  after incorporating partition form aspects. % (probably grand coalition may not be stable). 

%
%For example, \cite{cachon2003supply} study supply chain coordination through contract mechanisms.  

In \cite{nagarajan2009coalition}, authors study coalitional stability considering partition-form aspects.  However, as is mentioned in the same paper, they do not consider the worth of the coalition (based on partition), rather assume that all the players in the coalition to agree to quote a common (best) price. This (rather restricted) assumption facilitates in the derivation of the revenue generated by a single agent in any partition and thereby study  the stability aspects. In our study, we derive the utility of any coalition depending upon the   partition and then consider stability aspects based on the division of that worth and the anticipated utility of the `opposing coalition'. 

\ignore{
In \cite{nagarajan2009coalition}, authors study coalitional stability in a two echelon supply chain with $n$ suppliers at upper echelon and an assembler at lower echelon and also introduce some dynamic notitions of stbility such as Largest Consistent Set and Equilibrium Process of Coalition Formation.
The contrast between our study and that in \cite{nagarajan2009coalition} is that firstly they do not consider the fact that agents operate together when they form coalitions, they just agree upon a common price and thus they can determine the utility of each agent in a partition/coalition structure while in our study we include the costs related to the system and also consider the fact that agents operate together by pooling their resources and thus we  only know the worth of a coalition in a partition and we have to thus talk about the division of worth in a coalition for understanding stability. Also our study incoorporates the case of vertical cooperation among supplier and manufacturer. Although these factors were not accounted in \cite{nagarajan2009coalition} but their study provides us useful insights regarding coalitional stability in dynamic sense.
}

\ignore{
None of the existing strands of SC literature, including the above studies, to the best of our knowledge considers the fact that the worth of any coalition depends upon the arrangement of remaining members; such a game is known as a partition form game and the results considering this realistic aspect can be drastically different.  In all the above mentioned papers  the grand coalition turns out to be stable, however 
this is not always the reality.

These games are examples of partition form games, where the worth of any coalition depends on the arrangement of opponents. This framework is more realistic to practical applications  than the previous literature.
Partition form games and their stability aspects was extensively studied by authors in \cite{singhal2021coalition} in queuing systems which led to many interesting practical insights. The significance of this fact motivates us to study this aspect in SC and understand the insights by the consideration of this fact. As a result we interestingly find that Grand Coalition is not stable under certain conditions which contradicts the existing literature in SC where the stability of GC is proposed.}

\vspace{-5mm}
\section{Model} \label{sec:model}
\vspace{-2mm}
We consider a two-echelon supply chain (SC), with two manufacturers at the lower echelon and a single supplier at the upper echelon. The customers purchase the final product from the manufacturers depending upon various factors (price and the essentialness of the product, reputation of the manufacturer, etc.); while manufacturers obtain the required raw materials from the supplier depending upon their own customer demand and the price quoted by the supplier, production cost, etc. 

Any manufacturer can operate alone, or can collaborate with the other manufacturer, or with the supplier, or with both of them -- when both the manufacturers operate together, they choose the best among them for any aspect (e.g., influence, reputation, production capacity), while the supplier and manufacturer pair quote one price directly to the customers. 

We examine the impact of the interplay between cooperation and competition in the  above SC using a cooperative game-theoretic framework; in particular, our research aims to explore the potential for horizontal (within the same echelon) and vertical cooperation (across echelons) in an SC.
 We now describe the ingredients of this study in detail.

\vspace{-2mm}
\subsection{Coalitions and Partitions}
\vspace{-2mm}
All the agents or a  subset of them can operate together by forming coalitions. Basically, the agents within a coalition make joint decisions to generate a common revenue while facing competition from other coalitions or agents. One may have more than one coalition operating in the system. 
Any partition,  say~$\P = \{\C_1, \cdots, \C_k \}$, represents the operating arrangement of agents into distinct coalitions and satisfies the following:
\begin{align*}
\cup_m \C_m = \{M_1,  M_2, S\} ,  
\mbox{and } \C_m \cap \C_l = \emptyset \ \mbox{  if } m \ne l. 
\end{align*} 
The goal of this paper is to study the interactions between these coalitions and predict the emergence of stable partition(s) (if any). Prior to this, we need to understand the criteria for declaring a partition stable. Even prior to this, we need to derive the revenues generated by various coalitions in each partition -- we refer to these revenues as the worths of the coalitions, a term commonly used in the cooperative game theory literature~\cite{singhal2021coalition,shikshathesis,aumann1974cooperative}. The stability concepts are discussed in Section \ref{sec_coalition_formation_game}, while the worths related to various partitions are derived in various sections. For now, we discuss important and interesting partitions and coalitions.

When two manufacturers operate together, we have a coalition $\M = \{M_1, M_2\}$, with horizontal cooperation (HC) at the lower echelon. When the supplier and a manufacturer operate together, we have a coalition with vertical cooperation (VC), e.g., $\V_i = \{M_i, S\}$. When any agent operates alone, we have a coalition with a single player, e.g.,  $\Mi = \{M_i\}$ or $\S = \{S \}$. When all the agents operate together as in a centralized SC, we have a grand coalition (GC), represented by $\G = \{S, M_1, M_2\}$.

The partition $\GC = \{ \G\}$ where all the agents operate together is referred to as the GC partition. While  we have an ALC partition  $\ALC = \{\S,\M_1,\M_2 \}$, when all the agents operate alone. We also have VC (vertical cooperation) partition $\VC_i = \{ \V_i, \M_{-i} \}$ and HC (horizontal cooperation) partition $\CHC = \{ \S, \M \}$.

The worth, the revenue generated by any coalition must be \textit{shared appropriately among its members and this payoff division also influences the stability aspects} \cite{aumann1974cooperative,singhal2021coalition,shikshathesis}. Further, departing from a majority of the literature \cite{li2023pricing,zheng2021willingness}, the worth of any coalition depends upon the operating partition leading to a partition form cooperative game \cite{singhal2021coalition,shikshathesis}; as already mentioned, the analysis of such games is significantly complicated, and the results obtained by omission of this dependency can be misleading.

In all, as a result of the choices made by various agents in the system, each agent derives some revenue/share. The agents are selfish and aim to maximize their individual revenue/share, which drives their choices, including their collaboration attempts; the paper precisely works in identifying the \textit{`stable configurations' -- the partitions and the corresponding payoff divisions.}
\vspace{-2mm}
\subsection{Market Segmentation}
\label{sec_market_seg}
\vspace{-2mm}
In an SC, the manufacturers satisfy customers' demands and rely upon suppliers for raw materials or intermediate products. Customers choose one manufacturer (or none) based on the price, reputation, loyalty, and other factors. The demand segmentation is also influenced by the essentialness of the product, which we capture using a parameter $\gamma$ and a cross-linking factor $\varepsilon$ that also captures the customers' affinity to switch loyalties. When the manufacturers do not operate together, the market is segmented between them based on their selling prices~$p_i$ and essentialness factors~$(\gamma, \varepsilon)$ as in equation~\eqref{Eqn_Demand_Function} given below. This is inspired by the models commonly used in SC literature (see, e.g.,~\cite{zheng2021willingness,li2023pricing}. With $y^+ = \max\{0, y\}$, the demand derived by $M_i$ equals:
\begin{align}
 D_{\sMi} & =    \bigg ( {\dbar}_{\sMi}
 - \alpha_\sMi p_{i} +   \varepsilon \alpha_{\sM_{-i}}  p_{{-i}}\bigg )^+ ,  \label{Eqn_Demand_Function} \\
  \mbox{ with } \alpha_{\sMi} &:=  \tilde\alpha_{\sMi} (1-\gamma) , \mbox{ where, } \nonumber
\end{align}
\begin{itemize}
    \item $\dbar_{\sMi}$ is the dedicated market size of~$M_i$,
    \item $\alpha_\sMi p_{i}$ is the fraction of demand 
    lost by $M_i$ due to its price $p_i$, sensitized by parameter $\alpha_\sMi$,
    \item The essentialness factor $\gamma$ dictates the sensitivity of price $p_i$ on demand -- for example, when $\gamma \approx 1$, the product is highly essential, and the customers are insensitive to price,
    \item $\varepsilon \alpha_{\sM_{-i}}  p_{{-i}}$ is the fraction of customer base of $M_{-i}$ that rejected $M_{-i}$ and shifted to~$M_i$, 
    \item The demand is positive as long as the term inside $(\cdot)^+$ is positive; else, the demand is zero.
\end{itemize}
%The first term in the demand function, $\dbar_{\sMi}$, represents the maximum potential market demand for $M_i$'s product. The second term, $-\alpha_\sMi p_{i}$, represents the fraction of demand that is lost by $M_i$ due to its price. The third term, $\varepsilon \alpha_{\sM_{-i}} p_{{-i}}$, represents the fraction of $M_{-i}$'s customer base that is attracted by $M_i$. The demand function is positive as long as the total demand for $M_i$'s product is greater than zero. If the total demand is less than zero, then the demand function is zero.

The product is essential, either when $\gamma \approx 1$ or when $\varepsilon \approx 1$ and then almost all the customers buy the product (for these parameters, observe $D_{\sMo}  + D_{\sMt} \approx \dbar_\sMo + \dbar_\sMt$, the total market size). When $\varepsilon \approx 1$, the customers buy the product from one or the other manufacturer (need not be loyal); otherwise, they prefer to buy from their own manufacturer (are loyal).
Generally, the sum of demands of both manufacturers is strictly less than the total market size, and the gap depends upon the essentialness parameters $(\gamma, \varepsilon).$

%On the other hand, $\varepsilon \approx 1$ also represents essentialness but now the customers are willing to switch loyalties -- now each customer buys the product from one or the other manufacturer. In this case, $d_\sMi$ is the demand attracted by $M_i$, and $\dbar_\sMi$ can be interpreted as the dedicated customer base of $M_i$. Additionally, 

\ignore{
\begin{eqnarray}
\left(1 - \alpha_i(p_i) \right) \left(\alpha_{-i} p_{-i} + (1 - \alpha_{-i} p_{-i})   \frac{p_i}{p_i + p_{-i}} \right)   \\
= \frac{p_i}{p_i + p_{-i}} - p_i \alpha_i \frac{p_i}{p_i + p_{-i}} + \alpha_{-i} p_{-i} \frac{p_{-i}}{p_i + p_{-i}} + o(p_i^2)
\end{eqnarray}
\vspace{-2mm}
\subsection*{Illustrative Example} 
\vspace{-2mm}
Before we proceed, let us present a scenario that can be modeled by a demand function \eqref{Eqn_Demand_Function}. 
Consider two manufacturers with dedicated customer bases $d_1$ and $d_2$, respectively, and a common pool of $2d_c$. The total market size available to either manufacturer is $d_\sM = d_1 + d_2 + 2d_c$.
Customers in the dedicated fraction $d_i$ buy products only from $M_i$, and only if they are comfortable with the price $p_i$ quoted by $M_i$. This fraction can be modeled by $d_i - \alpha_{i} p_i$.
Customers in the common pool $2d_c$ can buy from either $M_1$ or $M_2$ based on prices, the essentialness factor $\gamma$, and the cross-linking factor $\varepsilon$. As in \cite{zheng2021willingness}, the proportion of the common pool that buys from manufacturer $M_i$ can be modeled using $d_{c_i} = d_c - \alpha_{i} p_i + \varepsilon \alpha_{{-i}} p_{-i}$. 
 Then, the demand attracted by manufacturer~$M_i$ is :
\begin{align*}
\tus{\dbar (p_i, p_{-i})} &= 
 d_i - \alpha_{i} p_i +  (d_c - \alpha_{i} p_i +  \varepsilon  \alpha_{-i} p_{-i}) \\
 &= \dbar_{\sMi} - 2 \alpha_{i}  p_i  + \epsilon \alpha_{-i} p_{-i}.    
\end{align*}
This is of the form as in  \eqref{Eqn_Demand_Function}, and $d_\sMi = d_i + d_c.$
In this case, the total market size that can be attracted by the joint coalition $\M$ is $\dbar_\sM = d_1 + d_2 + 2d_c$. 
The common pool will definitely buy a product when either the cross-linking is good ($\epsilon \approx 1$) or the essentialness is high (when $\gamma \approx 1$).
In fact, when $\gamma \approx 1$, $d_{c_i} = d_c$ and the common pool is equally divided. In this example, in some sense the cross-linking factor is also linked to essentialness in a sense: when the product is essential only for the common pool (which can now be represented by $\gamma \approx 1$), the customers from this pool of size $2d_c$ buy either from $M_1$ or $M_2$.

  }
 
\noindent{\bf HC Coalition:}
When both manufacturers operate together, as in $\M $ or $\G$, they can potentially attract both customer bases. 
Further, for manufacturing purposes, the coalition uses the methods of the manufacturer with the lowest manufacturing cost; thus, its per-unit manufacturing cost is $C_\sM =  \min_{M_i \in \C} C_{\sMi}$, where $C_\sMi$ is the per-unit manufacturing cost of the manufacturer $M_i$. As the best of the two capabilities are utilized, and as the customers are aware of it, we assume the reputation of the coalition equals that of the best. In all, we assume the demand function of coalition with horizontal cooperation to be:
\begin{eqnarray}
\label{Eqn_Both_demand}
D_{\sM} &=&  {\dbar}_{\sM}- \alpha_{\sM}p ,
\mbox{ with }   \alpha_\sM := \min \{ \alpha_\sMo, \alpha_\sMt\},  \nonumber  \\
 {\dbar}_{\sM} &=& {\dbar}_{\sMo} + {\dbar}_{\sMt}.
\end{eqnarray}
 There is obviously no cross-linking (shift of customers from one manufacturer to the other), or basically, the customers have no choice. In some cases, this can be fatal to the system, as the customers can get discouraged by the unavailability of options. The product may not appear essential anymore, and the customers may find solace in other related products.
 We observe this phenomenon has significant influence on stability results  of sections \ref{sec_gc_two_agents}-\ref{sec_chc_two_agents}.

 \ignore{
\newpage

 When  two manufacturers operate together in a coalition $\C$ (either $\C = \M$ $\C = \{S, M_1, M_2\}$), they quote a common price~$p_\sC$ by projecting themselves to the customers as a single unit mainly belonging to the manufacturer with best reputation, indexed by $\bst(\C) := \arg \max_{M_i \in \C} \alpha_{\sM_i}$. This allows them to attract the customers of each of their members. The market demand attracted by the coalition also depends on other operating coalitions in the same way. Specifically, if $\P = \{\C_1, \C_2, \cdots, \C_k\}$ is the operating partition, then the market attracted by a coalition $\C_i$ is given by: 
 \begin{align}
\label{Eqn_Demand_Function_coalition}
 D_{\sC_i}  & =    \left ( {\dbar}_{\sC_i}-\tilde\alpha_{\sC_i}(1-\gamma)p_{\sC_i}+\varepsilon \tilde\alpha_{\sC_{-i}} (1-\gamma)p_{\sC_{-i}} \right )^+, 
 \end{align}
 where, 
 \begin{align}
 {\dbar}_{\sC_i} &=  \sum_{k \in \sC_i} \dbar_{\sM_k }   \ 
 {\tilde \alpha}_{\sC_i}  \  :=  \  {\tilde \alpha}_{\sM_{\bst(\sC_i)}}, \mbox{ for any coaltion } \C_i,  \nonumber \\ 
 \bst(\C)  &:=
  \arg   \min_{M_k \in \C} \tilde\alpha_{\sM_k} ,  \mbox{ and, }  \varepsilon_{\sC_i, \sC_j} := \varepsilon_{\bst(\sC_i), \bst(\sC_j) }.  \nonumber
\end{align}
The combined market size attracted by coalition~$\M$  is given by (in similar lines as in \eqref{Eqn_Demand_Function}):
\begin{equation} \label{Eqn_Both_demand}
D_{\sM} =  {\dbar}_{\sM}- \alpha_{\sM}p ,
\mbox{ with }   \alpha_\sM := \min \{ \alpha_\sMo, \alpha_\sMt\},
\end{equation}
where~$\dbar_\sM$ is the maximum possible combined market size that can be captured.  We assume that the combined market size that $\M$ can attract is greater than the maximum market size that either manufacturer can attract, but less than the sum of the two market sizes:
$$\max\{\dbar_\sMo, \dbar_\sMt \} \le \dbar_\sM \le \dbar_\sMo + \dbar_\sMt.$$
}

\ignore{
\vspace{-2mm}
\subsection{Operating choices of a Coalition} 
\label{sec_operating_choices}
\vspace{-2mm}
The manufacturer choice $\bst(\C)$, defined in Equation \eqref{Eqn_Demand_Function_coalition}, is the manufacturer with the best reputation among the manufacturer-members of the coalition $\C$. This choice is used to project the coalition to the public as a single unit in order to attract the maximum possible demand. For manufacturing purposes, the coalition uses the methods of the manufacturer with the lowest manufacturing cost, $\arg \min_{M_i \in \C} C_{\sMi}$, where $C_\sMi$ is the per-unit manufacturing cost of the manufacturer $M_i$.

We now discuss the sequence of actions of agents in our model. Our model uses a Stackelberg game to represent the interactions between a supplier and manufacturers. In a Stackelberg game, the supplier acts as the leader and sets the per-unit wholesale price $q$ of raw materials. The manufacturers then act as the followers and decide whether to operate by setting a selling price $p$ for the product or not to operate, depending on their anticipated profit given the supplier-quoted price.
After procuring raw materials, the manufacturers produce the product and quote a selling price $p$ to customers. Customers, in turn, select manufacturers based on selling prices and the reputations of manufacturers and their selected suppliers. 

Our research aims to explore the potential for horizontal (within the same echelon) and vertical cooperation (across echelons) in the supply chain using cooperative game theoretic tools such as coalition formation games. We investigate the possibilities for different coalitions among the manufacturers and suppliers within the supply chain.
We now define the ingredients of the game precisely. 
}

\vspace{-2mm}
\subsection{Costs, Actions and the Utilities}\label{sec_cost_util}
\vspace{-2mm}

Any agent, supplier, manufacturer, or coalition has a fixed cost of operation. Therefore, if the agent/coalition does not generate sufficient profit, it incurs negative revenue and can choose not to operate. Let $n_o$ represent the choice of not operating. The utility of any agent/coalition is 0 when it chooses $n_o$.

The supplier can decide not to operate, or can quote a price~$q \in [0, \infty)$. Thus, the  action set of supplier when operating alone is $ \Aset_\sS  := \{n_o\} \cup [0, \infty)$, and its action $a_\sS  \in \Aset_\sS$. 
%\tus{(see Figure~\ref{fig:model})}.
Similarly, when manufacturer $M_i$ decides to operate alone, it quotes a selling price $p_{i} \in [0, \infty)$. Thus, the action and the action  set of manufacturer $M_i$ is $a_\sMi \in \Aset_\sMi := \{n_o\} \cup [0,\infty)$.

\begin{figure}[ht]
    \centering
\includegraphics[trim = {.7cm 0.3cm .7cm 0.3cm}, clip, scale = 0.23 ]{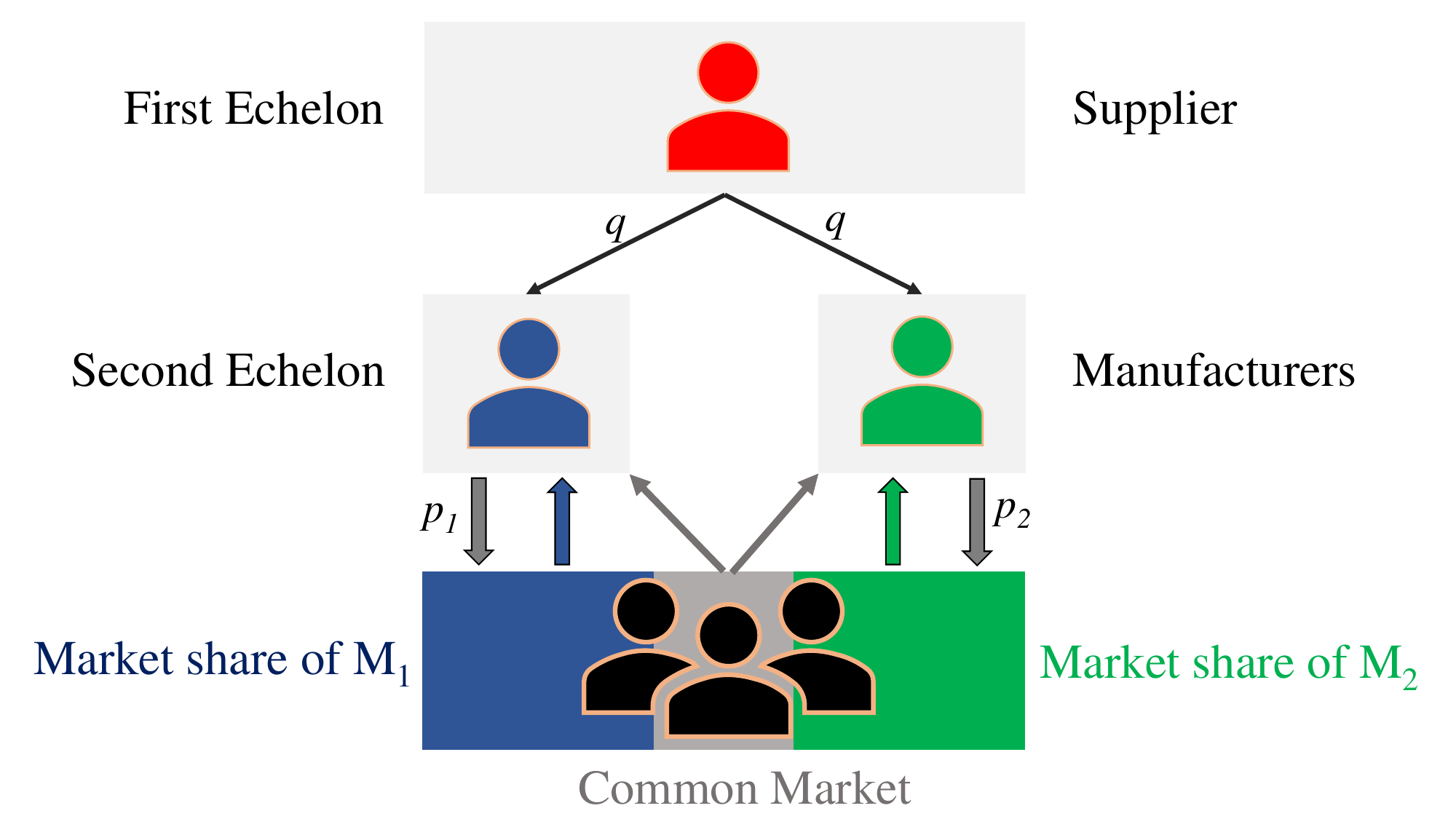}
{\caption{System model, when all agents operate alone.} 
\label{fig:model}}
\end{figure}

In a VC coalition $\V_i = \{S, M_i\}$, the coalition sets a price $q$ for supplying raw materials to the manufacturer outside the coalition, and sets a price $p_i$ directly to customers by jointly producing the final product.
In the grand coalition $\G = \{S, M_1, M_2\}$, which involves vertical and horizontal cooperation, the coalition directly quotes a price $p$ for customers and makes a combined effort to produce the final product. The respective actions are represented by $\a_\sV$ and $\a_\sG$ and the action sets by $\Aset_\sV$ and $\Aset_\sG$ (defined  as before).

\subsubsection{Utilities in ALC  Partition}
We begin with describing the utilities of various agents when all of them operate alone, i.e., when the partition is $\ALC$ (see Figure \ref{fig:model}). Let~$\a := (a_\sS, \a_\sM)$ represent the actions   of all the agents (the supplier, and both the manufacturers), where~$\a_\sM :=  (a_\sMo, a_\sMt)$.

\noindent {\bf Manufacturers' Utility:} 
The utility of manufacturer $M_i$ is zero either if it chooses not to operate or if the supplier does not operate. Otherwise, utility is the total profit gained minus the operating cost, where the former is the product of the demand attracted~$D_\sMi $~\eqref{Eqn_Demand_Function} and the profit gained per-unit:

\vspace{-4mm}
{\small \begin{equation}\label{Eqn_UtilityofManufacturer}
U_\sMi (\a   ) = \left ( D_{\sMi}( \a_\sM) (p_i - C_{\sMi} - q) \F_\sS - O_\sMi \right ) \F_\sMi ,
\end{equation}
}where~$C_{\sMi}$ is the per-unit production cost incurred by~$M_i$,~$O_\sMi$ is the fixed operating/setup cost, $\F_\sC = \indc{a_\sC \ne n_o}$ represents the flag that coalition $\C$ operates, and~$q$ denotes the wholesale price quoted by   supplier. 

\noindent {\bf Suppliers' Utility:}
The demand for the supplier's raw materials (at higher echelon) percolates from the lower echelon (manufacturers),  based on the choices of the manufacturers. This dictates the utility of the supplier, which is non-zero only if the supplier and at least one of the manufacturers operate. In all, the utility of the supplier $S$ when it operates alone equals:

\vspace{-3mm}
{\small\begin{eqnarray}\label{UtilityofSupplier}
    U_{\sS} (\a) = \left ( \left ( \sum_{i=1}^{2} D_\sMi (\a_\sM)   \F_\sMi  \right ) (q- C_\sS ) - O_\sS  \right ) \F_\sS ,
\end{eqnarray}}%
where~$C_{\sS}$ is the cost for procurement of a bundle of raw material required for producing one unit of product and~$O_{\sS}$ is the fixed operational cost of the supplier.

\subsubsection{Utility in General Partition}\label{subsec:utility_general_partition} In a general partition, the utility of a coalition is defined as the sum of the utilities of all the agents within the coalition. As in equation \eqref{Eqn_Both_demand} and as described in the corresponding sub-section, the coalition utilizes the best agent for each feature. Also, any VC-based coalition directly quotes a price to the customers. Thus, for example, the action and utility of the grand coalition $\G$ are: 
\begin{align*}
a_\sG &\in \{p \in [0, \infty)\} \cup \{n_o\}, \text{ and} \\
U_\sG (a_\sG) &=\left ( ( \dbar_\sM  -  \alpha_\sG p )
(p - C_\sG ) -O_\sG \right )  \F_\sG, \mbox{ where, }
\end{align*}
\begin{itemize}
    \item $O_\sG$ is the combined operational cost, defined as $O_\sG = \min \{ O_\sMo, O_\sMt\} + O_\sS$.
    \item $C_\sG$ is the combined production cost, defined as $C_\sG = \min\{C_\sMo, C_\sMt \} + C_\sS$.
    \item $\alpha_\sG$ is the price sensitivity of the grand coalition, defined as $\alpha_\sG = \min\{\alpha_\sMo, \alpha_\sMt\}$.
\end{itemize}
It is important to note here that the agents in the grand coalition share the revenue generated ($U_\sG^* = \sup_{a_\sG} U_\sG$), and there is no price per item to be paid between any subset of them. 
The definitions of utilities for other partitions follow similar logic and will be discussed in the respective sections.

We conclude this section by making an important assumption (inspired by commonly made choices in practical scenarios):

\begin{enumerate}[{\bf A}.1]
    \item If any agent, either supplier or manufacturer, is indifferent between the action~$a = n_o$ and an $a \ne n_o$,   the agent prefers operating choices.
\end{enumerate}

We begin with studying an SC with a single manufacturer and supplier, which provides the basis and benchmark for analysing the more generic SC (with two manufacturers) of     section~\ref{sec_sing_Supp_two_manu}.
The stability concepts of partition-form games are  in section~\ref{sec_coalition_formation_game}.

%% file: Survey.tex
\subsection {Understanding \cite{nagarajan2007stable}}

Here the collaborating players are not operating together. Any coalition just agrees to use common price. As a result of it, you know the individual 
utilities themselves once the partition is given to you.  In other words they directly know  $\nu_i^\P$ for every player $i$  once $\P$ is the operating partition.

While in our game we know $\nu_\sC^\P$, the worth/utility of each coalition $\C$ and then discuss the division of worth among players of $\C$ and this for each $\C \in \P.$

Because of such a structure, they can define an individual preferring a partition to other   --       $\P' <_i  \P$    player $i$ prefers $\P$ over $\P'$ if $\nu_i^{\P'} < \nu_i^{\P}$.
or group/coalition $S$ preferring      $\P' <_S \P$  when 
$\P' <_i \P$ for all $i \in S$.

A partition $\P \to_S \P'$ if $\P'$ is formed from $\P$ when $S$ deviates. 

A partition  $\P$ is dominated by $\P'$ directly, represented by $\P < \P'$, if there exists an $S$ such $\P \to_S \P'$ ($\P$ leads to $\P'$ via $S$) and further $S$ prefer the new partition, $\P <_S \P'$.

This is like our $S$ blocking $P$ (but with a difference, in their  case the new partition is a part of the definition).

They also have indirect blocking  $\P \ll \P'$ --  there  is a sequence of coalitions  $S_1, S_2, ... S_k$  the corresponding sequence $\{S_l\}$ shifts, we have $\P$ indirectly  leading to $\P'$ along with domination, i.e., 
\begin{eqnarray*}
   \P = \P_0 \to_{S_1}  \P_1 ,   \cdots,   \P_{l-1} \to_{S_{l}} \P_l, \cdots,  \to_{S_k}  \P_k = \P' \\
\mbox{and at each $l$, }  \P  <_{S_1} \P_1, \cdots,  \P_{l-1}  <_{S_l} P_l ..., \P_{k-1}  <_{S_k} \P'
\end{eqnarray*}

{\bf Consistent Set} A collection of partitions ${\cal Y}$ is called consistent if every $\P \in {\cal Y}$ has the following property:

i) take any $S$  and  $\P'$ which results from coalition shift $P \to_S {\P'} $   
\\
 a ) either if $\P' \in {\cal Y}$ then the original $\P$ is not dominated y $\P'$, i.e.,   $\P \nless_S \P'$
\\
b)  or there exists $\P'' \in {\cal Y}$ such that $\P' \ll \P''$ and $\P \nless_S \P''$.

Their proposal solution via thee largest Consistent set..
\subsection{Understanding \cite{nagarajan2009coalition}}
 This is similar to their 2007 paper but here they consider a two echelon supply chain with an assember in one  echelon and n suppliers in another echelon. They consider 3 models : Supplier Stackelberg , Assembler Stackelberg and Vertical Nash. Here suppliers can form coalitions and again when they form coalitions they just agree upon a common per unit price but they operate seperately. Later they also talk briefly about alliance formation (membership) cost and alliance defection (friction ) cost and analyse some of their results using the same.